\documentclass[acmsmall,nonacm]{acmart}

\usepackage{tikz}
\usetikzlibrary{shapes.geometric, arrows.meta, positioning}
\usepackage[utf8]{inputenc}
\usepackage{amsmath}

\usepackage[linesnumbered,ruled,vlined]{algorithm2e}
\usepackage{enumitem}
\usepackage{tcolorbox}
\tcbuselibrary{breakable}
\tcbset{breakable}
\usepackage[htt]{hyphenat}
\usepackage{multirow}
\usepackage{hyperref}
\usepackage{booktabs}
\usepackage{array}
\usepackage{pifont}
\usepackage[section]{placeins}
\usepackage{listings}
\begin{document}

\title[Revisiting the Survival of Agent-Attributed Code]{Revisiting the Survival of Agent-Attributed Code: How Measurement Choices Decide the Verdict}

\author{Musfiqur Rahman}
\affiliation{%
    \institution{Concordia University}
    \city{Montr\'eal}
    \country{Canada}
}
\email{musfiqur.rahman@mail.concordia.ca}

\author{Mahi Begoug}
\affiliation{%
    \institution{Concordia University}
    \city{Montr\'eal}
    \country{Canada}
}
\email{mahi.begoug@concordia.ca}

\author{Emad Shihab}
\affiliation{%
    \institution{Concordia University}
    \city{Montr\'eal}
    \country{Canada}
}
\email{emad.shihab@concordia.ca}

\renewcommand{\shortauthors}{Rahman et al.}

\begin{abstract}
Studies of how long code from AI coding agents survives after it is merged reach conflicting verdicts: agent code is reported to be retained longer than human code, about as long, or modified sooner. Each study makes defensible measurement choices, but how much these choices decide the verdict is unknown. We revisit the survival of agent-attributed code in our earlier study: 854,139 lines that 2,802 merged pull requests (PRs) added to 215 repositories of the AIDev dataset. We vary nine measurement choices in five research questions and build a specification curve of 88 survival analyses, tested under three variance estimators, several observation windows, and permuted labels. For code from human-supervised agents, mostly GitHub Copilot, the choices change the size of the effect but not its direction. For code from the fully-autonomous agent Devin, they decide the verdict. Across repositories, 15 of 22 specifications report it modified sooner when lines are treated as independent, but only 3 when standard errors are clustered by repository, no more than permuted labels produce. Within repositories, no specification detects a difference. LLM intent classifiers agree with human labels about as well as two humans agree with each other, whereas keyword rules do not ($\kappa = 0.69$ to $0.72$ against $0.38$), and the two reach different verdicts on why agent code is modified. Humans make most modifications to agent code under every unit and window. Re-run on our data, five published designs each change their result with a single choice. Whether agent code survives longer than human code is therefore decided as much by the measurement as by the code. We derive an operationalization checklist for post-merge studies.
\end{abstract}


\maketitle
\section{Introduction}
\label{sec:introduction}

Coding agents such as GitHub Copilot~\cite{githubCopilot}, OpenAI Codex~\cite{openaiCodex}, Cursor~\cite{cursorAgent}, Claude Code~\cite{claudeCode}, and Devin~\cite{devinDocs} now open pull requests (PRs) in open-source projects, and many of these PRs are reviewed and merged like any other contribution~\cite{li2025rise}. Benchmarks such as SWE-bench~\cite{jimenez2023swe} measure whether an agent can resolve a real issue at the moment the code is generated. However, once the code is merged, the project has to maintain it, and maintenance accounts for a large share of the total cost of software, around 70\% in Boehm's analysis~\cite{boehm1984software}. Therefore, a small but fast-growing body of work studies the \textit{post-merge} behavior of code from coding agents, that is, what happens to the code after it is merged into a repository~\cite{gitclearCodingCopilot,xia2026these,sawada2026extent,popescu2026investigating,takerngsaksiri2026finishes,kraishan2026agents}. Several of these studies ask how long code from agents stays unmodified after merge, which we call its \textit{survival} or \textit{retention}, and compare it with code from human developers.

However, these studies reach different answers about the same kind of code. For example, Xia and Miller~\cite{xia2026these} track the lines added in 182 repositories and find that agentic code is terminated at a rate similar to human code but receives more corrective maintenance, that is, more modifications that fix defects. In contrast, Popescu et al.~\cite{popescu2026investigating} find that lines written by humans survive consistently longer than lines from every agent they study. Kraishan~\cite{kraishan2026agents} finds that code from some agents is changed less often after merge (churns less) than human code, and code from others at the human level. The studies also disagree on who maintains the code of agents. Takerngsaksiri et al.~\cite{takerngsaksiri2026finishes} find that most follow-up fixes of agent PRs come from the same agent, whereas Sawada et al.~\cite{sawada2026extent} find that human developers do most of the maintenance of files created by agents.

These studies use different data, but they also measure differently. Each study makes a set of measurement choices, and each choice is defensible on its own. More specifically, a post-merge study must decide:
\begin{enumerate}[label=(\roman*),leftmargin=*,nosep]
    \item how a line is followed through later commits, by its \textit{location} (its line number) or by its \textit{identity} (the same line, followed through moves and renames);
    \item which files are kept, for example whether \textit{vendored} code (third-party code copied into the repository) and \textit{generated} code (code produced by tools) are removed;
    \item what the unit of analysis is (a line, a \textit{hunk}, that is, a block of consecutive added lines, a file, or a PR) and how much each unit counts;
    \item what counts as a modification, for example whether an edit that only changes whitespace or renames a variable counts;
    \item how the intent of a modification is inferred, for example whether it fixes a defect;
    \item whether code is compared across repositories or only within them;
    \item how the uncertainty of an estimate is computed, in particular whether lines from the same repository are treated as independent;
    \item how long the code is followed after merge (the observation window); and
    \item how the developer or agent who modifies the code is identified.
\end{enumerate}
None of these studies reports how its verdict would change under the choices of the others. Therefore, a reader cannot yet tell whether they disagree about the agents or about how the agents were measured.

Methods research in software engineering has shown that such choices can change conclusions, although mostly one choice at a time. For example, the SZZ algorithm traces a bug fix back to the commit that introduced the bug~\cite{sliwerski2005changes}, and its output changes for 6\% to 13\% of bug-fix commits with the diff algorithm alone~\cite{nugroho2020different}. Similarly, \texttt{git blame} reports the commit that last changed each line~\cite{gitBlame}, and it attributes lines to the wrong commit more often than blame at the level of tokens~\cite{german2019cregit}. What counts as a change matters as well, since refactorings make up a sizeable share of the lines that SZZ flags~\cite{neto2018impact} and up to 15.5\% of method updates are non-essential, such as renames~\cite{kawrykow2011non}. The same holds for choices made in the analysis. A published defect-prediction finding reverses when units are weighted by their size~\cite{kamei2010revisiting}, and a conclusion drawn at one level of aggregation need not hold at another~\cite{posnett2011ecological}. Labels are a further source of error. A third of bug reports are misclassified~\cite{herzig2013not}, and once keyword-label errors and project-level effects were modeled, the number of programming languages associated with defects fell from 11 to 4~\cite{berger2019impact}. When many choices are varied together, the effect can be larger still. In a multiverse analysis, which runs every defensible combination of analysis choices, fewer than 0.2\% of 3,072 pipelines for one published mining study reproduced its result~\cite{cassee2026exploring}. However, to the best of our knowledge, no study has measured how these choices, taken together, change the verdicts of post-merge studies of agent code. Without such a measurement, two studies that disagree cannot be reconciled, and the verdict of a single study cannot be told apart from an artifact of its measurement.

Our earlier study~\cite{rahman2026will} suggests that these choices do matter for agent code. It followed the lines that agent and human PRs of the AIDev dataset~\cite{li2025rise} added to 215 repositories, and found that the verdict on the same code changed with how the comparison was set up. Here, the \textit{verdict} is the direction of a significant difference: agent code retained longer than human code, modified sooner, or no difference. For example, agent lines were modified later than human lines in the same repository when each PR counted once, but not when each line counted once. However, the study reported one main design, so it could not say how far its verdict would move under the other choices that published studies make. To address this gap, in this paper we revisit the survival of the same code and measure how much each of the nine choices decides the verdict. We also identify which conclusions hold under all of them.

We start from the lines of our earlier study, the 854,139 lines that 2,802 merged PRs added to 215 open-source repositories of AIDev, and from that study's line tracker. We chose AIDev because it attributes each PR to a specific coding agent, and because prior post-merge work uses it~\cite{sawada2026extent}, which makes our results comparable. In this study, the term \textit{agent-attributed code}, or \textit{agent code} for short, refers to the code added by a merged PR that AIDev attributes to a coding agent. We say ``attributed'' because the attribution comes from PR metadata and not from ground-truth authorship (Section~\ref{sec:validation}). Knowing the agent of each PR lets us separate two \textit{tiers}: \textit{human-supervised agents} (Copilot, Cursor, Claude Code, and Codex), which a developer directs and supervises to varying degrees, and the \textit{fully-autonomous agent} Devin. We compare each tier with \textit{human code}, the code that human PRs added to the same repositories, which serves as the \textit{human baseline}.

To estimate how long code is retained, we use Kaplan-Meier curves~\cite{kaplan1958nonparametric}, which show the share of lines still unmodified over time, and Cox proportional hazards regression~\cite{cox1972regression}. Software engineering studies have long used these methods for the lifetime of code lines, warnings, projects, and developers~\cite{spinellis2021software,aman2019survival,samoladas2010survival,lin2017developer}, and recently for code from coding agents~\cite{xia2026these,rahman2026will}. Cox regression estimates a \textit{hazard ratio} (HR), which compares how likely agent and human lines are to be modified at any moment after merge. An HR below 1 means that agent lines are modified later than human lines, which we call \textit{retained longer}, and an HR above 1 means that they are \textit{modified sooner}. Lines from the same repository are often modified by the same commits, so we report \textit{standard errors clustered by repository}, which allow for this dependence~\cite{cameron2015practitioner}, next to the \textit{model-based} errors that treat every line as independent. RQ1 to RQ5 change one or two choices at a time, starting from the units and the tracker of our earlier study (Section~\ref{sec:earlier}). We then combine the choices in a \textit{specification curve}, which reports the estimate under every \textit{specification}, that is, every combination of choices~\cite{simonsohn2020specification}. We also re-run the designs of five recent post-merge studies on our data.

Specifically, we aim to answer the following five research questions, which follow the order in which a post-merge study makes its measurement choices. RQ1 to RQ3 concern the survival verdict, that is, whether agent code is modified sooner than human code. RQ4 and RQ5 concern its interpretation, that is, why and by whom agent code is modified.

\smallskip
\noindent\textbf{RQ1 (Tracking and Unit):} \textit{How do line tracking and the unit of analysis change the verdict?}
We follow the same lines with three trackers, with and without vendored and generated files, and count them per line, per PR, and per hunk. These choices move the hazard ratio of Devin code from below 1 to above 1, whereas the estimate for human-supervised code stays below 1. Once uncertainty is estimated per repository, however, few of these differences are significant.

\smallskip
\noindent\textbf{RQ2 (Modification Definition):} \textit{Does the definition of a modification change the verdict?}
We compare a syntactic definition, which counts every change to a line, with a structural one, which ignores changes to names, values, and formatting. The structural definition changes the size of the effects but never their direction.

\smallskip
\noindent\textbf{RQ3 (Comparison Scope and Uncertainty):} \textit{How much do the comparison scope and the variance estimator change the verdict?}
We compare agent code with human code across all repositories and within the same repositories, under three ways of estimating uncertainty. Human-supervised code is retained longer than human code more often than chance would give, in both scopes. For Devin code, the way uncertainty is estimated decides whether it looks modified sooner, and within repositories no difference can be detected.

\smallskip
\noindent\textbf{RQ4 (Intent Classification):} \textit{Does the intent classifier change the verdict on why agent code is modified?}
We classify the commits that modify the code with keyword rules and with three LLMs, and compare both with commits that we labeled by hand. The LLMs agree with the human labels about as well as two people agree with each other, whereas keyword rules do not. Only keyword rules find that Devin code is modified for different reasons than human code.

\smallskip
\noindent\textbf{RQ5 (Maintainer Identification):} \textit{Do measurement choices change the verdict on who maintains agent code?}
We identify who makes each modification of agent code, and vary how follow-up work is counted and for how long. Human developers make most of the modifications under every choice, and when an agent modifies agent code, it is almost always the agent that wrote it.

\smallskip
Our findings imply that the measurement choices matter much more for Devin code than for human-supervised code. For human-supervised code, mostly from Copilot, the choices change the size of the effect but not its direction. Across all 88 specifications, none reports it modified sooner, except with hunk units under line-number tracking. For Devin code, the variance estimator and the comparison scope decide the verdict, and with errors clustered by repository its ``modified sooner'' verdicts are no more frequent than under shuffled labels. When we re-run five published designs on our data, we find, like Xia and Miller~\cite{xia2026these}, no significant difference for pooled agent code within repositories. However, their corrective-maintenance result depends on the intent classifier.

\smallskip
\noindent

\textbf{\underline{Paper Contributions.}} In summary, this paper makes the following contributions:

\begin{itemize}[leftmargin=*,nosep]
    \item \textbf{A sensitivity map of post-merge measurement.} To the best of our knowledge, this is the first study that measures how the measurement choices of post-merge studies, taken together, change their verdicts about code from coding agents. Its specification curve of 88 analyses covers two agent tiers and is evaluated under three variance estimators, several observation windows, and shuffled labels.

    \item \textbf{Published designs re-run on one dataset.} We re-run the designs of five recent post-merge studies on our data, report where we reproduce their headline result, and show which single measurement choice moves it.

    \item \textbf{Results that hold across specifications.} Human-supervised code, mostly from Copilot, is retained longer than human code in more specifications than chance would give, both across and within repositories. Devin code shows no difference that survives the variance estimator, and human developers make most of the modifications to agent code.

    \item \textbf{LLM classifiers as a replacement for keyword rules.} Validated on 618 human labels, three LLM classifiers agree with people about as well as two human raters agree with each other, whereas keyword rules create a Devin difference that no human label confirms.
    \item \textbf{An operationalization checklist.} We derive a checklist for future post-merge studies, and tie each item to the size of the effect it controls on our data.
\end{itemize}

\smallskip
\noindent

\textbf{\underline{Open Science.}} We provide a replication package with the identity-tracking miner, the data, every specification of the curve, the human intent labels, and the analysis scripts, available online~\cite{replicationpackage}.

\smallskip
\noindent

\textbf{\underline{Paper Organization.}}
The remainder of this paper is organized as follows. Section~\ref{sec:related} positions our work within the related literature and motivates the methodological design. Section~\ref{sec:earlier} summarizes the design and findings of our earlier study that this paper builds on. Section~\ref{sec:methodology} details our dataset, agent taxonomy, and analysis methods. Sections~\ref{sec:rq1} to~\ref{sec:rq5} answer the five research questions, and Section~\ref{sec:synthesis} presents the specification curve and the re-run published designs. Section~\ref{sec:discussion} discusses the operationalization checklist and implications for researchers and practitioners. Section~\ref{sec:threats} addresses threats to validity. Section~\ref{sec:conclusion} concludes.

\section{Related Work and Motivation}
\label{sec:related}

We first review studies of the quality of code from coding agents before it is merged, and then the recent studies of what happens to that code after merge, together with their measurement choices. We then review methods research showing that such choices can change the conclusions of mining studies, and close with the three gaps that our study addresses.

\subsection{Quality of AI-Generated Code (Pre-Merge)}

AI tools are now widely used in software development: 76\% of the respondents to the 2024 Stack Overflow Developer Survey use or plan to use them~\cite{stackoverflow2024Stack}. Research on the quality of AI-generated code has focused primarily on correctness at generation time~\cite{chen2021evaluating,yeticstiren2023evaluating,pearce2025asleep}. First, Chen et al.~\cite{chen2021evaluating} established Pass@k as the standard benchmark metric, measuring the probability that at least one of $k$ generated samples passes unit tests. Similarly, Ren et al.~\cite{ren2020codebleu} proposed CodeBLEU, which compares generated code with a reference solution. Then, Yeti\c{s}tiren et al.~\cite{yeticstiren2023evaluating} evaluated Copilot, CodeWhisperer, and ChatGPT on several code quality dimensions, including code correctness, compilability, and maintainability, at snapshot time. Other studies examined security, and found that AI-generated code frequently introduces vulnerabilities~\cite{pearce2025asleep,sandoval2023lost}, although Asare et al.~\cite{asare2023github} found it is not demonstrably worse than human code in this regard.

Recently, Ehsani et al.~\cite{ehsani2026aicodingagentsfail} studied 33,000 agent-authored PRs on GitHub, finding that unmerged agent PRs involve larger code changes, fail CI/CD validation more often, and face rejection due to socio-technical factors. Several other studies use the AIDev dataset~\cite{li2025rise} to examine PRs from coding agents from different angles: their characteristics and review patterns~\cite{mazloomzadeh2026aicodingagentscontribute}, agentic refactoring~\cite{horikawa2025agentic}, security-related PRs~\cite{siddiq2026security}, failure modes~\cite{hindi2026coding}, code review agents~\cite{chowdhury2026industry}, and inconsistencies between messages and code~\cite{gong2026analyzing}. \textit{Unlike these pre-merge studies, our work examines what happens to code from coding agents after it is accepted into the codebase.}

\subsection{Post-Merge Studies of Code from Coding Agents}

While pre-merge studies assess code at the moment it is generated or reviewed, a growing body of work tracks what happens to code from coding agents after it is merged~\cite{gitclearCodingCopilot,popescu2026investigating,sawada2026extent,xia2026these,kraishan2026agents,takerngsaksiri2026finishes,rahman2026will}. We present these studies in the order they appeared and summarize their measurement choices in Table~\ref{tab:prior_designs}.

GitClear's industry reports~\cite{gitclearCodingCopilot,gitclearCopilotCode} gave the earliest signal, measuring aggregate code churn across millions of changed lines and suggesting that code written with AI assistance is revised more often. Popescu et al.~\cite{popescu2026investigating} then studied five coding agents (Codex, Claude Code, Copilot, Jules, and Devin). For the first commit of each PR, they checked whether the lines it added were still unchanged in the nearest later commit within three days, one week, and three weeks. Human commits keep all their added lines more often than commits from every agent, with small effects. \textit{Their unit is the commit, they sample commits stratified by repository popularity without stating how repositories enter the comparison, and lines are compared by content in the next commit.}

Sawada et al.~\cite{sawada2026extent} studied the maintenance of files created in PRs of the AIDev dataset~\cite{li2025rise}. They classified the intent of later commits with a classifier based on Conventional Commits~\cite{conventionalCommits}, a convention of message prefixes such as \texttt{fix:} and \texttt{feat:}. They find that AI-generated files are most often modified to extend features, whereas human files are most often modified to fix bugs, and that human developers do about 83\% of the maintenance of AI-generated files. \textit{Their unit is the file, and intent is inferred from commit-message conventions, which developers themselves often confuse~\cite{zeng2025first}.}

Afterwards, Xia and Miller~\cite{xia2026these} followed every line added in 182 repositories from May 2025 to May 2026. They replay later diffs to find the commit that terminates each line, and then verify it with \texttt{git blame}~\cite{gitBlame}. They estimate Cox models stratified by repository, and Fine-Gray models~\cite{fine1999proportional}, which treat termination for different reasons as competing events, with standard errors clustered by repository. They find that agentic code is terminated at a rate similar to human code (HR\,=\,1.11, not significant) but receives more corrective maintenance (HR\,=\,1.49). They also find that projects differ almost completely from one another ($I^2 = 100\%$, where $I^2$ is the share of the variation that comes from differences between projects~\cite{higgins2002quantifying}). Their study is the closest to ours, and it already compares code within repositories. \textit{It pools all agents under one label. It counts any later diff that touches a line as its termination, without stating how renames, moves, or whitespace edits are handled. It infers intent with keyword and Conventional Commit rules, with an LLM fallback for ambiguous commits.}

Most recently, two studies compared individual agents and follow-up work at the level of PRs. Kraishan~\cite{kraishan2026agents} measured churn as the number of follow-up commits that touch a PR's most-changed file within 90 days, relative to the lines the PR changed, against a human baseline from the same repositories. Claude Code shows the lowest churn, Copilot and Cursor lower churn than humans, and Codex and Devin churn at the human level. \textit{The comparison uses non-parametric tests on pooled PRs, without a model of the repository.} Takerngsaksiri et al.~\cite{takerngsaksiri2026finishes} asked whether a PR from a coding agent is followed, within 30 days, by a verified fix PR that edits the same file. They use an odds ratio pooled over 218 repositories (the Mantel-Haenszel estimator~\cite{mantel1959statistical}). They find that agent PRs are more likely to need such a fix (OR\,=\,1.62), and that most of these fixes come from the same agent (69.6\%). \textit{Their unit is the PR, their outcome is a verified fix rather than any modification, and agents are pooled.}

This paper extends our earlier study~\cite{rahman2026will}, which followed every line added by agent and human PRs of the AIDev dataset and found agent code retained longer than human code. Its verdict moved with the operationalization, but it reported one primary combination of choices, so how far each choice moves the conclusion was left open. Section~\ref{sec:earlier} summarizes its design and findings.

\begin{table*}[t]
\centering
\caption{Measurement choices of published post-merge studies of code from coding agents, as reported in each paper. ``Not stated'' means we could not find the choice in the paper.}
\label{tab:prior_designs}
\scriptsize
\setlength{\tabcolsep}{3pt}
\begin{tabular}{@{}>{\raggedright\arraybackslash}p{1.55cm}>{\raggedright\arraybackslash}p{1.8cm}>{\raggedright\arraybackslash}p{1.6cm}>{\raggedright\arraybackslash}p{1.8cm}>{\raggedright\arraybackslash}p{1.75cm}>{\raggedright\arraybackslash}p{1.85cm}>{\raggedright\arraybackslash}p{2.3cm}@{}}
\toprule
\textbf{Choice} & \textbf{Popescu et al.~\cite{popescu2026investigating}} & \textbf{Sawada et al.~\cite{sawada2026extent}} & \textbf{Xia and Miller~\cite{xia2026these}} & \textbf{Kraishan \cite{kraishan2026agents}} & \textbf{Takerng\-saksiri et al.~\cite{takerngsaksiri2026finishes}} & \textbf{This study} \\
\midrule
Data, window & Agent and human commits; 3 days, 1 week, 3 weeks & AIDev; first 6 months & 182 repositories; May 2025 to May 2026 & Agent and same-repository human PRs; 90 days & 218 repositories; 30 days & AIDev, 215 repositories; at least 5 months \\
Unit & Commit (all added lines), per agent & File & Line, agents pooled & PR, per agent & PR, agents pooled & Line, hunk, PR; two tiers \\
Tracking & Line content in the nearest later commit & File history & Diff replay, verified with blame & Most-changed file & File co-edited by a later PR & Line identity (reverse blame with \texttt{-M -C}), compared with two other trackers \\
Modification & Line changed & Later commit to the file & Any diff touching the line & Follow-up commit to the file & Verified fix PR & Syntactic and AST-structural \\
Intent & None & Conventional Commits & Rules, LLM fallback & None & LLM and human fix verification & Keyword, 3 LLMs, combinations; 618 human labels \\
Comparison & Not stated (sample stratified by repository stars) & Same-repository sample, pooled & Within repositories & Same-repository baseline, pooled tests & Within repositories (Mantel-Haenszel) & Across and within repositories \\
Uncertainty & Mann-Whitney U and Cliff's $\delta$; no clustering stated & Not stated & Standard errors clustered by repository & Non-parametric tests; no clustering stated & Mantel-Haenszel estimate; clustering not stated & Clustered by repository (primary), by PR, and model-based; permutation test \\
Maintainer & Not studied & Commit authors: human share of later commits & Authors and co-author trailers of terminating commits & Not studied & Share of fixes made by the same agent & Commit author, with and without co-author trailers; unit and window varied \\
\bottomrule
\end{tabular}
\end{table*}

\subsection{Measurement Choices That Change Conclusions}

The studies above differ in every row of Table~\ref{tab:prior_designs}. Methods research on mining software repositories has shown, one choice at a time, that each of these rows can change a conclusion. We review this evidence in the order of the choices listed in Section~\ref{sec:introduction}.

\textit{Line tracking.} Code retention and churn are established quality proxies: relative code churn predicts defect density~\cite{nagappan2005use}, and churn-based features outperform static code metrics for defect prediction~\cite{moser2008comparative}. How a line is followed through history, however, is itself a source of error. Switching the diff algorithm in Git changes code-churn values in 1.7\% to 8.2\% of commits, and the output of the SZZ algorithm~\cite{sliwerski2005changes} for 6\% to 13\% of bug-fix commits~\cite{nugroho2020different}. Token-level blame identifies the commit that introduced a token in 94.5\% to 99.2\% of cases, against 75\% to 91\% for line-level blame~\cite{german2019cregit}, and refactoring-aware trackers recover histories that tools based on line positions lose~\cite{jodavi2022accurate}.

\textit{Files kept, unit, and weighting.} Vendored and generated files are regenerated or replaced as a whole, so every line in them is modified at once, however the code around them evolves (Section~\ref{sec:rq1}, Step~2). Which units remain, and how much each one counts, matters as well. The unit of analysis and how much each unit counts can decide a result. A published defect-prediction finding, that package-level predictions outperform file-level ones, no longer holds when units are weighted by the effort they require~\cite{kamei2010revisiting}. Conclusions drawn at one level of aggregation need not hold at another, an instance of the ecological fallacy~\cite{posnett2011ecological}. Large commits differ in kind from small ones, being more often perfective~\cite{hindle2008large}, so when every line counts once, a few large changes can dominate an estimate.

\textit{What counts as a modification.} Up to 15.5\% of method updates in version histories are non-essential, such as renames and local refactorings~\cite{kawrykow2011non}. Refactorings make up 6.5\% of the lines that SZZ flags as bug-introducing and 19.9\% of the lines removed in bug fixes~\cite{neto2018impact}, and up to 15\% of bug fixes are tangled with unrelated changes~\cite{herzig2013impact}. A definition that counts every touched line therefore mixes substantive and cosmetic edits.

\textit{Intent classification.} Mockus and Votta~\cite{mockus2000identifying} introduced keyword-based classification of commits into the maintenance categories of Swanson~\cite{swanson1976dimensions}, and Hindle et al.~\cite{hindle2009automatic} showed that large changes can be classified automatically. Levin and Yehudai~\cite{levin2017boosting} later found that classifiers using the commit message alone are much weaker across projects than classifiers that also use the source-code change. Labels from such classifiers carry errors that change conclusions: 33.8\% of bug reports in five projects were misclassified~\cite{herzig2013not}, and keyword labels of bug fixes had 36\% false positives in a widely cited study of programming languages~\cite{berger2019impact}. Because mining studies classify thousands of commits, too many to label by hand, keyword rules remain the default. LLMs offer an alternative. Following the use of LLMs as judges of open-ended answers~\cite{zheng2023judging}, software engineering research now uses them to evaluate code, summaries, and patches in place of human raters~\cite{wang2025can,he2025code}, and to annotate artifacts~\cite{hou2024large}. Outside software engineering, an LLM has matched or exceeded crowd workers on text annotation~\cite{gilardi2023chatgpt}. Ahmed et al.~\cite{ahmed2025can} replaced one of two human raters with an LLM on ten annotation tasks from five software engineering datasets. For some tasks, agreement stayed equal or close to the agreement between the two humans, and agreement between models signaled where the replacement was safe. For commits, Sazid et al.~\cite{sazid2024commit} classified commits into maintenance activities with an LLM and no training data ($\kappa = 0.62$), and H\"artel~\cite{hartel2026agentic} evaluated LLMs and LLM agents on the same task. However, whether LLM classifiers can replace keyword rules for commit intent at the scale of a post-merge study, and whether this changes its conclusions, has not been tested.

\textit{Comparing across or within repositories.} Projects can differ more from one another than the effects of interest. Models fitted to local subsets of projects describe software data better than global ones, whose trends are too general~\cite{bettenburg2012think}, and Xia and Miller find that almost all the variation in termination rates comes from differences between projects~\cite{xia2026these}. Survival analysis offers a direct remedy: stratifying the baseline hazard by repository, so that code is compared only with code from the same project~\cite{therneau2000cox}.

\textit{Uncertainty with clustered data.} When observations come in groups, such as the lines of one repository, they are not independent, and standard errors that ignore the grouping are too small~\cite{cameron2015practitioner}. Clustered standard errors~\cite{lin1989robust} and resampling whole clusters~\cite{cameron2015practitioner} account for the grouping. With few clusters, even these become unreliable, so the number of clusters limits what a comparison can detect~\cite{cameron2015practitioner}.

\textit{Observation window.} Prior studies follow code for windows from three days~\cite{popescu2026investigating} to about a year~\cite{xia2026these} (Table~\ref{tab:prior_designs}). A window decides which contributions can be followed at all, and for how long. In a cohort with a fixed end date, a longer window can be applied only to contributions merged earlier, so it changes the cohort as well as the follow-up.

\textit{Who maintains the code.} Ownership research shows that who changes a component matters for its quality: components changed by many minor contributors have more defects~\cite{bird2011don}, and a replication on further Microsoft products largely confirmed this relationship~\cite{greiler2015code}. Who counts as the developer of a change depends on how authorship is recorded, and coding agents add new ways to record it, such as bot accounts and co-author trailers in commit messages~\cite{li2025rise}.

\textit{All choices together.} When several choices are made at once, their effects can compound. In defect prediction, the research group explains far more of the variance in reported performance than the choice of classifier~\cite{shepperd2014researcher}. In a multiverse of 3,072 defensible pipelines for one published mining study, fewer than 0.2\% reproduced its result, and many reached qualitatively different conclusions~\cite{cassee2026exploring}. A specification curve, which reports the result of every defensible combination of choices, makes this dependence visible instead of hiding it behind one chosen pipeline.

\subsection{Positioning and Motivation}

Each design of the post-merge studies reviewed above is defensible. However, three things remain unknown.

Firstly, methods research shows that line tracking, the files kept, unit weighting, the modification definition, intent classification, the comparison across or within repositories, the variance estimator, the observation window, and how the modifier is identified can each change conclusions. However, to the best of our knowledge, no prior work has measured how these choices, taken together, shape the conclusions of post-merge studies of code from coding agents. No study has reported its result over more than one combination of them. Whether validated LLM classifiers can replace keyword rules for commit intent has not been tested either.

Secondly, the published post-merge studies reach different verdicts, but each uses its own data and its own design, so it is not known whether they disagree about the agents or about the measurement. No study has re-run their designs on the same data.

Thirdly, despite the growing number of post-merge studies, there is no evidence-based guidance on how to operationalize them, that is, which choices matter most and by how much.

To summarize, this work differs from existing studies, including our earlier one~\cite{rahman2026will}, in three ways. \textit{(1)}~We vary nine measurement choices together on one dataset and report the full specification curve instead of a single pipeline. \textit{(2)}~We re-run the designs of five published studies on the same data, and identify the single choice that moves each of their headline results. \textit{(3)}~From the size of each effect on our data, we derive an operationalization checklist for future post-merge studies.

\section{Background: Our Earlier Study}
\label{sec:earlier}

This paper starts from the cohort, the code units, and the findings of our earlier study~\cite{rahman2026will}. This section summarizes the parts of that study that the rest of the paper builds on, so that the paper can be read on its own. It reports the design and findings of that study and adds no new experiments, except for a check, at the end of the section, that our pipeline reproduces its estimates on the same units.

\subsection{Question and Scope}
The earlier study tested what it called the ``disposable code'' hypothesis, under which code from coding agents is merged quickly but modified or deleted soon afterwards. It was motivated by GitClear's report of rising two-week churn in AI-assisted code~\cite{gitclearCodingCopilot}. It asked whether agent code is modified sooner than human code merged into the same repositories (\textit{survival}), why agent and human code are modified (\textit{intent}), and whether the later modification of agent code can be predicted when it is merged (\textit{forecasting}). This paper revisits the first two. We do not revisit forecasting, which is out of the scope of this paper and requires its own study of how its labels change with the tracker.

\subsection{Data, Cohort, and Attribution}
The earlier study used the AIDev dataset~\cite{li2025rise}, which attributes PRs to an agent by PR-level signals, such as the head-branch prefix, the bot account that opens the PR, or a co-author line. It kept the repositories that contain both agent and human PRs and applied the filtering criteria of Xiao et al.~\cite{xiao2025self} for engineered software projects (Section~\ref{sec:filtering}). It then mined every merged PR whose merge commit it could locate on the default branch and that adds at least one line to a tracked source file. The mined cohort, which we also use, has 2,802 merged PRs in 215 repositories: 1,470 agent PRs and 1,332 human PRs.

The study grouped the agents into two tiers, which we also use (Section~\ref{sec:taxonomy}): \textit{human-supervised agents} (Copilot, Codex, Cursor, and Claude Code), which a developer directs and reviews, and the \textit{fully-autonomous agent} Devin, to which developers delegate whole tasks. The human-supervised tier has 1,058 PRs and Devin 412.


Because AIDev's signals show why a PR was collected but not who wrote its code, the study audited 325 merged agent PRs, the sample size for a proportion at 95\% confidence and a $\pm$5\% margin~\cite{cochran1977sampling}, stratified by agent. The first author labeled each PR \textit{agent}, \textit{mixed}, \textit{human}, or \textit{unclear} by whether the agent made its substantive code changes, where a change is non-substantive only if it belongs to the non-essential changes of Kawrykow and Robillard~\cite{kawrykow2011non}. The agent made substantive changes in 88.9\% of the 325 PRs (67.1\% labeled agent, 21.8\% mixed), 6.2\% were labeled human, and none of these came from Devin. Of the 221 audited PRs in the mined cohort, 16 were labeled human, and excluding them, or the 28 PRs labeled human or unclear, left the primary estimate unchanged (HR\,=\,0.84 and 0.83). Section~\ref{sec:validation} adds a review of the human baseline.

\subsection{Operationalization}
A unit was \textit{born} when its PR was merged into the default branch, and \textit{modified} when a later commit on the default branch changed or deleted it. Units still unmodified on December 31, 2025 were right-censored, that is, the study only knows that they stayed unmodified at least until then. Because AIDev's PR cutoff is August 1, 2025, every unit was observed for at least five months. A modification was treated as neutral, since code is also modified to add features or adapt to its environment.

The primary unit was the line: every line that a PR adds to a source file, with no sampling. Each line was followed with \texttt{git blame --reverse} from the merge commit to the last default-branch commit on or before the end of the window. The blame reports the last commit in which the line still exists, and the line was counted as modified by the next commit that touches its file. Hunks (blocks of added lines, modified when any of their lines is modified) and files (modified when any later commit touches them) were reported for comparison. A \textit{substantial modification} event marked a hunk or file as modified only once a share $\theta$ of its added lines had changed, with $\theta \in \{25, 50, 75, 100\}\%$.

Two further choices defined the primary design. First, lines in vendored and generated paths (e.g., \texttt{vendor}, \texttt{node\_modules}, \texttt{dist}, build output, minified bundles, generated protocol buffers) were excluded, because regenerating such a file modifies all its lines at once. This leaves 648,959 of the 854,139 added lines. Second, a few very large PRs contribute most of the lines: the largest human PR alone adds more than 34,000 lines outside vendored and generated paths. Each unit was therefore weighted by the inverse of the number of units in its PR, so that every PR counts once. The unweighted analysis, the analysis with vendored and generated paths, and a sampled design that follows at most 20 source files per PR and 100 random lines per file were reported as sensitivity analyses.

\subsection{Analysis}
The study estimated Kaplan-Meier retention curves~\cite{kaplan1958nonparametric} and compared them with the log-rank test~\cite{mantel1966evaluation}, which tests whether two retention curves differ. Its primary model was a Cox model~\cite{cox1972regression} stratified by repository, so that agent code is compared only with human code in the same repository, adjusted for the PR's added lines and changed files, with standard errors clustered by PR. Further models replaced the agent indicator with one indicator per tier and one per agent. Because the hazard ratio was not constant over time (the proportional hazards assumption did not hold), the study also split the follow-up into periods and reported the restricted mean survival time~\cite{royston2013restricted} at 90 and 150 days, with a bootstrap that resamples whole repositories. The restricted mean survival time at 90 days is the average number of days a line stays unmodified within the first 90 days. A Cox model pooled across repositories showed how much a comparison across repositories differs from one within them.

Intent was classified from the message of each modifying commit into Swanson's maintenance categories~\cite{swanson1976dimensions}, extended with preventive changes: corrective, perfective, adaptive, preventive, and other. The classifier used keyword rules in the tradition of Mockus and Votta~\cite{mockus2000identifying}, with a fixed priority when several categories match, and matched whole words only. Each modifying commit counted once per author group that it modified, and agent and human distributions were compared with a $\chi^2$ test and Cram\'er's $V$~\cite{cramer1946mathematical}, an effect size for categorical data for which values around 0.1 count as small~\cite{cohen1988statistical}.

\subsection{Findings}
Table~\ref{tab:earlier_findings} summarizes the survival results. With each PR counted once and within repositories, agent lines were modified later than human lines (HR\,=\,0.83, $p = 0.001$). The effect came from the human-supervised agents (HR\,=\,0.76, $p < 0.001$), mainly Copilot and Codex, whereas Devin code did not differ from human code (HR\,=\,0.95, $p = 0.53$). The gap built up over the window rather than early: in the first 14 days the agent HR was 1.01, and the 90-day restricted mean survival time did not differ significantly (+0.8 days). Within 14 days of merge, 8.3\% of human-supervised lines, 12.0\% of Devin lines, and 9.6\% of human lines were modified.

\begin{table}[t]
\centering
\caption{Survival results of our earlier study at the line level, as reported there~\cite{rahman2026will}: Cox hazard ratio of agent code against human code, overall and per tier. HR $<1$: agent code is modified later. Bold: $p < 0.05$.}
\label{tab:earlier_findings}
\footnotesize
\begin{tabular}{lrrrrr}
\toprule
\textbf{Design} & \textbf{Lines} & \textbf{Mod.\ \% A / H} & \textbf{Agents} & \textbf{Supervised} & \textbf{Devin} \\
\midrule
\multicolumn{6}{l}{\textit{Within repositories}} \\
Primary: no vendored, PR-weighted & 648,959 & 28.0 / 34.5 & \textbf{0.83} & \textbf{0.76} & 0.95 \\
With vendored, PR-weighted & 854,139 & 28.1 / 34.8 & \textbf{0.83} & \textbf{0.76} & 0.95 \\
No vendored, unweighted & 648,959 & 23.4 / 30.1 & 0.91 & 0.80 & 1.06 \\
With vendored, unweighted & 854,139 & 23.3 / 50.3 & 0.92 & 0.86 & 1.00 \\
Sampled, PR-weighted & 277,283 & 28.1 / 34.7 & \textbf{0.83} & \textbf{0.76} & 0.95 \\
\midrule
\multicolumn{6}{l}{\textit{Pooled across repositories}} \\
Primary design & 648,959 & 28.0 / 34.5 & 0.89 & & \\
With vendored, unweighted & 854,139 & 23.3 / 50.3 & 0.76 & & \\
\bottomrule
\end{tabular}
\parbox{\linewidth}{\vspace{0.3em}\footnotesize Mod.\ \%: share of lines modified by December 31, 2025, agent / human. Vendored: lines in vendored and generated paths. PR-weighted: each PR counts once. Sampled: at most 20 files per PR and 100 lines per file. Supervised: human-supervised agents.}
\end{table}

The verdict, however, depended on how the comparison was set up. Without PR weighting, no difference was significant, because a few very large PRs then dominate. With vendored code, 34.1\% of human lines but under 4\% of agent lines lie in vendored paths, and the unweighted human modification rate rose from 30.1\% to 50.3\%. Pooled across repositories, the agent estimate ranged from 0.76 to 0.89 depending on the design, so a comparison across repositories was no substitute for one within them. At the hunk level, human-supervised hunks were modified later (HR\,=\,0.82), and Devin hunks were modified sooner without weighting (HR\,=\,1.24, $p = 0.049$) but not with it (1.11). Under the substantial-modification event, agent hunks and files were modified later at every threshold (HR 0.80 to 0.86), again because of the human-supervised agents (0.72 to 0.77), while Devin never differed (0.93 to 1.01). The difference between the tiers may also partly reflect when their PRs were merged: among PRs merged on or after May 16, 2025, Devin code was modified later than human code too (HR\,=\,0.61).

For intent, the commits that modify agent and human code had almost the same profile (Cram\'er's $V = 0.080$), with a slightly higher corrective share for agent code (35.6\% against 32.5\%). This result also depended on the classifier. With substring instead of whole-word matching, the keyword \textit{auth} matched the \texttt{Co-authored-by} trailer that agents add to their commits. The preventive share then became 29.7\% for agent code against 18.8\% for human code, and the effect size doubled ($V = 0.155$).

The study concluded that the disposable-code hypothesis is not supported: agent code was not modified sooner than human code in the same repositories in any of its designs. The longer retention of agent code, however, held only with each PR counted once and only for the human-supervised agents. It recommended that such studies compare code within repositories, state how they weight units, and check their intent classifier. Section~\ref{sec:disc_holds} revisits this conclusion with the results of this paper.

\subsection{What the Earlier Study Left Open}
The earlier study varied its measurement only as sensitivity checks around one primary design, one choice at a time. Its checks did not all point the same way, and it left five questions open. First, it used one tracker, which counts a line as modified by the next commit that touches its file, and it never compared that tracker with others on the same lines (RQ1). Second, it varied the definition of a modification only by its size, through $\theta$, and not by its kind, such as whitespace-only or renaming edits (RQ2). Third, it contrasted the comparison within and across repositories with a single pooled model and one variance estimator (RQ3). Fourth, it classified intent with keyword rules only, without validating them against human labels (RQ4). Fifth, it did not ask who makes the modifications (RQ5). Finally, it left open how these choices act together, and whether the conflicting verdicts of other post-merge studies come from the same choices (Section~\ref{sec:synthesis}).

\subsection{Reproduction on Our Pipeline}
Before changing any choice, we checked that our pipeline reproduces the earlier study on its own units. We used that study's tracker, all its lines including vendored and generated ones, each PR counted once, and the comparison within repositories. The hazard ratio is then 0.761 for human-supervised code and 0.940 for Devin code, against the reported 0.76 and 0.95 for the same setup. The counts of PRs, lines, and repositories match exactly, and so do the modification rates with and without PR weighting (28.1\% and 34.8\%, and 23.3\% and 50.3\%). This reproduction is the starting point of RQ1, which then changes the choices one step at a time (the \textit{measurement ladder}, Section~\ref{sec:rq1}). RQ2 to RQ5 also start from these units and this tracker.

\subsection{What This Paper Adds}
Table~\ref{tab:delta} lists, for each part of the design, what this paper reuses from the earlier study and what it adds.

\begin{table}[t]
\centering
\caption{What this paper reuses from our earlier study~\cite{rahman2026will} and what it adds.}
\label{tab:delta}
\footnotesize
\begin{tabular}{@{}>{\raggedright\arraybackslash}p{2.6cm}>{\raggedright\arraybackslash}p{4.6cm}>{\raggedright\arraybackslash}p{4.6cm}@{}}
\toprule
\textbf{Part of the design} & \textbf{Earlier study} & \textbf{This paper} \\
\midrule
Cohort & 2,802 PRs in 215 repositories & Reused \\
Line tracker & Reverse blame, next commit touching the file & Reused as the first step; new: identity tracking, line-number tracking, and their comparison (RQ1) \\
Vendored and generated code & Excluded from the primary analysis & Reused; new: 4,102 further generated lines removed, and the removal as an explicit step (RQ1) \\
Unit and weighting & Line, PR-weighted; hunk and file for comparison & Reused; new: line, PR-weighted line, and hunk compared across all other choices (RQ1) \\
Modification definition & Any change; substantial share $\theta$ of hunk or file & New: structural definition based on the abstract syntax tree (AST), with a manual check (RQ2) \\
Intent classifier & Keyword rules & New: three LLMs, three ways of combining them, 618 human labels, corrected shares (RQ4) \\
Comparison scope & Within repositories; one pooled model & New: across vs within for every choice, repository overlap (RQ3) \\
Modifier identity & Not studied & New: who modifies agent code, under three units, three windows, and two identification rules (RQ5) \\
Standard errors & Clustered by PR in the within-repository model & New: clustered by repository in every model, compared with PR-clustered and model-based errors, and a permutation test of the curve (RQ3) \\
Observation window & One window & New: longer windows for a single pipeline (RQ1) and across the specification curve (Section~\ref{sec:synthesis}) \\
Choices combined & One choice at a time & New: specification curve of 88 analyses (Section~\ref{sec:synthesis}) \\
Published designs & Not re-run & New: five designs re-run on our data (Section~\ref{sec:synthesis}) \\
Forecasting & Studied & Not revisited \\
\bottomrule
\end{tabular}
\end{table}

\section{Methodology}
\label{sec:methodology}

The \textit{goal} of this study is to measure how much the measurement choices of post-merge studies change their conclusions about code from coding agents, and to identify which conclusions hold under all of them. This goal addresses the three gaps of Section~\ref{sec:related}. We therefore keep the data fixed, starting from the code units of our earlier study~\cite{rahman2026will}. We then vary the nine choices of Section~\ref{sec:introduction}. Three concern the data: how lines are tracked, which files are kept, and the unit of analysis and its weighting. Two concern the outcome: the definition of a modification and the intent classifier. The other four are whether code is compared across or within repositories, the variance estimator, the observation window, and how the modifier is identified. Each choice has the alternatives used by the studies in Table~\ref{tab:prior_designs}.

\begin{figure*}[t]
\centering
\includegraphics[width=\textwidth]{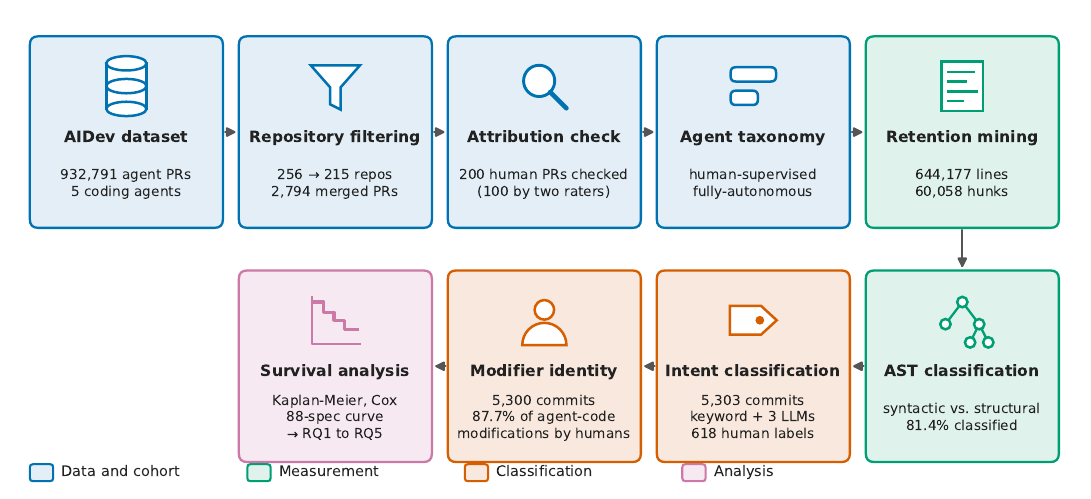}
\caption{Study pipeline. Stages run left to right on the top row and right to left on the bottom row. The text in each box gives the key output of that stage, and the box color marks the type of stage (legend). The counts are those of the primary analysis set: removing vendored and generated code leaves 644,177 of the 854,139 lines, from 2,794 of the 2,802 PRs. Of the 5,303 modification commits, 5,300 can be mapped to a tier.}
\label{fig:pipeline}
\end{figure*}

Figure~\ref{fig:pipeline} gives an overview of our study. We start from the AIDev dataset (Section~\ref{sec:dataset}) and the cohort of repositories and PRs of our earlier study (Section~\ref{sec:filtering}), check the attribution of the human baseline (Section~\ref{sec:validation}), and group the agents into two tiers (Section~\ref{sec:taxonomy}). We then follow every line that the earlier study recorded as added by a PR (Section~\ref{sec:operationalization}), classify each modification as syntactic or structural, classify the intent of each modifying commit (Section~\ref{sec:intent_method}), and identify who made it (Section~\ref{sec:modifier_method}). Finally, we estimate retention with survival analysis and combine the choices into a specification curve (Section~\ref{sec:stats}).

\subsection{Source Dataset: AIDev}
\label{sec:dataset}

We use the AIDev dataset~\cite{li2025rise,li2026aidev}, which captures 932,791 PRs contributed by five AI coding agents (GitHub Copilot, Cursor, Claude Code, OpenAI Codex, and Devin) across public GitHub repositories. AIDev identifies agent-attributed PRs via metadata heuristics: branch prefixes, bot usernames, and Co-Authored-By strings in commit messages.

\subsection{Repository Filtering and Cohort Construction}
\label{sec:filtering}

We use the cohort of our earlier study~\cite{rahman2026will}, whose filtering funnel adapts the criteria of Xiao et al.~\cite{xiao2025self} for selecting a stable, comparable cohort of engineered software projects. From the AIDev PRs, 816 repositories contain both agent-attributed and human-attributed PRs. The funnel first retains repositories with a software license (666). It then keeps those that are not archived, publish releases or tags, and have at least two contributors (566). Per language, it removes repositories in the bottom quartile of PR count, open issues, or size (266 remain). Finally, it removes repositories whose code ratio lies outside the 97\% interval of their language (256 remain). Merged PRs exist in 246 of these repositories. The earlier study located the merge commit of each merged PR on the default branch and kept the PRs that add at least one tracked source line: 2,802 PRs in 215 repositories, which add 854,139 lines. We take these PRs and their lines unchanged as our starting point. PRs that were never merged (no merge timestamp in AIDev) are not part of the cohort.

The earlier study tracked only files with a source-code extension, and excluded lines in vendored and generated paths from its primary analysis. We also remove these lines, because they are not maintained like the code around them. The earlier study identifies 205,180 lines as vendored or generated by their paths (e.g., \texttt{/dist/}, \texttt{/vendor/}, \texttt{node\_modules}, build output, protobuf-generated files, minified or bundled JavaScript). Of these, 96\% lie in human PRs, most of them in one bundled file. Our own rule recognizes a further 4,102 lines in generated paths (e.g., \texttt{zz\_generated} files). Eight PRs add only such lines. Results with the vendored lines kept are reported as a sensitivity analysis (Section~\ref{sec:threats}). Our cohort thus contains 2,794 merged PRs in 215 repositories: 1,327 human PRs, 1,055 PRs from human-supervised agents, and 412 PRs from Devin. One human PR of 46 lines, which an earlier version of our pipeline excluded as a size outlier, is left out of the hunk-level, sampling, intent, and modifier analyses. The results with and without it agree to the third decimal.

\subsection{Attribution Validation}
\label{sec:validation}

AIDev's metadata heuristics provide agent attribution signals, not ground-truth authorship. We validate the human baseline on 200 human-baseline PRs. We sampled them from the 142 repositories that an earlier mining of our cohort covered, in proportion to each repository's share of human PRs (seed 42). An automated pre-screen assigned each PR to a signal group (109 with no AI signal, 73 suspicious, 9 with explicit AI attribution, 9 with generated content). Rules then labeled the 35 PRs opened by a bot account, the 30 PRs with an empty description, and the PRs with no AI signal, which they counted as human. The first author reviewed the remaining 26 PRs by hand. To check these labels, the first author labeled 100 of the 200 PRs from their pages on GitHub as human, bot, AI-attributed, or not assessable. The 100 PRs are all 26 PRs reviewed by hand, and the 15 bot accounts, 15 empty descriptions, and 44 PRs with no AI signal that have the shortest descriptions.


Weighted to the strata of the 200 PRs, 21.0\% (two-stage bootstrap 95\% CI [14.7, 27.4]) come from an automation bot, mostly dependency updaters (Dependabot, Renovate) and CI bots, including three bots that the rules missed. In 5.5\% (CI [2.8, 8.9]), the PR shows that an AI tool wrote its code or its description although AIDev did not flag it. All 11 such PRs are among the 26 reviewed by hand. We call this \textit{undetected agent attribution}. No PR was left unassessed: the PRs with an empty description could be judged from their commits, and 14 of the 15 labeled ones are human. AI assistance used locally without any trace in the PR cannot be detected by this or any metadata-based check, so the human baseline may still contain AI-assisted code.

We therefore frame our comparison as \textit{explicitly agent-attributed contributions vs.\ the standard developer workflow baseline}, and we bound the effect of undetected AI assistance in a sensitivity analysis. Because contaminated lines cannot be identified, this analysis refits no model. It treats a share $c$ of the human lines as hidden agent code. We set $c$ to the lower bound, the point estimate, and the upper bound of the rate above (2.8\%, 5.5\%, and 8.9\%). We assume that this hidden code is modified never, as often as human-supervised code, or as often as human code. The analysis adjusts the human modification rate accordingly, $MR_{\text{human}}' = (1 - c)\,MR_{\text{human}} + c\,MR_{\text{hidden}}$. It then converts modification rates to an approximate hazard ratio under a constant hazard over the window, $\text{HR} \approx \log(1 - MR_{\text{tier}}) / \log(1 - MR_{\text{human}}')$ (Section~\ref{sec:threats}).

\noindent\textbf{Bot exclusion robustness check.} To assess whether automation bots in the human baseline affect the results, we re-ran the analysis after excluding human-baseline PRs authored by bot accounts. Only 676 of the 378,756 human-baseline lines come from bot-authored PRs. The pooled hazard ratio changes by less than 0.001 (0.896 in both cases), and the repository-stratified hazard ratios by less than 0.005 (human-supervised 0.807 to 0.811, Devin 1.059 to 1.060), so bots in the baseline do not affect any result.

\subsection{Agent Taxonomy}
\label{sec:taxonomy}

Prior post-merge studies either pool all coding agents under one label or report each agent separately (Table~\ref{tab:prior_designs}). We use a middle ground that keeps the comparison interpretable: a two-tier taxonomy based on what the AIDev metadata can reliably confirm (Table~\ref{tab:taxonomy}). We also report per-agent results where prior designs require them (Section~\ref{sec:synthesis}).

\begin{table}[t]
\centering
\caption{Agent taxonomy. The tier is the primary analytical unit.}
\label{tab:taxonomy}
\begin{tabular}{lll}
\toprule
\textbf{Tier} & \textbf{Agents} & \textbf{Collaboration Mode} \\
\midrule
Human-supervised & Copilot, Cursor, & Variably human-supervised; \\
                 & Claude Code, Codex & degree of oversight unverifiable \\
                 &                    & from PR-level metadata \\[4pt]
Fully-autonomous & Devin & Whole tasks delegated to the \\
& & agent, which submits PRs as \\
& & an independent contributor \\
\bottomrule
\end{tabular}
\end{table}

The human-supervised tier groups four coding agents because the AIDev metadata does not record which collaboration mode was used per PR, making finer-grained attribution unverifiable. The fully-autonomous tier consists of Devin alone, which its vendor describes as an autonomous software engineer to which developers delegate whole tasks~\cite{devinDocs}. Because this tier contains only Devin, differences at this tier conflate tool-level and tier-level effects, which we acknowledge in Section~\ref{sec:threats}.

In our cohort, the human-supervised tier contributes 177,787 lines (GitHub Copilot 67,398, OpenAI Codex 48,867, Claude Code 48,182, and Cursor 13,340), Devin contributes 87,634 lines, and the human baseline 378,756 lines.

\subsection{Retention Mining}
\label{sec:operationalization}

We follow each line that a merged PR adds to the default branch, from the moment the PR is merged until the line is first modified or the observation window ends. This subsection describes the lines a PR adds, how we follow them, and how we define a modification. Because line tracking and the unit of analysis are two of the choices we vary, we also describe the alternatives we compare them with.

\noindent\textbf{Identifying the merge commit.} We take the merge commit that our earlier study identified for each PR~\cite{rahman2026will}. It lies on the first-parent history of the default branch, that is, on the commits made directly on the default branch, not on merged side branches. The 2,794 PRs fall into four cases. For most PRs (2,488), this is the commit whose committer time lies within 180 seconds of the PR's merge time. When several commits qualify, we take the one whose message names the PR. When no message names the PR, we take the commit at the merge time (70 PRs). For PRs merged by rebasing, we take the run of consecutive commits that share the merge time (49 PRs). When no commit lies within 180 seconds, we take the only commit within three days whose message names the PR (187 PRs). For all 2,794 PRs, the selected commit lies within one day of the merge time that AIDev records.

\noindent\textbf{Birth units.} A PR's units are all source lines that its merged change adds to the default branch, from the parent of its first commit to its last commit (\texttt{git diff -M} against that base), as recorded by our earlier study. This covers squash and rebase merges (2,575 PRs) as well as true merge commits (219 PRs). We keep every added line, without sampling or caps, which gives 644,177 verified line units after the exclusions above (644,857 tracked lines minus 680 unverified ones). For the analyses by hunk, a hunk is a run of consecutive added lines in one hunk of that diff, which gives 60,058 hunk units.

\noindent\textbf{Observation window.} We include PRs merged before August 1, 2025, whose lines were added between December 28, 2024 and July 30, 2025, and observe modifications until December 31, 2025, 23:59:59 UTC. Every line is therefore observed for at least five months. Lines still unmodified at the cutoff are censored there.

\noindent\textbf{Identity tracking.} We follow each line by its identity rather than by its position. For each file, we run \texttt{git blame --reverse}~\cite{gitBlame} from the merge commit to the end snapshot, the last first-parent commit of the default branch at or before the cutoff. We run it along the first-parent chain and with rename and copy detection (\texttt{-M -C}). For every line, reverse blame reports the last commit in which the line still existed, following the line across renames, moves, and insertions above it. If that commit is the end snapshot, the line was retained until the cutoff. Otherwise, the next first-parent commit is the modifying commit, and we verify in its diff that the line lies inside a changed hunk. We record whether the commit changed the line's content, deleted the line, or deleted the file, and whether the change touched only whitespace. For 680 lines (0.11\%), the next commit does not visibly change the line, and we exclude these unverified lines rather than count them as modifications or as retained. The cohort contains 149,468 verified modification events.

\noindent\textbf{Two trackers for comparison.} We keep, for every line, the result of the tracker of our earlier study. That tracker runs \texttt{git blame --reverse} without rename or copy detection and without restricting it to the first-parent chain. It records as the modifying commit the first later commit that touches the file. Several other prior pipelines follow a line by its position. We therefore also run a position-based miner. It checks each line with \texttt{git blame} on the line number it had at birth, and treats a file path missing at the end as a deletion of all its lines. It samples at most 20 files per PR and 100 lines per file, which gives 276,621 sampled units. In RQ1, we compare the three trackers on the same 276,314 sampled units, so that only the tracking differs, and we compare the sampled units with all lines.

\noindent\textbf{Tracker accuracy.} To estimate how accurate each tracker is, the first author labeled 300 of the units that all three trackers cover, each from the diffs of every later first-parent commit that touches its file, without seeing the trackers' claims. The sample is stratified by the pattern of the three trackers' claims and restricted to files with at most 12 such commits of at most 300 changed lines each, so that each history can be read in full. Estimates are weighted to the covered units that meet this restriction. The strata that carry most of the weight are unanimous (for example, none of 50 lines modified). A bootstrap that resamples the labeled lines within strata would therefore not vary there and would give intervals that are too narrow. We therefore place a Jeffreys prior on the proportion of each stratum~\cite{brown2001interval}, a standard way to give an interval for a proportion that is 0 or 1 in the sample, and report the weighted posterior mean with its 95\% interval, together with the unweighted counts in the sample. The counts differ from the weighted estimates because the sample over-represents the strata in which the trackers disagree. A line moved to another file is followed there. One line that moved a second time is left out, which leaves 299 lines.

Identity tracking is right in 98.7\% of its ``modified'' claims (95\% CI [95.2, 99.9], 99 of 100 in the sample) and finds 92.8\% ([83.3, 96.9], 99 of 133) of the modified lines. Of the 34 modified lines it misses in the sample, 24 had an identical line elsewhere in the file when they were merged, against 19 of the 99 that it finds, because reverse blame can confuse a line with an identical line elsewhere in the file~\cite{german2019cregit}. The tracker of our earlier study finds almost every modified line (96.1\%, [86.1, 99.9], 131 of 133) but is right in 93.3\% ([90.0, 95.0], 131 of 197) of its ``modified'' claims. Line-number tracking is right in only 35.9\% ([34.2, 38.5], 69 of 150) of them and finds 84.0\% ([75.9, 86.8], 69 of 133).
When both a tracker and the label record a modification, identity tracking names the same first commit in 97 of 99 cases, the earlier tracker in 120 of 131, and line-number tracking in 47 of 69. Identity tracking therefore slightly understates how often lines are modified, whereas line-number tracking overstates it by a factor of about 2.4. Counting a move to another file as a modification instead lowers the recall of identity tracking to 88.1\% and raises the precision of the earlier tracker to 98.5\%, because that tracker counts such moves as modifications.

We apply two operationalizations of \textit{modification}:

\noindent\textbf{Definition 1: Syntactic (current norm).} A line is modified at its first verified change: a change to its content, the deletion of the line, or the deletion of its file. This includes whitespace edits, variable renames, and literal changes alongside structural refactors, as in prior post-merge work~\cite{xia2026these,popescu2026investigating}.

\begin{figure}[t]
\centering
\resizebox{\textwidth}{!}{%
\begin{tikzpicture}[
    n/.style={draw, rounded corners, fill=blue!8, font=\scriptsize\ttfamily, inner sep=2.5pt, minimum height=0.45cm},
    new/.style={n, fill=red!20, draw=red!60!black, thick},
    lf/.style={font=\scriptsize\ttfamily, text=black!70},
    ttl/.style={font=\small\bfseries},
    verdict/.style={font=\scriptsize, align=center},
    e/.style={-Stealth, black!70}
]
\newcommand{\hlf}[1]{\colorbox{yellow!50}{\strut\textbf{#1}}}
\node[ttl] at (0,1.0) {(a) Original};
\node[n] (fa) at (0,0.3) {function\_definition};
\node[n] (ba) at (0,-0.6) {block};
\node[n] (aa) at (-1.15,-1.5) {assignment};
\node[n] (ra) at (1.15,-1.5) {return\_statement};
\node[lf] at (-1.15,-2.1) {x\enspace price\enspace 0.9};
\node[lf] at (1.15,-2.1) {x};
\draw[e] (fa)--(ba); \draw[e] (ba)--(aa); \draw[e] (ba)--(ra);
\node[verdict] at (0,-3.2) {Before either edit};
\begin{scope}[xshift=5.2cm]
\node[ttl] at (0,1.0) {(b) After Listing~\ref{lst:leaf}};
\node[n] (fb) at (0,0.3) {function\_definition};
\node[n] (bb) at (0,-0.6) {block};
\node[n] (ab) at (-1.15,-1.5) {assignment};
\node[n] (rb) at (1.15,-1.5) {return\_statement};
\node[lf, inner sep=0pt] at (-1.15,-2.15) {\hlf{discounted}\,price\,\hlf{0.8}};
\node[lf, inner sep=0pt] at (1.15,-2.15) {\hlf{discounted}};
\draw[e] (fb)--(bb); \draw[e] (bb)--(ab); \draw[e] (bb)--(rb);
\node[verdict] at (0,-3.2) {Definition~1: modified\\Definition~2: \textbf{not modified}};
\end{scope}
\begin{scope}[xshift=10.4cm]
\node[ttl] at (0,1.0) {(c) After Listing~\ref{lst:struct}};
\node[n] (fc) at (0,0.3) {function\_definition};
\node[n] (bc) at (0,-0.6) {block};
\node[new] (ic) at (-1.15,-1.5) {if\_statement};
\node[n] (rc) at (1.15,-1.5) {return\_statement};
\node[new] (ric) at (-1.15,-2.4) {return\_statement};
\node[lf] at (1.15,-2.1) {price\enspace 0.9};
\draw[e] (fc)--(bc); \draw[e] (bc)--(ic); \draw[e] (bc)--(rc); \draw[e] (ic)--(ric);
\node[verdict] at (0,-3.2) {Definition~1: modified\\Definition~2: \textbf{modified}};
\end{scope}
\end{tikzpicture}}
\caption{ASTs of the function in Listings~\ref{lst:leaf} and~\ref{lst:struct}, with tree-sitter's Python node names: (a)~before either edit, (b)~after the leaf-only change, in which only leaves (highlighted in yellow) change and the tree keeps its shape, and (c)~after the structural change, in which new nodes (red) change the shape of the tree. Grey text under a node gives its leaves. For readability, the trees omit the nodes that wrap an expression and the nodes of the arithmetic and the comparison.}
\label{fig:ast_example}
\end{figure}
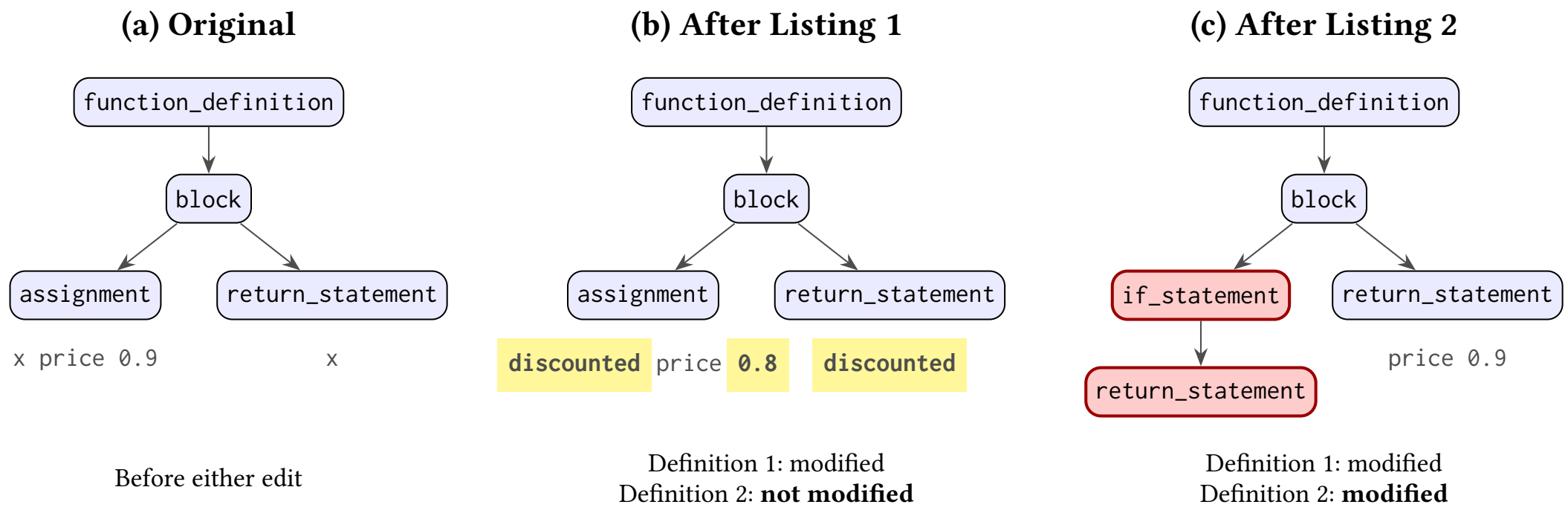

\noindent\textbf{Definition 2: AST-structural.} A line is modified only when the edit changes the structure of the function, method, or class that contains it, that is, the shape of its abstract syntax tree (AST, Figure~\ref{fig:ast_example}). An edit that changes only leaves, such as names, values, or formatting, is a \textit{leaf-only} change and is not counted.

Listings~\ref{lst:leaf} and~\ref{lst:struct} show two edits of the same Python function, and Figure~\ref{fig:ast_example} shows the same edits as ASTs. In Listing~\ref{lst:leaf}, the developer renames the variable \texttt{x} to \texttt{discounted} and changes the discount from 0.9 to 0.8. The function still does the same thing in the same order: it computes one value and returns it. In Listing~\ref{lst:struct}, the developer adds a condition, so that prices above 100 get a larger discount. The function now takes a different path depending on its input.

\begin{lstlisting}[language=Python, caption={A leaf-only change: a rename and a new value.}, label=lst:leaf]
 def apply_discount(price):
-    x = price * 0.9
-    return x
+    discounted = price * 0.8
+    return discounted
\end{lstlisting}

\begin{lstlisting}[language=Python, caption={A structural change: a new condition.}, label=lst:struct]
 def apply_discount(price):
-    x = price * 0.9
-    return x
+    if price > 100:
+        return price * 0.8
+    return price * 0.9
\end{lstlisting}

An AST represents code as a tree of its building blocks, such as a function, a block of statements, an assignment, or a condition. The names and values in the code, such as \texttt{x} or \texttt{0.9}, are the leaves of this tree. Figure~\ref{fig:ast_example}a shows the tree of the original function. After the edit in Listing~\ref{lst:leaf}, only the leaves change, and the tree keeps its shape (Figure~\ref{fig:ast_example}b). After the edit in Listing~\ref{lst:struct}, the assignment is replaced by an \texttt{if\_statement} node, and the shape of the tree changes (Figure~\ref{fig:ast_example}c). Definition~1 counts both edits as modifications, because both change the text of the lines. Definition~2 counts only the second, because only it changes the structure of the function.

We use {\tt tree-sitter}~\cite{treesitterIntroductionTreesitter} with language-specific grammars covering 17 languages to parse each modified file. For each modified line, we locate the smallest enclosing named AST node in the parse tree at the merge commit (function, method, or class definition), match it by name in the modification parse tree, and compare structural signatures. Lines with no enclosing named node (e.g., top-level comments or blank lines) receive no AST status.

AST classification is computed automatically and deterministically by tree-sitter. It is available for 81.4\% of the 149,468 modification events. The rest have no AST status because the modified line has no enclosing named node, the file could not be parsed, or its language is not supported. To check the rule, we drew a sample of 383 modifications with an AST status (95\% confidence, 5\% margin of error) from an earlier mining of 142 of the cohort's repositories. The sample is stratified by tier and by the rule's label, and matches the population's language mix. The first author labeled each sampled modification as leaf-only or structural, without access to the rule's label, following a written guide that asks whether the commit changed the structure of the enclosing unit or only its leaf tokens. The rule agrees with the human labels on 93.7\% of the sample (bootstrap 95\% CI [91.4, 96.1], Cohen's $\kappa = 0.60$ [0.45, 0.74]). Cohen's $\kappa$ is lower than the raw agreement because most modifications are structural, so two raters would often agree by chance.

Because proportional strata leave only 26 modifications with the leaf-only label, we labeled 70 more leaf-only modifications, shuffled with 30 structural ones so that the label could not be inferred, which gives 32 leaf-only modifications per tier. When the rule reports a leaf-only change, the human label agrees in 86.5\% of cases (Wilson 95\% CI~\cite{wilson1927probable} [78.2, 91.9]), and when it reports a structural change, in 94.7\% [91.8, 96.6]. Agreement is high for named functions, methods, and classes (95.7\%) and lower for anonymous functions (73.0\%) and module-level statements (58.3\%), which the rule identifies by their text, so that reformatting them is counted as structural.



Under Definition~2, a modified line without an AST status is not counted as a structural modification and is censored at the time of that modification, so both definitions are fitted on the same 644,177 units. Dropping these lines instead changes the hazard ratios by at most 0.006 (human-supervised 0.919 to 0.914, Devin 0.847 to 0.841). Because the AST status is unknown for some modifications, raw Definition~2 modification rates are a lower bound and are compared across tiers, not with Definition~1 rates.

\subsection{Intent Classification}
\label{sec:intent_method}

Knowing \textit{why} code from coding agents is modified matters as much as knowing \textit{how often}. A modification rate driven by defect fixes signals lower quality than the same rate driven by feature extensions. Prior post-merge studies infer intent with keyword or Conventional Commit rules, sometimes with an LLM fallback (Table~\ref{tab:prior_designs}). We compare both families of classifier on the same commits, and validate them against human labels.

We classify the intent of each of the 5,303 modification commits, the commits that first modify at least one line of the cohort. We use the three maintenance categories of Swanson~\cite{swanson1976dimensions}, corrective, adaptive, and perfective, extended with preventive maintenance as in ISO/IEC/IEEE 14764~\cite{iso14764}. Following Mockus and Votta~\cite{mockus2000identifying}, we treat the addition of new functionality as adaptive. Our four categories are therefore \textit{corrective (fixing a defect)}, \textit{perfective (improving existing functionality)}, \textit{adaptive (adding new functionality or adapting to requirements)}, and \textit{preventive (preventing future problems)}. In the following, we describe each classification approach:

\noindent\textbf{Approach 1: Keyword-based classification (baseline).} Keyword-based commit classification was pioneered by Mockus and Votta~\cite{mockus2000identifying} and has been widely adopted in subsequent MSR work~\cite{hindle2009automatic,levin2017boosting,mazloomzadeh2026aicodingagentscontribute,horikawa2025agentic}. We follow this established approach as our baseline, with a rule-based classifier in three stages:
\begin{enumerate}[leftmargin=*,nosep]
    \item Conventional Commits prefixes~\cite{conventionalCommits} on the first line of the message (\texttt{fix:}, \texttt{feat:}, \texttt{chore:}) are mapped to the taxonomy with a lookup table.
    \item Bracket-style prefixes on the first line (\texttt{[fix]}, \texttt{[improve]}) are mapped in the same way. If that line also contains defect-fix language, the commit is labeled corrective.
    \item For messages without a recognized prefix, 86 regular expressions are matched against the full message in a fixed priority order: corrective, perfective, adaptive, and preventive.
\end{enumerate}
 We put corrective first because a defect-fix signal usually marks the main purpose of a commit, whereas Mockus and Votta~\cite{mockus2000identifying} resolve conflicts by counting keywords per category. The classifier returns the first matching category in this order.

\noindent\textbf{Approach 2: LLM-based classification.} Every other manual step in this paper checks an automated measurement on a sample: the attribution of PRs, the tracking of lines, and the AST rule. Intent is different. Classifying it means reading each commit and judging its main purpose, and the 5,303 modification commits are too many to judge by hand. Mining studies have used keyword rules for this reason (Approach~1). LLMs are now used in place of human annotators for software engineering artifacts~\cite{hou2024large,ahmed2025can}. We therefore test whether LLM classifiers can take the place of keyword rules, and validate both against a human-labeled sample (below). Because a single model's labels can reflect that model's idiosyncrasies, we use three LLMs from different model families. We called them through Together AI~\cite{togetherServerless,togetherModels} between 27 October and 3 November 2026, and on 6 November 2026 for the message-only check in RQ4. Together AI's serverless model names do not refer to dated snapshots. Classifying the 5,303 commits with all three models used about 216 million input and 4.5 million output tokens, which cost about US\$47 for the three models together at Together AI's list prices (accessed 16 November 2026)~\cite{togetherModels}.
 The three models are DeepSeek-V4.1-Flash (\texttt{deepseek-ai/DeepSeek-V4.1-Flash}), gpt-oss-120b (\texttt{openai/gpt-oss-120b}), and MiniMax-M3.1-Flash (\texttt{MiniMaxAI/MiniMax-M3.1-Flash}). All three receive the same prompt, which defines the four categories and instructs the model to identify the single dominant purpose of the commit, weighted by the scope and magnitude of each change, rather than an incidental secondary purpose. Each model receives the full commit message, the complete file list, and the diff. Diffs exceeding 200,000 characters are truncated to their first and last 100,000 characters, which preserves both the beginning and the end of large diffs. We set the temperature to 0. All three models reason before answering, so we allow up to 4,000 output tokens and retry once with 8,000 when an answer is truncated or empty. We cache every call's raw output, finish reason, and token counts. The full prompt is available in our replication package~\cite{replicationpackage}.

\noindent\textbf{Combining LLM labels.} Beyond the three individual classifiers, we evaluate three ways of combining their labels. The first is an unweighted majority vote. When all three models disagree, the vote is left undecided, and in a sensitivity analysis we assign it to the model that agrees best with the human labels. The second is a vote weighted by each model's agreement with the human labels. We estimate these weights on 9 of 10 folds and apply them to the remaining fold, so that no commit is scored with weights estimated from its own label. The third is Dawid-Skene aggregation~\cite{dawid1979maximum}, a statistical model that estimates each model's error rates from the agreement between the models alone, without human labels. We designate no approach as primary, and RQ4 reports intent results under every approach.

\noindent\textbf{Validation.} We validate the classifiers against commits that the first author labeled by primary intent, following a written annotation guide that uses the same category definitions as the LLM prompt, without access to any classifier output. The labels come in three rounds. First, before we mined the final 215-repository cohort, we drew a stratified random sample of 385 modification commits, with about equal allocation across the AIDev agent groups and the human baseline. This is the minimum for 95\% confidence and a 5\% margin of error. Second, we drew 148 commits (seed 42) from modification commits that the first sample could not contain. Both samples came from the 142 repositories of an earlier mining of the cohort. Third, to cover the other repositories, we drew 256 commits from the 2,202 modification commits of the other 55 repositories that have modification commits, in proportion to their tiers (seed 20261103). Of the 789 labeled commits, 618 are modification commits of our corpus (224, 144, and 250 from the three rounds), 368 in the 142 repositories and 250 in the others, the same 59/41 split as the corpus. The other 171 lie in cohort repositories but are not first modifications of a line in our corpus. Of these, 164 modify no line that the cohort's PRs added under identity tracking, because the first two rounds were drawn from an earlier mining that tracked line numbers. The other 7 modify only vendored lines.
For a proportion estimated from these strata, the worst-case margin of error is $\pm 3.71\%$ at 95\% and $\pm 4.87\%$ at 99\% confidence, against the 590 labels that Cochran's formula~\cite{cochran1977sampling} requires for $\pm 5\%$ at 99\%. We measure agreement with Cohen's $\kappa$~\cite{cohen1960coefficient} and report bootstrap 95\% confidence intervals, both unweighted and weighted to the strata of the corpus.



\noindent\textbf{Correcting shares for classifier error.} Even a good classifier misclassifies some commits, and its errors need not be the same for every category. We therefore also report intent shares corrected for each classifier's error. On the 618 labeled commits, we estimate the classifier's confusion matrix, the table of how often each true intent is labeled as each category. We invert it to recover the shares that the classifier would report without error. We then bootstrap the labeled commits and the corpus to obtain 95\% confidence intervals.

\subsection{Modifier Identity}
\label{sec:modifier_method}

Beyond who writes agent-attributed code, we examine who subsequently modifies it. For each modification commit (the commit that first changes a code unit after its PR is merged), we query the GitHub REST API for commits~\cite{githubRestCommits} and extract the \texttt{author.login} field. We classify a modifier as an \textit{agent} if the login is a GitHub App account (ending in \texttt{[bot]}) or ends in \texttt{-bot}. We also classify as agents the known coding-agent and automation accounts, such as \texttt{Copilot}, \texttt{cursoragent}, and \texttt{devin-ai-integration[bot]}. The full list is in our replication package~\cite{replicationpackage}. Every other login is \textit{human}. The coding-agent accounts are those through which the five AIDev agents~\cite{li2025rise} commit, so modifier attribution uses the same agent identities as the attribution of the original PR. We deliberately do not match vendor names as substrings: doing so labels human staff accounts whose login merely contains a vendor name (for example, a login ending in \texttt{-openai}) as agents. We audited every distinct modifier login in the corpus against this rule.

\noindent\textbf{Variants.} Prior studies count follow-up work in different ways (Table~\ref{tab:prior_designs}), so we vary three choices in RQ5. Firstly, the unit: every first modifying commit (our default), every modified line, and every follow-up PR, that is, a later PR that contains at least one commit modifying the agent PR's lines, which is the unit of Takerngsaksiri et al.~\cite{takerngsaksiri2026finishes}. We map modifying commits to PRs with the GitHub REST API (\texttt{GET /repos/\{owner\}/\{repo\}/commits/\{sha\}/pulls})~\cite{githubRestCommits}. Secondly, the window: modifications within 7 days, within 30 days, and within the full observation window of each line. Thirdly, the identification rule: the commit author (our default) and the commit author together with any \texttt{Co-authored-by} trailer that names a coding agent. We also considered the committer. However, 4,693 of the 5,291 commits with metadata were committed by GitHub's \texttt{web-flow} account, under which GitHub records and signs the commits made through its web interface, such as PR merges~\cite{githubWebSigning}. No commit has an agent as committer, so this rule cannot separate agents from humans. We include the trailer rule because coding agents such as Claude Code add such a trailer by default when a developer commits their changes~\cite{claudeCodeSettings}. A commit authored by a human may therefore contain agent-written changes. A commit that modifies lines of more than one group (530 of the 5,303 commits, 10.0\%) is counted under the group whose lines it mostly modifies. As a sensitivity check, we also count it under every group it modifies, and, separately, leave it out.

\subsection{Statistical Analysis}
\label{sec:stats}

Having defined the code units, the modification events, their intent, and who made them, we now describe how we compare retention between groups. We use survival analysis because many lines are still unmodified at the end of the window: their time to modification is censored, and survival methods use these lines correctly instead of dropping them or treating them as never modified~\cite{kalbfleisch2002statistical}. Survival and lifetime analyses are established in software engineering: they have been used for the lifetime of code lines and tokens~\cite{spinellis2021software}, statically detected vulnerabilities~\cite{dipenta2009life}, static-analysis warnings~\cite{aman2019survival}, open-source projects~\cite{samoladas2010survival}, and developers~\cite{lin2017developer}, and for code from coding agents~\cite{xia2026these,rahman2026will}. We follow this line of work. Every model compares one agent tier with the human baseline, on the line or hunk units of Section~\ref{sec:operationalization}.

\paragraph{Kaplan-Meier Estimation} Kaplan-Meier estimation~\cite{kaplan1958nonparametric} is the standard non-parametric method for estimating time-to-event distributions, widely used in SE studies of code evolution~\cite{aman2019survival,spinellis2021software}. We estimate the retention function $S(t)$, the probability that a code unit remains unmodified beyond time $t$:
\begin{equation}
    \hat{S}(t) = \prod_{t_i \le t} \left(1 - \frac{d_i}{n_i}\right)
\end{equation}
where $d_i$ is the number of modifications at time $t_i$ and $n_i$ is the number of unmodified units just prior to $t_i$.

\paragraph{Log-Rank Test} We use the log-rank test~\cite{mantel1966evaluation}, the standard non-parametric test for comparing survival distributions between groups, previously applied in SE contexts by Aman et al.~\cite{aman2019survival} and Lin et al.~\cite{lin2017developer}. We test whether retention distributions differ significantly between groups:
\begin{align*}
H_0 &: S_{\text{group}}(t) = S_{\text{human}}(t) \quad \text{for all } t \\
H_1 &: S_{\text{group}}(t) \neq S_{\text{human}}(t) \quad \text{for some } t
\end{align*}

\noindent\textbf{Cox regression across repositories.} We fit a Cox proportional hazards model~\cite{cox1972regression,therneau2000cox} with a single indicator for the agent tier, against the human reference. Its hazard ratio (HR) compares how quickly agent and human lines are modified: an HR below 1 means that agent lines are modified later or less often than human lines. In this model, a Devin line in one repository is compared directly with human lines in any of the 215 repositories. This is the comparison \textit{across repositories} that pooled designs make~\cite{popescu2026investigating}.

\noindent\textbf{How to read a hazard ratio.} Let us take an arbitrary example. Suppose that on a given day, 1 in 1,000 unmodified human lines is modified. A hazard ratio of 0.8 for agent code then means that 0.8 in 1,000 unmodified agent lines are modified on the same day. Agent code is thus 20\% less likely to be modified at any moment. A hazard ratio of 1.5 means 1.5 in 1,000, that is, 50\% more likely. A 95\% confidence interval that contains 1 means that the data are consistent with no difference.

\noindent\textbf{Clustered standard errors.} Lines of the same repository, and of the same PR, are not independent observations, so we cluster the standard errors of every across-repository model by repository~\cite{lin1989robust,cameron2015practitioner}. Model-based standard errors treat each of the 644,177 lines as independent. For the Devin line model, the standard error of the log hazard ratio is 0.008 when lines are treated as independent and 0.38 when they are clustered by repository. Lines of one repository are modified by the same commits, so 644,177 lines carry far less information than 644,177 independent observations would. Treating them as independent makes the intervals about 50 times too narrow, so that almost any difference looks significant. We report errors clustered by PR as a sensitivity check.

\noindent\textbf{Cox regression within repositories.} A comparison across repositories mixes the effect of the tier with the effect of the project: if Devin happens to be used in projects that change code quickly, Devin code will look short-lived even if it is not. To compare code only with code from the same project, we give each repository its own baseline hazard (stratified partial likelihood~\cite{kalbfleisch2002statistical,therneau2000cox}), so that at each modification time, a line is compared only with lines of the same repository. Repositories that contain only one of the two groups contribute nothing to this comparison. In total, 128 repositories contribute to the human-supervised comparison and 31 to the Devin comparison. We cluster the standard errors by repository~\cite{lin1989robust,cameron2015practitioner}, because lines of the same project are not independent. Xia and Miller use the same stratification for post-merge code~\cite{xia2026these}.

\noindent\textbf{Cluster bootstrap.} With only 31 repositories that contain both Devin and human code, the analytic standard errors of the within-repository model can be unreliable. For the headline within-repository estimates, we therefore resample whole repositories with replacement 500 times (seed 42), refit the model on each resample, and report the 2.5th and 97.5th percentiles of the hazard ratio as the 95\% confidence interval~\cite{cameron2015practitioner}. Resampling repositories rather than lines keeps each project's lines together, so the interval reflects how much the result depends on which projects are in the sample.

\noindent\textbf{Multiple comparisons.} Because each analysis compares two tiers with the human baseline, we judge significance after a Bonferroni correction for two comparisons~\cite{dunn1961multiple}, which divides the significance level by the number of comparisons ($\alpha = 0.05 / 2 = 0.025$).

\noindent\textbf{Weighting.} When every line counts once, a PR with thousands of lines outweighs hundreds of small PRs. To give each PR the same weight, we also fit the models with each line weighted by one over the number of lines in its PR, with robust standard errors clustered by repository. At the hunk level, each hunk counts once, and a hunk is modified at the first verified modification of any of its lines.

\noindent\textbf{Proportional hazards assumption.} A Cox hazard ratio is a single average over time. Schoenfeld residual tests~\cite{schoenfeld1982partial} reject a constant hazard ratio in every model ($p < 0.001$). With 644,177 line units, the test detects even small deviations~\cite{therneau2000modeling}, but the deviations here are not small: the hazard ratio changes with the time since merge. We therefore report Cox hazard ratios as average effects over the window. We also fit piecewise Cox models~\cite{therneau2000modeling}, which estimate a separate hazard ratio for 0 to 30, 30 to 90, and more than 90 days after merge, with standard errors clustered by repository. A time window chosen after seeing the data can produce a spurious effect. We therefore test any window-specific effect on held-out repositories. We split the repositories at random into two halves, choose the window on one half, and test it on the other, over five random splits.

\noindent\textbf{Specification curve.} To show how the choices act together, we fit the tier-versus-human Cox model under every combination of choices that our data allow~\cite{simonsohn2020specification,cassee2026exploring}. The combinations cover five line sets, the unit (line, line weighted by PR, or hunk), the modification definition (syntactic or AST-structural, for lines tracked by identity or by line number), and the comparison (across or within repositories). Each line set pairs a tracker with the lines it is applied to. Two cover all lines including vendored code, one with the tracker of our earlier study and one with identity tracking. One applies identity tracking to all lines without vendored and generated code. The last two apply identity and line-number tracking to the sampled units. Some combinations cannot be built, for example the AST-structural definition for hunks or for the tracker of our earlier study. This leaves 44 specifications per tier and 88 in total. Of these, 56 use the three line sets without vendored code. For each specification, we record the hazard ratio and its 95\% confidence interval.

We classify the verdict as \textit{retained longer} (the interval lies below 1), \textit{modified sooner} (above 1), or \textit{no difference}. The intervals of every specification use analytic standard errors clustered by repository, because bootstrapping all 88 specifications is not feasible. The headline within-repository results use the cluster bootstrap.

As sensitivity checks, we recount the verdicts with standard errors clustered by PR and with errors that treat units as independent, which is what most pooled designs report. These are model-based errors for unweighted units and unclustered robust errors for units weighted by PR, whose weights are not frequency counts.

We also vary the observation window for Devin, the only tier with enough PRs to do so. The first variant is the full window, with at least five months of follow-up for every PR. The second keeps only PRs with at least eight months of follow-up, and the third censors the same PRs at 240 days. The fourth keeps only PRs with at least ten months, censored at 300 days. We added the window dimension after seeing the results of the other dimensions, so it is not a pre-planned choice. We therefore test the window variants against permuted labels, like the rest of the curve, and add two variants that separate the follow-up length from the cohort (Section~\ref{sec:rq1} and Section~\ref{sec:synthesis}).

\noindent\textbf{Inference on the curve.} A curve of many specifications produces some significant verdicts by chance, and the specifications are not independent. Following Simonsohn et al.~\cite{simonsohn2020specification}, we therefore compare the observed curve with curves under the null hypothesis of no tier difference. We permute (shuffle) the tier labels among PRs 500 times, across repositories among all PRs and within repositories among the PRs of each repository. We refit every specification on each permutation. We then report how often the permuted curves give at least as many significant verdicts in one direction, and a median hazard ratio at least as far from 1, as the observed curve. We also report the share of specifications whose point estimate lies below 1. For the headline within-repository estimates, we report the minimum detectable hazard ratio~\cite{schoenfeld1983sample}, the smallest effect that the comparison would detect with 80\% probability at $\alpha = 0.05$, computed from the bootstrap standard error.
The re-runs of published designs are described with their results in Section~\ref{sec:synthesis}.

\section{RQ1: How Do Line Tracking and the Unit of Analysis Change the Verdict?}
\label{sec:rq1}


\noindent\textbf{Objective.}

Every post-merge study must decide how to follow a line through later commits and what to count. Prior studies follow lines through diffs or by their position, and count lines, files, or PRs (Table~\ref{tab:prior_designs}). If these choices change the verdict, then two studies of the same agents can disagree without either being wrong, and a single reported hazard ratio says as much about the pipeline as about the agents. The goal of RQ1 is to measure how much line tracking and the unit of analysis, each defensible on its own, move the verdict for the same code.

\subsection{Approach}

To isolate these choices, we hold the definition (syntactic) and the comparison (across repositories) fixed and fit one unadjusted Cox model per agent tier against the human baseline (Section~\ref{sec:stats}). We start from the units and the tracker of our earlier study (Section~\ref{sec:earlier}) and change one or two choices at each step. First, we compare the three trackers of Section~\ref{sec:operationalization} on the same 276,314 sampled units, so that only the tracking differs. We also compare the tracker of our earlier study with identity tracking line by line on all 854,139 lines. Second, on all lines tracked by identity, we remove the vendored and generated lines. Third, we compare the sampled units with all lines. Fourth, on all lines, we change the unit: each line counts once, each PR counts once (each line weighted by one over the lines in its PR), or each hunk counts once. We also split hunks by size into single lines, small blocks (2 to 10 lines), and large blocks (11 or more lines), because large commits differ in kind from small ones~\cite{hindle2008large}. A single line is typically a minimal fix or tweak, while a block of more than ten lines is typically a whole function or a new feature. Fifth, because the observation window decides which PRs enter the data, we keep only PRs with at least eight months of follow-up. We then separate the effect of this cohort from that of the longer follow-up (Section~\ref{sec:operationalization}).

\subsection{Findings}

\begin{figure}[t]
    \centering
    \includegraphics[width=\textwidth]{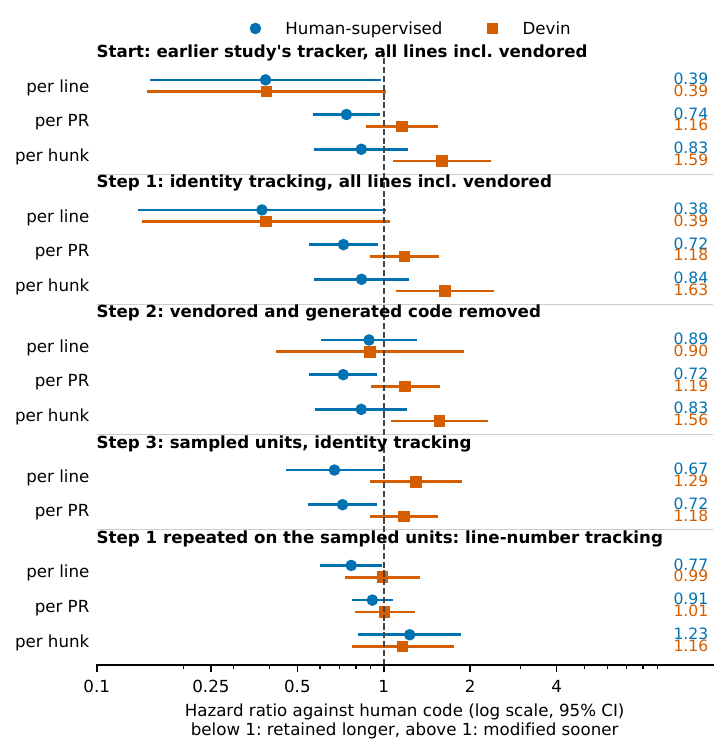}
    \caption{The measurement ladder of RQ1, starting from the units and tracker of our earlier study (Section~\ref{sec:earlier}). Hazard ratio of each tier against human code (syntactic definition, across repositories), with 95\% confidence intervals from standard errors clustered by repository. Each block title names the step of the approach that the block adds, and the rows within a block change the unit (Step~4). The last block repeats the tracker change of Step~1 on the sampled units. The number at the right of each interval is its hazard ratio. Per PR: each PR counts once, whatever its number of lines. Sampled units: at most 20 files per PR and 100 lines per file. Hunks were not built for the sampled units under identity tracking, so Step~3 has no per-hunk row. Values below 1 mean that agent code is retained longer.}
    \label{fig:rq1_ladder}
\end{figure}

Figure~\ref{fig:rq1_ladder} shows how the hazard ratio of each tier moves along the first four steps, starting from the units and tracker of our earlier study. Step~5, the observation window, is reported below the ladder. We first describe the retention of all lines under our full measurement, and then each step.

\subsubsection*{Retention under full measurement}

\begin{figure}[t]
    \centering
    \includegraphics[width=\textwidth]{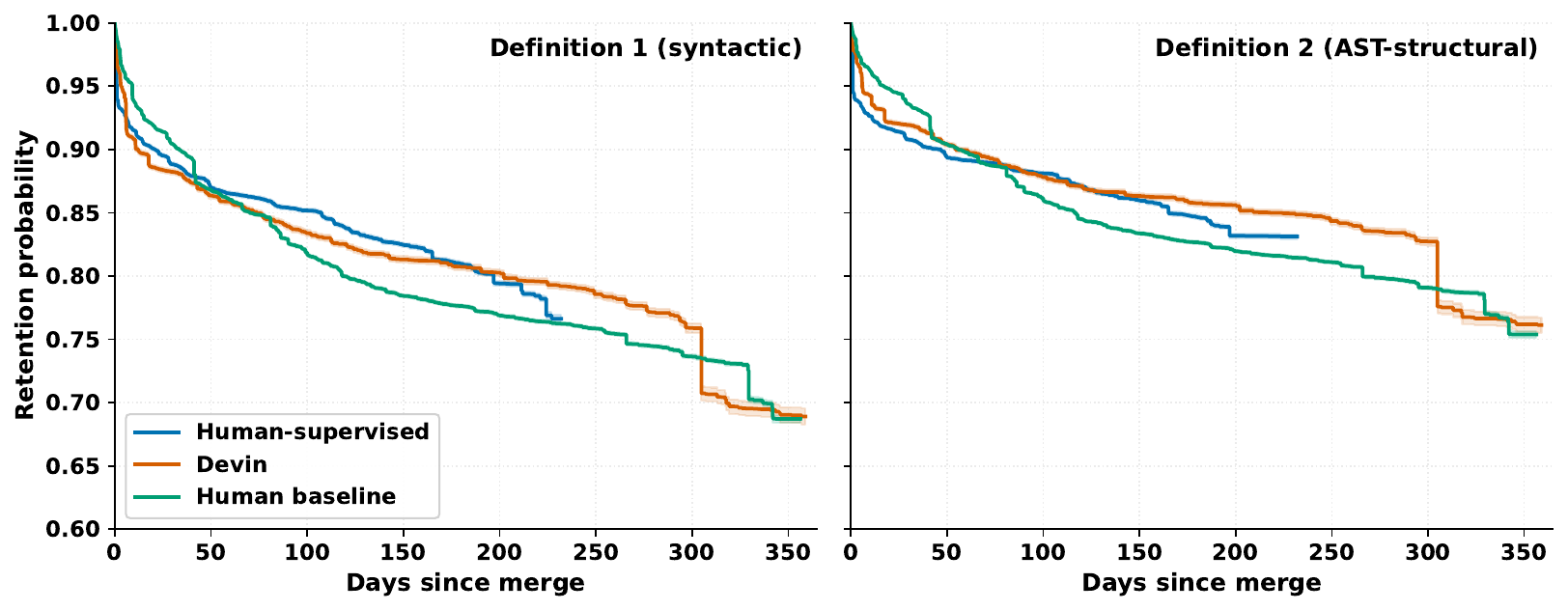}
    \caption{Kaplan-Meier retention of all lines by group, identity tracking. Left: syntactic definition (Definition~1, RQ1). Right: AST-structural definition (Definition~2, RQ2). Retention probability is the estimated share of a group's lines not yet modified. Higher curves mean longer time to first modification, and shaded regions are 95\% confidence intervals. The y-axis starts at 0.6. Each curve stops where fewer than 1\% of the group's lines remain at risk: day 232 for human-supervised, 359 for Devin, and 356 for the human baseline. The human-supervised curve stops earliest because 1,053 of its 1,055 PRs were merged from May to July 2025, so none of them has a full year of follow-up before the end of the data (31 December 2025).}
    \label{fig:km_def1}
\end{figure}

With identity tracking on all lines, 20.1\% of the lines from human-supervised agents, 21.5\% of the lines from Devin, and 25.0\% of human lines are modified within the window (Figure~\ref{fig:km_def1}). Fewer than half of the lines in any group are modified, so no median retention time is reached. The point estimates show both tiers retained slightly longer than human code (human-supervised HR\,=\,0.888, Devin HR\,=\,0.896): at any point in time, an agent line is about 10\% less likely to be modified than a human line. Neither difference is significant once the standard errors are clustered by repository (95\% CI [0.612, 1.289] and [0.425, 1.889]). With model-based errors, which treat every line as independent, both would be significant ($p < 0.001$, intervals [0.877, 0.899] and [0.882, 0.910]). The curves also show that the hazard is not constant over time. In the first weeks after merge, lines of both tiers are modified slightly sooner than human lines: 30 days after merge, 88.8\% of human-supervised lines, 88.2\% of Devin lines, and 90.4\% of human lines are still unmodified. The curves cross about 45 to 60 days after merge, and for most of the following months the agent curves lie above the human one. RQ3 returns to this pattern with piecewise models. The steps below show that the point estimates, too, move with how they are measured.

\subsubsection*{Step 1: three trackers on the same units}

On the same units, line-number tracking records about twice as many modifications as identity tracking: 64.5\% instead of 29.0\% of human lines, 63.8\% instead of 34.9\% of Devin lines, and 46.8\% instead of 18.3\% of human-supervised lines (Table~\ref{tab:rq1_tracking}). Line-number tracking inflates the modifications of human code more than those of Devin code (by 35.5 against 28.9 points), so Devin code looks relatively less modified, and its hazard ratio falls from 1.294 under identity tracking to 0.992. Neither estimate is significant with standard errors clustered by repository (95\% CI [0.904, 1.853] and [0.741, 1.327]), so the tracker changes Devin's point estimate but not its verdict. For human-supervised code, the direction holds and both estimates are significant (0.770 [0.607, 0.976] under line-number tracking, and 0.673 [0.463, 0.979] under identity tracking). The tracker of our earlier study is close to identity tracking: it records slightly more modifications (32.4\%, 37.7\%, and 21.4\%) and gives similar hazard ratios (1.260 for Devin, 0.707 for human-supervised code). It agrees with identity tracking on whether a unit is modified for 96.8\% of the units, against 62.1\% for line-number and identity tracking. A manual check of 300 units confirms that most of the extra modifications under line-number tracking are spurious. Only 35.9\% of its ``modified'' claims are correct, against 98.7\% for identity tracking and 93.3\% for the tracker of our earlier study (Section~\ref{sec:operationalization}). About two in three modifications that line-number tracking reports therefore never happened.

On all 854,139 lines, the tracker of our earlier study and identity tracking give the same status for 96.8\% of the lines that identity tracking verifies. They name the same modifying commit for 88.9\% of the lines that both record as modified. The disagreements go almost entirely one way: 26,963 lines that the earlier tracker records as modified are retained under identity tracking, against 121 in the other direction. We inspected a random sample of 1,000 of them. Of these, 56.3\% are lines whose file was renamed or moved, 20.5\% lines moved within their file, 11.8\% lines moved or copied from another file, and 2.9\% lines changed only on a side branch. The earlier tracker counts each of these as a modification.

\begin{table}[t]
\centering
\caption{Three trackers on the same 276,314 sampled units: share of lines modified within the window and unadjusted hazard ratio against human code. Earlier: the tracker of our earlier study (Section~\ref{sec:earlier}).}
\label{tab:rq1_tracking}
\begin{tabular}{lrrrrrr}
\toprule
& \multicolumn{2}{c}{\textbf{Line number}} & \multicolumn{2}{c}{\textbf{Identity}} & \multicolumn{2}{c}{\textbf{Earlier}} \\
\cmidrule(lr){2-3}\cmidrule(lr){4-5}\cmidrule(lr){6-7}
\textbf{Group} & \textbf{Modified} & \textbf{HR} & \textbf{Modified} & \textbf{HR} & \textbf{Modified} & \textbf{HR} \\
\midrule
Human-supervised & 46.8\% & 0.770 & 18.3\% & 0.673 & 21.4\% & 0.707 \\
Fully-autonomous (Devin) & 63.8\% & 0.992 & 34.9\% & 1.294 & 37.7\% & 1.260 \\
Human baseline & 64.5\% & 1.000 & 29.0\% & 1.000 & 32.4\% & 1.000 \\
\bottomrule
\end{tabular}
\par\smallskip\parbox{\linewidth}{\footnotesize 95\% CIs with standard errors clustered by repository: line number, human-supervised [0.607, 0.976], Devin [0.741, 1.327]. Identity: human-supervised [0.463, 0.979], Devin [0.904, 1.853]. Identity tracking excludes 261 lines whose modification could not be verified. These are the part of the 680 unverified lines (Section~\ref{sec:operationalization}) that lies in the sampled units.}
\end{table}

Most of the extra modifications come from two sources. Under line-number tracking, 43.4\% of all recorded modifications are deletions of a whole file. Of these 70,254 file deletions, 16,537 are renames or moves, in which the file survives under a new path. Another 17,171 occur in commits whose file list was truncated, so the file only appeared to be missing. A further 46.9\% leave the line's content unchanged, because the line was only moved or its file renamed, which identity tracking follows and line-number tracking counts as a modification.

\subsubsection*{Step 2: removing vendored and generated code}

Our earlier study removed vendored and generated lines from its primary analysis but tracked them, and reported them as a sensitivity analysis. With identity tracking on all lines including them, the point estimates show both agent tiers retained much longer than human code (human-supervised HR\,=\,0.376, Devin 0.389). The reason is that 46.5\% of human lines are modified, against 20.4\% and 21.7\% of agent lines. Most of these modifications rewrite bundled or generated files in human PRs: 96\% of the 205,180 vendored lines come from human PRs, most of them from one bundled file. With vendored files kept, human code thus looks modified almost three times as often as agent code, simply because projects regenerate their bundled files. Because these lines sit in a few repositories, the clustered intervals are wide and just include 1 ([0.140, 1.009] and [0.146, 1.040]). The tracker of our earlier study gives the same picture on the lines that include vendored code (0.388 and 0.390). Removing these lines, together with 4,102 lines in generated paths, lowers the human modification rate to 25.0\% and moves both hazard ratios to about 0.9 (0.888 and 0.896). Within repositories, the removal matters much less: with these files kept, the hazard ratio is 0.778 (95\% CI [0.510, 1.119]) for human-supervised code and 0.715 ([0.382, 1.713]) for Devin code, neither significant. A study that keeps such files across repositories compares how often projects regenerate their build output, not how long agent and human code is retained.

\subsubsection*{Step 3: sampled units or all lines}

The position-based miner samples at most 20 files per PR and 100 lines per file, a common way to keep mining tractable. The cap drops 57.1\% of Devin lines, 61.2\% of human lines, and 48.1\% of human-supervised lines. Re-applying it to the identity-tracked data moves Devin's hazard ratio from 0.896 to 1.292 (1.289 to 1.293 over five random samples) and human-supervised from 0.888 to 0.674. Sampling within PRs is therefore not neutral: it removes most of the lines of the largest PRs, which, as the next step shows, behave differently from the rest.

\subsubsection*{Step 4: what each unit counts}

On all lines, the verdict for Devin depends on what counts once. When each line counts once, the point estimate shows Devin code retained slightly longer than human code (HR\,=\,0.896). When each PR counts once, it shows Devin code modified sooner (HR\,=\,1.187), and when each hunk counts once, sooner still (HR\,=\,1.56). With clustered standard errors, only the hunk estimate is significant (95\% CI [1.07, 2.28], against [0.425, 1.889] per line and [0.910, 1.548] per PR). At any moment, a block of Devin code is thus about 1.6 times as likely to be modified as a block of human code. For human-supervised code, all three units give HR below 1 (0.888, 0.722, 0.834), and the PR-weighted estimate is significant ([0.556, 0.939]).

One PR explains most of the difference. Devin's PR \#1059 in \texttt{pdfme/pdfme} adds 36,537 lines, 41.7\% of all Devin lines in the cohort, and only 4.8\% of them are modified within the window. The five largest Devin PRs hold 50.2\% of Devin lines, and 5.5\% of their lines are modified, against 37.5\% in the other 407 Devin PRs. When each line counts once, these few PRs set the verdict for Devin. When each PR counts once, they weigh as much as any other PR. Human and human-supervised code are less concentrated: their five largest PRs hold 25.0\% and 23.9\% of their lines. Half of the human-supervised lines come from 40 PRs, against 5 for Devin. The five largest human-supervised PRs are also modified as often as the others (20.2\% against 20.1\% of their lines). The repository with the most human-supervised lines holds 20.6\% of them. No single PR or repository therefore sets the human-supervised verdict.

Hunks show the same pattern (Table~\ref{tab:hunk_stratified}). The point estimates show Devin hunks of every size modified sooner than human hunks of the same size, most clearly small blocks of 2 to 10 lines (HR\,=\,1.736). With standard errors clustered by repository, the Devin estimates for small and large blocks are significant (95\% CI [1.11, 2.71] and [1.01, 2.05]), but the one for single lines is not ([0.86, 2.22]). The human-supervised point estimates lie below 1 for single lines and small blocks (HR\,=\,0.776 and 0.893) and close to 1 for large blocks (0.963), and none of them is significant. A hunk-level analysis gives the many small Devin contributions the weight that a line-level analysis gives to a few very large ones.

\begin{table}[t]
\centering
\caption{Hunk-level unadjusted hazard ratio against human hunks of the same size, identity tracking, syntactic definition, with 95\% CIs from standard errors clustered by repository. Agent hunk counts in parentheses.}
\label{tab:hunk_stratified}
\resizebox{\linewidth}{!}{%
\begin{tabular}{lrrr}
\toprule
\textbf{Tier} & \textbf{Single-line} & \textbf{Small block} & \textbf{Large block} \\
& \textbf{(1 line)} & \textbf{(2 to 10 lines)} & \textbf{(11+ lines)} \\
\midrule
Human-supervised & 0.776 [0.55, 1.10] (10,657) & 0.893 [0.57, 1.39] (5,384) & 0.963 [0.64, 1.45] (1,867) \\
Fully-autonomous (Devin) & 1.385 [0.86, 2.22] (2,932) & 1.736 [1.11, 2.71] (1,595) & 1.440 [1.01, 2.05] (968) \\
\midrule
Human baseline (N) & 21,051 & 11,646 & 3,934 \\
\bottomrule
\multicolumn{4}{l}{\footnotesize The pooled Devin hunk estimate (1.56) is also significant with repository-clustered errors.} \\
\end{tabular}}
\end{table}

\subsubsection*{Step 5: the observation window}

With the full window, every PR has at least five months of follow-up (Section~\ref{sec:operationalization}). Requiring at least eight months removes 142 of the 412 Devin PRs, which hold 61.7\% of Devin's lines, including pdfme PR \#1059. On this cohort, Devin's line-level estimate moves from 0.896 to 1.704 (95\% CI [1.00, 2.90]), and the lower bound of its interval touches 1. The restriction changes two things at once: it lengthens the follow-up, and it drops the early PRs. Two further estimates separate the two. Dropping only pdfme PR \#1059 from the full window gives 1.423 ([0.866, 2.339]). Removing this single PR of 36,537 lines thus turns Devin's estimate from ``slightly longer retained'' to ``about 40\% more likely to be modified'', although neither is significant. Dropping the 142 PRs but censoring the remaining ones at 150 days (five months), the follow-up of the full window, gives 1.717 ([1.010, 2.916]), the same as with eight months of follow-up. The shift therefore comes from which PRs enter the cohort, not from the longer follow-up, and it acts through the same concentration of lines as the unit in Step~4. The human-supervised tier cannot be tested this way, because only 3 of its PRs have eight months of follow-up. The Synthesis shows how the window acts on the whole specification curve, including a six-month cohort for human-supervised code.

\subsection{Comparison with Prior Work}

These results extend, to post-merge studies of agent code, what methods research has found for other mining tasks. As with the choice of diff algorithm for SZZ~\cite{nugroho2020different}, the way lines are followed changes the outcome, here the size and direction of the point estimate. As with effort-aware defect prediction~\cite{kamei2010revisiting}, weighting units by size can reverse a finding. They also help reconcile the published studies. Line-level designs that pool lines across PRs~\cite{xia2026these}, commit-level designs~\cite{popescu2026investigating}, and PR-level designs~\cite{takerngsaksiri2026finishes,kraishan2026agents} weigh the same code differently. Their verdicts therefore need not agree, even on the same agents. A design that keeps vendored files, moreover, measures those files as much as the code. Because neither weighting is wrong, we report both in the rest of the paper.

\begin{tcolorbox}[colback=gray!10, colframe=gray!50, title=RQ1 Finding]
  Line tracking, the files kept, and the unit move Devin's hazard ratio from below 1 to above 1 (0.90 per line, 1.56 per hunk), whereas the estimate for human-supervised code stays below 1 under every tracker and unit. With errors clustered by repository, few of these estimates are significant.
\end{tcolorbox}

\section{RQ2: Does the Definition of a Modification Change the Verdict?}
\label{sec:rq2}

\noindent\textbf{Objective.}

From RQ1, the verdict depends on how lines are followed and counted. Once a line is followed correctly, a study must still decide which of its changes count as a modification. Prior post-merge studies count any change that touches a line (Table~\ref{tab:prior_designs}), so a formatter run or an identifier rename counts as much as a rewrite of the logic. Methods research shows that such non-essential changes are common in version histories~\cite{kawrykow2011non} and inflate change-based measures such as SZZ~\cite{sliwerski2005changes,neto2018impact}. If code from one group receives more of them, a study that counts every touch would report a difference in formatting habits as a difference in how long code lasts. The goal of RQ2 is to measure whether the definition of a modification changes the verdict.

\subsection{Approach}

We hold the tracking (identity), the units (all lines), and the models of RQ1 fixed, and change only which events count as modifications. We report each estimate both across and within repositories, a choice that RQ3 examines in detail.

\noindent\textbf{Syntactic vs.\ AST-structural.} Starting from the line-level models of RQ1, which use Definition~1 (syntactic), we refit them with Definition~2 (AST-structural, Section~\ref{sec:operationalization}) as the event, both across and within repositories. Under Definition~2, a first modification that changes only the leaves of the syntax tree, such as a rename, a literal, or formatting, is not counted, and the line is censored at that point. Because only the event changes, any difference between the fits comes from the definition alone. The distinction follows prior work showing that syntax-tree differencing separates such edits from structural ones~\cite{fluri2007change}. We also refit Definition~2 after dropping the modified lines without an AST status, to check that they do not drive the result. Censoring a line at its first leaf-only edit removes it from the comparison even if its structure changes later. We therefore also follow the 8,317 lines whose first modification is leaf-only to their first structural modification or to the end of the window, and refit the models on these continued histories.

\noindent\textbf{Types of modification.} To see what each definition counts, we break the verified modification events of each group down by type: a change to the line's content, the deletion of the line, or the deletion of its file. We also report the share of content changes that touch only whitespace, which both definitions in prior work count as modifications.

\subsection{Findings}

We first compare the verdicts under the two definitions, and then show what kinds of change each group receives.

\subsubsection*{Syntactic vs.\ AST-structural definition}

\begin{figure}[t]
    \centering
    \includegraphics[width=\textwidth]{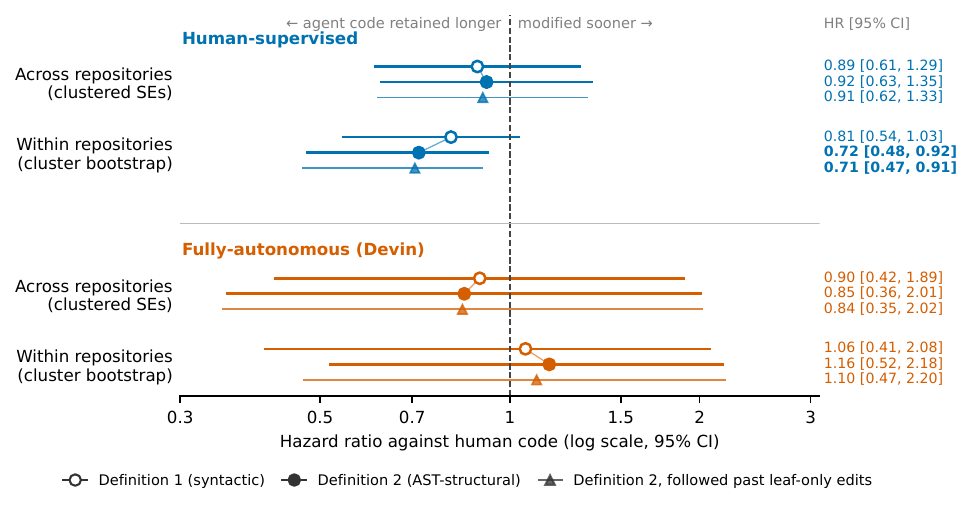}
    \caption{Hazard ratios against human code across and within repositories, under Definition~1 (syntactic, open circles) and Definition~2 (AST-structural, filled circles), with 95\% CIs (log scale), identity tracking, all lines. Triangles: Definition~2 with lines followed past their leaf-only edits. Across repositories, standard errors are clustered by repository. Within repositories, the CIs come from a 500-resample cluster bootstrap. Blue: human-supervised code. Orange: Devin code. The numbers on the right are the hazard ratios with their CIs, in bold where the CI excludes 1. Estimates below 1 mean agent code is retained longer than human code.}
    \label{fig:rq2_def}
\end{figure}

Figure~\ref{fig:rq2_def} shows the hazard ratios under both definitions, and Table~\ref{tab:rq2_def_comparison} the share of lines each definition counts as modified. The right panel of Figure~\ref{fig:km_def1} shows the Kaplan-Meier curves under Definition~2. Across repositories, the AST-structural definition moves the two point estimates in opposite directions: human-supervised code moves toward human code, and Devin code away from it. None of the four across-repository estimates is significant with standard errors clustered by repository.

Within repositories, the definition decides whether the human-supervised difference is detected. It is not significant under the syntactic definition, but significant under the structural one (HR\,=\,0.717, 95\% CI [0.477, 0.921]). At any point after merge, a line from a human-supervised agent is thus about 28\% less likely to be structurally modified than a human line in the same repository. The structural estimate stays significant after the Bonferroni correction for the two tiers (97.5\% CI [0.456, 0.950]). It also stays significant when lines are followed past their leaf-only edits, and the across-repository estimates hardly move when they are. Devin code does not differ from human code under either definition.

Over the specification curve, 24 specifications can be fitted under both definitions, and the two definitions give the same verdict in 19 of them. All five exceptions concern human-supervised code: in four, the structural definition turns ``no difference'' into ``retained longer'', and in one (line-number tracking, per line, across repositories) it turns ``retained longer'' into ``no difference''. For Devin, the definition never changes the verdict. The definition changes how large the effects are and, near the threshold, whether they are detected, but it reverses no direction.

\begin{table}[t]
\centering
\caption{Share of lines modified within the window under Definition~1 (syntactic) and Definition~2 (AST-structural), identity tracking, all lines. The hazard ratios are in Figure~\ref{fig:rq2_def}.}
\label{tab:rq2_def_comparison}
\small
\begin{tabular}{lrr}
\toprule
\textbf{Group} & \textbf{Syntactic} & \textbf{AST-structural} \\
\midrule
Human-supervised & 20.1\% & 15.7\% \\
Fully-autonomous (Devin) & 21.5\% & 15.3\% \\
Human baseline & 25.0\% & 18.9\% \\
\bottomrule
\end{tabular}
\end{table}

Devin's estimate moves most because its code receives the most leaf-only changes: among modified lines with a known AST status, 14.2\% of Devin modifications are leaf-only, against 7.0\% for human-supervised code and 5.8\% for human code. Our manual validation found the rule's leaf-only label correct in 86.5\% of cases (Section~\ref{sec:operationalization}). Its precision does not differ by tier (84.4\% for human-supervised, 93.8\% for Devin, and 81.2\% for human code, Fisher's exact test~\cite{fisher1922interpretation}, $p \ge 0.26$), so this ordering does not come from the rule's errors. The structural label is less precise for Devin than for human code (87.0\% against 96.7\%, $p = 0.011$). If anything, this understates Devin's share of leaf-only edits.

One commit in \texttt{lingodotdev/lingo.dev}, \textit{``fix: update PHP SDK code style to fix failing workflow (\#689)''}, shows the mechanism. It is the first modification of 128 lines written by Devin, and it only restyles them: 111 of the changes are leaf-only and 17 structural. Definition~1 records 128 modifications of Devin code, and Definition~2 records 17. Devin code is often restyled after merge, for example to pass a project's formatter or linter, and a syntactic definition counts each such line as a modification.

Definition~2 cannot classify every modification: 16.1\% of modified human-supervised lines, 16.9\% of Devin lines, and 19.8\% of human lines have no AST status. Dropping these lines changes the estimates by less than 0.01 (human-supervised HR\,=\,0.914 and Devin HR\,=\,0.841, or 0.723 and 1.170 within repositories).

\subsubsection*{What kinds of change each group receives}

The groups also differ in the kind of modification they receive (Table~\ref{tab:rq2_event_mix}). Human code and Devin code are often modified by deleting their whole file (27.8\% and 29.2\% of modification events), whereas code from human-supervised agents is more often modified by deleting individual lines (20.8\%, against 12.9\% and 10.3\%). Whitespace-only edits make up 18.8\% of all content modifications: 21.0\% for Devin code, 18.9\% for human-supervised code, and 18.4\% for human code. A study that counts every touch therefore counts all groups as modified more often than their substantive changes justify, Devin code slightly more than the others.

\begin{table}[t]
\centering
\caption{Modification events by type (\% of each group's modification events) and the share of content changes that touch only whitespace.}
\label{tab:rq2_event_mix}
\small
\resizebox{\linewidth}{!}{%
\begin{tabular}{lrrrr}
\toprule
\textbf{Group} & \textbf{Content changed} & \textbf{Line deleted} & \textbf{File deleted} & \textbf{Whitespace-only}$^{a}$ \\
\midrule
Human-supervised & 63.7 & 20.8 & 15.5 & 18.9 \\
Fully-autonomous (Devin) & 60.5 & 10.3 & 29.2 & 21.0 \\
Human baseline & 59.3 & 12.9 & 27.8 & 18.4 \\
\bottomrule
\end{tabular}}
\par\smallskip\parbox{\linewidth}{\footnotesize $^{a}$ Share of content changes in which the old and the new line differ only in whitespace.}
\end{table}

\subsection{Comparison with Prior Work}
These results agree with the methods literature on non-essential changes. Kawrykow and Robillard~\cite{kawrykow2011non} found that up to 15.5\% of method updates are non-essential. We find that 18.8\% of content modifications touch only whitespace, before counting renames and literal edits. Neto et al.~\cite{neto2018impact} showed that removing refactorings changes the output of SZZ. Here, removing leaf-only edits strengthens the across-repository estimate for Devin and weakens it for human-supervised code. It also makes the within-repository human-supervised difference significant, which none of the across-repository estimates is once errors are clustered. For post-merge studies, the implication is that a syntactic definition is defensible for the direction of an effect. Its size, and whether it is detected, should be reported alongside a structural definition.

\begin{tcolorbox}[colback=gray!10, colframe=gray!50, title=RQ2 Finding]
The definition of a modification changes how large the effects are but not their direction. Within repositories, the structural definition makes the human-supervised difference significant (HR\,=\,0.72).
\end{tcolorbox}

\section{RQ3: How Much Do the Comparison Scope and the Variance Estimator Change the Verdict?}
\label{sec:rq3}

\noindent\textbf{Objective.}

RQ1 and RQ2 mostly compare agent code with human code from all repositories at once. However, agent code and human code need not come from the same projects, and projects can differ more from one another than the effects of interest~\cite{bettenburg2012think,xia2026these}. For example, if either agent tier is used mostly in projects that rewrite their code quickly, its code will look short-lived whatever its quality. The same structure affects how certain an estimate looks. Lines from one repository, and especially from one PR, are modified by the same commits, so a model that treats each of the 644,177 lines as an independent observation reports intervals that are far too narrow. Prior post-merge studies differ in both choices (Table~\ref{tab:prior_designs}). Some pool all code across repositories, others compare within repositories, and few state how their standard errors are computed. The goal of RQ3 is to measure how much the comparison scope and the variance estimator change the verdict, for each tier.

\subsection{Approach}

We proceed in four steps, holding the tracking (identity) and the units (all lines) of RQ1 fixed and reporting both definitions of RQ2.

\noindent\textbf{Who shares a repository.} We count the repositories that contain each group, and compare the characteristics of repositories with and without each tier with Mann-Whitney U tests~\cite{mann1947test}, which test whether values in one group tend to be larger than in the other: stars, forks, size, open issues, contributors, and lines of code. Projects that differ on these characteristics may also differ in how often their code changes.

\noindent\textbf{Across vs.\ within repositories.} We contrast each across-repository estimate of RQ1 and RQ2 with its repository-stratified counterpart (Section~\ref{sec:stats}), with 500-resample cluster-bootstrap intervals. Stratification changes two things at once. It compares code only with code from the same repository, and it uses only the repositories that contain both groups (128 for human-supervised code and 31 for Devin code). To separate the two, we also fit the pooled model on these shared repositories only. We report the minimum detectable hazard ratio of each within-repository comparison, because with few shared repositories, ``no difference'' may only mean that the design cannot detect one.

\noindent\textbf{Variance estimator.} We count the verdicts of every across- and within-repository specification of the curve (Section~\ref{sec:stats}) under three variance estimators. Model-based standard errors treat each unit as independent, the second estimator clusters the errors by PR, and our primary estimator clusters them by repository. We then test whether the counts exceed chance by refitting the curve on 500 permutations of the tier labels among PRs~\cite{simonsohn2020specification}.

\noindent\textbf{What drives the within-repository estimates.} Within a repository, tiers may still touch different files. Therefore, we add four file-level covariates to the stratified model. Two are the age of the file at merge and the number of commits to the file in the 90 days before merge, both log-transformed. The other two record whether the PR created the file and whether the file is a test file. We also estimate the within-repository model for each of the five agents separately, because the human-supervised tier groups four agents whose effects may differ. Finally, because the proportional hazards assumption fails (Section~\ref{sec:stats}), we fit piecewise models by time since merge and test any window-specific effect on held-out repositories.

\subsection{Findings}

We first show that the tiers come from different repositories, then how the scope and the estimator change the verdict for each tier, and finally what drives the within-repository estimates.

\subsubsection*{The tiers come from different repositories}

The agent tiers rarely share a repository (Table~\ref{tab:repo_coverage}). Human-supervised code appears in 150 of the 215 repositories and Devin code in 31, but only 10 repositories contain both. Their Jaccard overlap~\cite{jaccard1912distribution}, the number of shared repositories divided by the number of repositories with either tier, is 0.058. In total, 161 repositories (74.9\%) contain exactly one agent tier. Repositories with and without human-supervised code do not differ on any characteristic we measured. Repositories with Devin code, however, have more contributors than the rest (median 9 against 4, $p < 0.001$). Therefore, a comparison across repositories compares Devin's projects with other projects as much as Devin's code with human code, whereas this is not the case for human-supervised code.

\begin{table}[t]
\centering
\caption{Repository coverage of each group in the 215-repository cohort.}
\label{tab:repo_coverage}
\small
\begin{tabular}{lr}
\toprule
\textbf{Repositories containing} & \textbf{Count (\%)} \\
\midrule
Human-supervised code & 150 (69.8\%) \\
Fully-autonomous (Devin) code & 31 (14.4\%) \\
Human-baseline code & 193 (89.8\%) \\
\midrule
Both agent tiers & 10 (4.7\%) \\
Only human-supervised & 140 (65.1\%) \\
Only fully-autonomous & 21 (9.8\%) \\
No agent tier & 44 (20.5\%) \\
\bottomrule
\end{tabular}
\end{table}

\subsubsection*{Human-supervised code: the direction holds under every scope and estimator}

Figure~\ref{fig:rq2_def} (Section~\ref{sec:rq2}) contrasts the two scopes for the all-lines estimates, and Table~\ref{tab:spec_summary} counts the verdicts of the curve under each estimator. With standard errors clustered by repository, human-supervised code is retained longer than human code in 13 of the 22 across-repository specifications and in 11 of the 22 within-repository specifications. In each scope, 21 of the 22 point estimates lie below 1. Both counts exceed what permuted labels produce ($p = 0.008$ for both, Figure~\ref{fig:perm_null}a), and so does the within-repository median hazard ratio of 0.77 ($p = 0.002$).

On identity-tracked data with all lines, neither the across-repository nor the within-repository line-level estimate is significant (Figure~\ref{fig:rq2_def}). Within repositories, however, human-supervised code is retained longer when each PR counts once (HR\,=\,0.765) and under the structural definition. The scope therefore changes how large the human-supervised effect looks and, near the threshold, whether it is detected, but not its direction. Model-based errors would make almost every specification significant (21 of 22 ``longer'' in each scope). With 128 shared repositories, the design can detect hazard ratios below 0.63 or above 1.58. The single ``sooner'' verdict within repositories combines hunk units with line-number tracking.

\subsubsection*{Devin code: the estimator decides the verdict}

\begin{figure}[t]
    \centering
    \includegraphics[width=\textwidth]{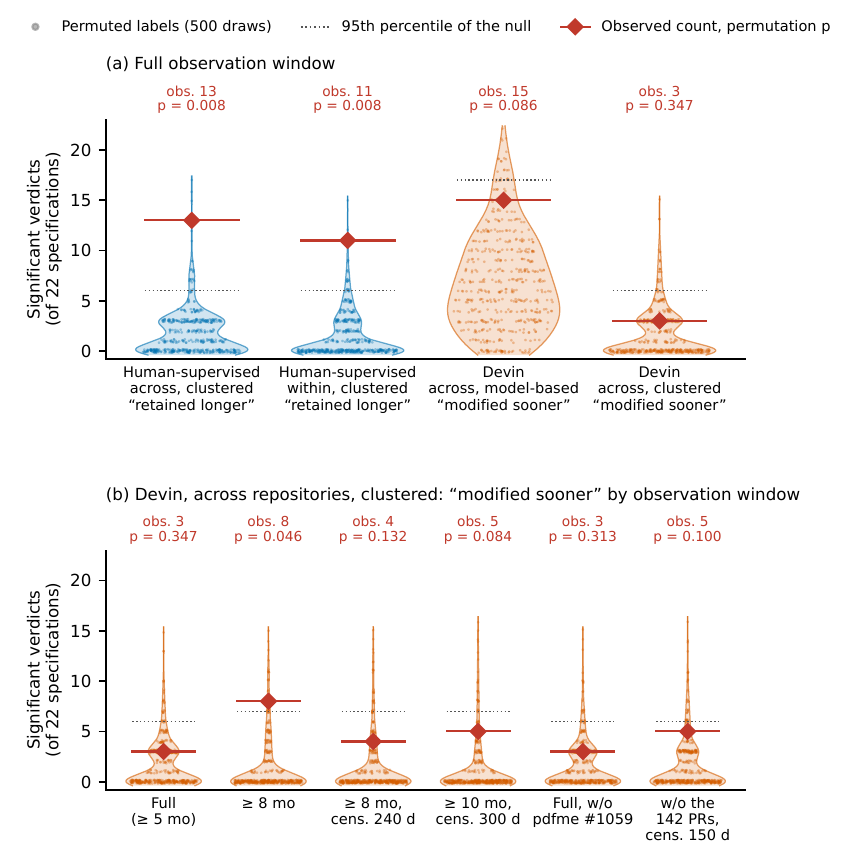}
    \caption{Counts of significant verdicts among the 22 specifications of a tier and comparison, under 500 permutations of the tier labels among PRs. Each dot is one permutation, spread sideways only so that equal counts do not overlap. Violins show the distribution of the counts, dotted lines their 95th percentile, and red diamonds the observed count with its permutation $p$-value. Blue: human-supervised code. Orange: Devin code. In (b), ``cens.'' means censored at the given number of days, and the last two windows drop pdfme PR \#1059 or the 142 PRs without eight months of follow-up, censoring the rest at 150 days. Devin code within repositories is left out of (a) because none of its specifications is significant under repository clustering.}
    \label{fig:perm_null}
\end{figure}

For Devin code, the estimator matters more than the scope. Across repositories, 15 of the 22 specifications report Devin code modified sooner than human code with model-based errors, 9 with errors clustered by PR, and 3 with errors clustered by repository, all of them hunk-level (HR about 1.6). The single specification that reports Devin code retained longer under repository clustering uses identity tracking on all lines including vendored code, with line units and the structural definition (HR\,=\,0.32, 95\% CI [0.10, 0.99]). Neither the model-based nor the repository-clustered count of ``sooner'' verdicts exceeds what permuted labels produce ($p = 0.086$ and $0.35$, Figure~\ref{fig:perm_null}a). With errors clustered by PR, the 9 ``sooner'' verdicts exceed chance ($p = 0.018$), but PR clustering ignores the correlation among PRs of the same repository, which repository clustering accounts for. Under the primary estimator, Devin's ``sooner'' verdicts are therefore what chance alone would give. What remains is a difference in point estimates. Across repositories, the median Devin hazard ratio is 1.16, at the edge of the null range ($p = 0.052$), and only 6 of the 22 estimates lie below 1.

Within repositories, all 22 Devin specifications report no difference under repository clustering, whereas model-based errors would report 11 ``longer'' and 7 ``sooner'' verdicts. In other words, a study that reports ``modified sooner across repositories and no difference within'' may be comparing two estimators as much as two scopes. Of the 22 within-repository point estimates, 15 lie below 1 (median 0.91, $p = 0.21$). The line-level estimate, 1.059 with interval [0.409, 2.076] (Figure~\ref{fig:rq2_def}), shows how little the data can tell. It is consistent with Devin code being modified anywhere from about 60\% less often to about twice as often as human code. With only 31 shared repositories, we would detect a difference only if Devin code were modified at less than a third, or more than three times, the rate of human code (hazard ratios below 0.32 or above 3.13). Therefore, ``no difference'' here means that a difference smaller than about threefold cannot be detected, not that there is none.

Part of the gap between the scopes comes from which repositories are compared, not from the stratification itself. Fitted only on the shared repositories, the pooled model gives 0.792 for human-supervised code ([0.513, 1.224]) instead of 0.888, and 0.624 for Devin code ([0.289, 1.348]) instead of 0.896. On the 31 shared repositories, 18 of Devin's 22 pooled point estimates lie below 1.

\begin{table}[t]
\centering
\caption{Verdicts of the 88 specifications (22 per tier and comparison) under three variance estimators, as longer / none / sooner. Longer: agent code retained longer than human code. None: no difference. Sooner: agent code modified sooner. Repository-clustered errors are primary. The 56 specifications on the three line sets without vendored code are in parentheses. Model-based errors treat every unit as independent. For units weighted by PR, unclustered robust errors give the same counts.}
\label{tab:spec_summary}
\small
\resizebox{\linewidth}{!}{%
\begin{tabular}{llccc}
\toprule
\textbf{Tier} & \textbf{Comparison} & \textbf{Clustered by repository} & \textbf{Clustered by PR} & \textbf{Model-based} \\
\midrule
Human-supervised & Across & 13 / 9 / 0 \enskip (8 / 6 / 0) & 13 / 9 / 0 & 21 / 0 / 1 \\
 & Within & 11 / 10 / 1 \enskip (7 / 6 / 1) & 12 / 9 / 1 & 21 / 0 / 1 \\
Fully-autonomous (Devin) & Across & 1 / 18 / 3 \enskip (0 / 13 / 1) & 0 / 13 / 9 & 5 / 2 / 15 \\
 & Within & 0 / 22 / 0 \enskip (0 / 14 / 0) & 1 / 20 / 1 & 11 / 4 / 7 \\
\bottomrule
\end{tabular}}
\end{table}

\begin{figure}[t]
    \centering
    \includegraphics[width=0.85\textwidth]{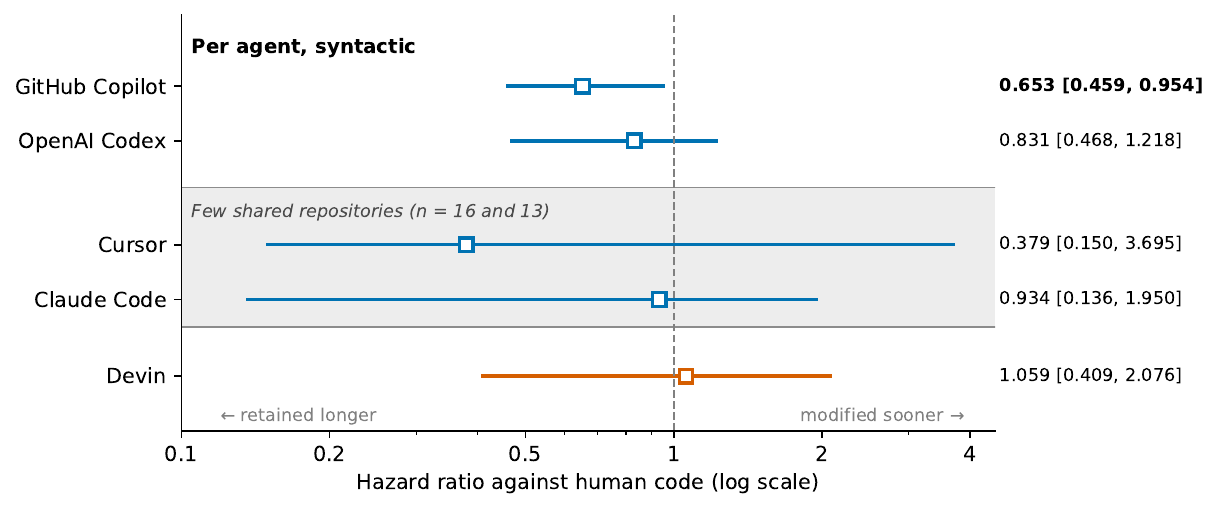}
    \caption{Within-repository hazard ratios against human code per agent, with 500-resample cluster-bootstrap 95\% CIs (log scale), identity tracking, all lines. The tier estimates are in Figure~\ref{fig:rq2_def}. Squares mark the point estimates. Blue: human-supervised agents. Orange: Devin. The numbers on the right are the hazard ratios with their CIs, in bold where the CI excludes 1. The shaded band marks Cursor and Claude Code, which share few repositories with human code (16 and 13), so their intervals are wide. Values from Table~\ref{tab:rq3_per_agent}.}
    \label{fig:rq3_forest}
\end{figure}

\subsubsection*{What drives the within-repository estimates}

The human-supervised effect is carried by GitHub Copilot (Table~\ref{tab:rq3_per_agent} and Figure~\ref{fig:rq3_forest}). Copilot code is retained longer than human code in the same 53 repositories (0.653, [0.459, 0.954]). Copilot lines are about a third less likely to be modified than human lines in the same repositories, whereas the intervals of OpenAI Codex (0.831), Cursor (0.379), and Claude Code (0.934) all include 1. Cursor's interval, from only 16 repositories, is too wide to be informative. Copilot also contributes the most human-supervised lines (67,398 of 177,787) and PRs (489 of 1,055), so the tier estimate leans on it. Devin's per-agent estimate is its tier estimate (1.059).

The human-supervised estimate does not come from the kind of files that agents touch. Lines in files that the PR created, and lines in test files, are modified later (covariate HR 0.57 and 0.59 under the syntactic definition), and human-supervised code has more of both. With the four file covariates, the structural estimate stays significant (0.742, [0.487, 0.967]), and the syntactic one stays below 1 (0.839, [0.581, 1.099]). Devin's within-repository estimates hardly move with the covariates (1.053 and 1.141).

\begin{table}[t]
\centering
\caption{Within-repository hazard ratio of each agent's code against human code (identity tracking, all lines, syntactic definition, each line counts once), with 500-resample cluster-bootstrap 95\% CIs.}
\label{tab:rq3_per_agent}
\small
\begin{tabular}{lrrrr}
\toprule
\textbf{Agent} & \textbf{Lines} & \textbf{PRs} & \textbf{Repositories}$^{a}$ & \textbf{HR [95\% CI]} \\
\midrule
GitHub Copilot & 67,398 & 489 & 53 & 0.653 [0.459, 0.954] \\
OpenAI Codex & 48,867 & 491 & 63 & 0.831 [0.468, 1.218] \\
Cursor & 13,340 & 44 & 16 & 0.379 [0.150, 3.695] \\
Claude Code & 48,182 & 31 & 13 & 0.934 [0.136, 1.950] \\
Devin & 87,634 & 412 & 31 & 1.059 [0.409, 2.076] \\
\bottomrule
\end{tabular}
\par\smallskip\parbox{\linewidth}{\footnotesize $^{a}$ Repositories that contain both the agent's code and human code.}
\end{table}

Piecewise models by time since merge do not show a robust within-repository difference either. In the first 30 days, human-supervised code does not differ from human code across repositories (HR\,=\,1.20, 95\% CI [0.67, 2.15]). Between 30 and 90 days, piecewise models with standard errors clustered by repository suggest that human-supervised code is modified later. This holds both across repositories (HR\,=\,0.45, 95\% CI [0.23, 0.90]) and within them (0.40, [0.16, 1.00], $p = 0.049$). However, the within-repository estimate does not survive a Bonferroni correction for the six tier-window comparisons. None of the three windows shows a Devin difference in either scope ($p \ge 0.68$ within repositories). When we look at the first week alone, a window we chose after seeing the curves, Devin code is modified sooner within repositories (HR\,=\,1.72, $p = 0.007$). We therefore chose the early window on one random half of the 31 repositories with Devin and human code, and tested it on the other half. The effect replicates in none of 5 such splits. We therefore report it as a hypothesis for future work, not as a finding.

\subsection{Comparison with Prior Work}
These results agree with prior work on the role of the repository. Xia and Miller~\cite{xia2026these} already compare code within repositories and find no overall difference for pooled agentic code. Our within-repository estimates agree, and Section~\ref{sec:synthesis} reproduces their pooled result on our data. The strong cross-project heterogeneity that they report ($I^2 = 100\%$~\cite{higgins2002quantifying}) matches the near-disjoint repository sets we observe. Cameron and Miller~\cite{cameron2015practitioner} warn that standard errors that ignore clustering are too small. Our results show how much this matters for post-merge studies: for Devin, the choice among three estimators changes the number of ``sooner'' verdicts from 15 to 3.

\begin{tcolorbox}[colback=gray!10, colframe=gray!50, title=RQ3 Finding]
Human-supervised code is retained longer than human code in more specifications than chance would give, in both scopes (13 and 11 of 22). For Devin code, the variance estimator decides the verdict, and within repositories no difference can be detected.
\end{tcolorbox}

\section{RQ4: Does the Intent Classifier Change the Verdict on Why Agent Code Is Modified?}
\label{sec:rq4}

\noindent\textbf{Objective.}

RQ1 to RQ3 measure when agent code is modified, but not why, and show that the measurement decides how large an effect looks. Yet only the reason shows whether a modification signals a problem: code that is modified to fix a defect says something different about quality than code that is extended with a new feature. Post-merge studies therefore classify the intent of each modifying commit, and some of their strongest claims rest on it, such as more corrective maintenance for agentic code~\cite{xia2026these}. Prior studies infer intent from commit messages with keyword or Conventional Commit rules (Table~\ref{tab:prior_designs}), whose labels are known to carry errors that change conclusions~\cite{herzig2013not,berger2019impact}. These studies use rules because their commits are too many to label by hand. LLM classifiers offer an alternative that reads the commit as a person would~\cite{ahmed2025can}, but post-merge studies use them at most as a fallback~\cite{xia2026these}. If the classifier, rather than the code, decides which group is modified for different reasons, then intent claims about agent code cannot be compared across studies. The goal of RQ4 is to measure whether LLM classifiers can reliably replace keyword rules for intent, and whether the choice between them changes the verdict.

\subsection{Approach}
We apply the seven approaches of Section~\ref{sec:intent_method} to the same 5,303 modification commits: keyword rules, three LLM classifiers (DeepSeek-V4.1-Flash, gpt-oss-120b, MiniMax-M3.1-Flash), and three combinations of their labels (majority vote, weighted vote, and Dawid-Skene aggregation). We designate no approach as primary. Of the 5,303 commits, 5,300 are mapped to a tier. The other 3 modify only lines of the one human PR that the intent analyses leave out (Section~\ref{sec:filtering}), and mapping them changes no $p$-value of the intent tests by more than 0.01. For each approach, we measure its agreement with the 618 human labels. We compare the intent mix of each tier with the human baseline with a $\chi^2$ test and Cram\'er's $V$~\cite{cramer1946mathematical}. We then ask the same question of the human labels themselves. Each labeled commit is weighted to the corpus by its stratum and tier, and we use a $\chi^2$ test corrected for these unequal sampling weights (the Rao-Scott test~\cite{rao1984chi}). We also correct each LLM's shares for its classification error (Section~\ref{sec:intent_method}). A correction with one confusion matrix for all commits assumes that a classifier errs in the same way for every tier. We therefore test whether each approach's accuracy differs by tier (Fisher's exact test~\cite{fisher1922interpretation}), and repeat the correction with a separate confusion matrix per tier. To see how much of the LLMs' agreement comes from the code change, we also rerun the three LLMs on the 618 labeled commits with the commit message only, without the diff.

\subsection{Findings}

We report how well each approach agrees with the human labels, which verdict each one supports, and what remains once classification error is accounted for.

\subsubsection*{Agreement with human labels}

\begin{figure}[t]
    \centering
    \includegraphics[width=0.85\textwidth]{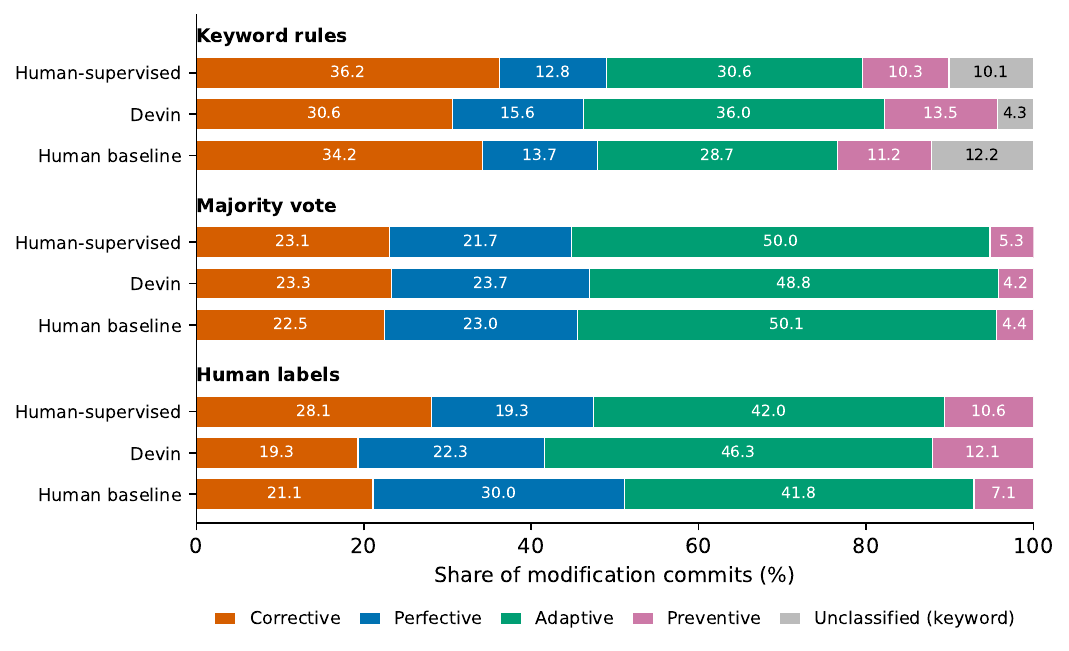}
    \caption{Intent shares of modification commits (\%) by group under keyword rules, the majority vote of the three LLMs, and the human labels (weighted to the corpus). Each bar sums to 100\%, and the numbers in the segments are the shares. Majority vote: the label that at least two of the three LLMs give. Values from Table~\ref{tab:rq4_intent}. Only keyword rules leave commits unclassified.}
    \label{fig:rq4_intent}
\end{figure}

Keyword rules agree with the 618 human labels at $\kappa = 0.38$ (95\% CI [0.33, 0.42]), the three LLMs at 0.69 to 0.71, and their combinations at 0.71 to 0.72 (Table~\ref{tab:rq4_intent}). Figure~\ref{fig:rq4_intent} shows the intent shares under keyword rules, the majority vote, and the human labels. Weighted to the strata of the corpus, the values are the same to the third decimal. Dawid-Skene aggregation (0.72) is only marginally above the best single LLM (MiniMax, 0.71), well within the confidence intervals. Keyword rules thus agree with a human annotator only fairly, and the LLMs substantially.

On the 100 commits that both humans labeled, the LLMs agree with the first author at $\kappa = 0.63$ to $0.65$ and with the second rater at $0.70$ to $0.75$. This is around the $\kappa = 0.68$ at which the two humans agree with each other (Section~\ref{sec:intent_method}). Keyword rules reach 0.34 and 0.38. On the 76 of these commits on which the two humans agree, the LLMs give the same label for 84\% to 87\% of them, and keyword rules for 57\%. The LLMs are therefore close to the ceiling that noisy human labels allow, and a higher agreement could not be told apart from agreement with the first author's own errors.

The diff does not improve agreement. With the commit message only, the LLMs agree with the human labels at $\kappa = 0.68$ to $0.74$ (MiniMax 0.70, DeepSeek 0.74, gpt-oss 0.68), and they give the same label as with the diff for 85\% to 88\% of the commits. The gap over keyword rules therefore comes from the classifier, not from the extra input.

The keyword errors follow the rules' priority order, which labels a commit corrective whenever its message carries a defect-fix signal. Across the corpus, 584 commits that the keyword rules label corrective are labeled adaptive by all three LLMs or perfective by all three, against 57 in the opposite direction. For example, \textit{``fix: update PHP SDK code style to fix failing workflow (\#689)''} in \texttt{lingodotdev/lingo.dev}, the commit that restyles 128 Devin lines in RQ2, is corrective under keyword rules because of its \texttt{fix:} prefix. Two of the three LLMs label it perfective.

\noindent\textit{Agreement between the models as a signal.} Ahmed et al.~\cite{ahmed2025can} recommend using the agreement between models to decide which items still need a human. The three LLMs agree with each other at Fleiss' $\kappa = 0.84$ (95\% CI [0.80, 0.86])~\cite{fleiss1971measuring}, more than any of them agrees with the human labels. All three give the same label to 521 of the 618 labeled commits (84.3\%). That label matches the human one for 86.0\% of them (95\% Wilson CI [82.7, 88.7]~\cite{wilson1927probable}). Two of the three agree on 94 commits (15.2\%). On these, the majority label matches the human one for only 54.3\% [44.2, 64.0], and the dissenting model's label for 36.2\%. All three differ on the remaining 3 commits. Weighted to the strata of the corpus, these shares change by at most 0.1 points. The models therefore share many of their errors, which is why combining them adds little over the best single model. Disagreement between the models thus marks the commits that most need a human: about one in six, on which a majority vote is right only about half of the time.

\subsubsection*{Which verdict each approach supports}

The two families of classifier reach different verdicts (Table~\ref{tab:rq4_intent}). For human-supervised code, neither family finds a difference from human code: keyword rules give $p = 0.10$, and every LLM approach $p \ge 0.31$. For Devin code, keyword rules find that it is modified for different reasons than human code ($p < 0.001$, $V = 0.11$), because they label fewer of its modifications corrective and more of them adaptive or preventive. Even this difference is small: the intent mixes of Devin and human code differ by a few percentage points per category. Moreover, no LLM approach confirms it ($p \ge 0.37$). Across both tiers, the LLM effect sizes are negligible ($V \le 0.03$). The choice among LLMs and among ways of combining them barely matters: their shares differ by at most about four percentage points, and none reaches significance. The choice between keyword rules and LLMs decides the verdict. Split by agent, no LLM approach finds a difference from human code for Copilot, Codex, or Claude Code ($p \ge 0.11$), although the keyword rules do for Copilot and Codex ($p < 0.001$). Cursor's 92 modifications are more often preventive than human ones under MiniMax and the majority vote (10.9\% against 4.4\% under the majority vote, $p = 0.022$ and $0.021$). This difference does not survive a Bonferroni correction for the four agents.

The human labels show a different result. Weighted to the corpus, they show a difference between human-supervised and human code ($p = 0.038$, $V = 0.13$). Human-supervised code receives more corrective (28.1\% against 21.1\%) and fewer perfective modifications (19.3\% against 30.0\%). This difference is not significant after a Bonferroni correction for the two tier comparisons ($p = 0.076$), and Devin code does not differ ($p = 0.37$). The difference lies in the labeled commits rather than in the classifiers: on the same 618 commits, the LLM approaches find it too ($p = 0.006$ to $0.058$).

In the 4,155 unlabeled commits of the two groups, however, the LLM majority vote finds no difference ($p = 0.88$, $V = 0.01$). Weighting the labeled commits to the stratum and tier composition of the corpus does not close the gap. Its design effect is 1.05, that is, the weighting widens the intervals by only 5\%. On the labeled commits, the majority vote shows a human-supervised corrective share 6.6 percentage points above the human one, close to the 8.1 points of the human labels. On the unlabeled commits, the gap is $-0.3$ points. Because two human raters agree only substantially on intent ($\kappa = 0.68$), the human labels are themselves noisy, which makes a difference of $p = 0.038$ in them fragile. The labeled and unlabeled commits do not differ in repository, month, or message length ($p = 0.29$, $0.38$, and $0.25$). We therefore read the human-label difference as a property of the labeled sample, which covers 524 of the 4,679 commits of the two groups that the majority vote labels, and not as evidence about the corpus.

\begin{table*}[t]
\centering
\caption{Modification intent by tier under seven classification approaches (full corpus: 5,300 modification commits mapped to a tier: 1,453 human-supervised, 617 fully-autonomous, 3,230 human) and under the human labels (618 labeled commits, weighted to the corpus). Shares are corrective / perfective / adaptive / preventive (\%). $\kappa$: agreement with the 618 human labels. Tests: $\chi^2$ $p$ of each tier against the human baseline (design-corrected for the human labels), with Cram\'er's $V$ in parentheses.}
\label{tab:rq4_intent}
\small
\resizebox{\textwidth}{!}{%
\begin{tabular}{lrlllll}
\toprule
\textbf{Approach} & \textbf{$\kappa$} & \textbf{Human-supervised} & \textbf{Fully-autonomous} & \textbf{Human baseline} & \textbf{Human-supervised vs.\ human} & \textbf{Devin vs.\ human} \\
\midrule
Keyword$^{a}$ & 0.38 & 36.2 / 12.8 / 30.6 / 10.3 & 30.6 / 15.6 / 36.0 / 13.5 & 34.2 / 13.7 / 28.7 / 11.2 & 0.10 (0.04) & $<$0.001 (0.11) \\
DeepSeek & 0.70 & 22.7 / 22.2 / 49.6 / 5.4 & 22.4 / 25.8 / 47.3 / 4.5 & 22.1 / 23.7 / 49.4 / 4.8 & 0.62 (0.02) & 0.67 (0.02) \\
gpt-oss & 0.69 & 21.2 / 22.4 / 51.5 / 4.9 & 22.2 / 24.0 / 48.6 / 5.2 & 20.9 / 23.2 / 51.8 / 4.1 & 0.59 (0.02) & 0.37 (0.03) \\
MiniMax & 0.71 & 24.8 / 21.6 / 48.2 / 5.4 & 24.5 / 22.7 / 48.5 / 4.4 & 24.3 / 22.6 / 48.8 / 4.2 & 0.34 (0.03) & 0.998 (0.003) \\
Majority vote$^{b}$ & 0.71 & 23.1 / 21.7 / 50.0 / 5.3 & 23.3 / 23.7 / 48.8 / 4.2 & 22.5 / 23.0 / 50.1 / 4.4 & 0.43 (0.02) & 0.93 (0.01) \\
Weighted vote & 0.71 & 23.3 / 21.8 / 49.6 / 5.4 & 23.3 / 23.7 / 48.6 / 4.4 & 22.7 / 23.0 / 49.9 / 4.4 & 0.40 (0.03) & 0.95 (0.01) \\
Dawid-Skene & 0.72 & 22.9 / 21.7 / 49.7 / 5.6 & 23.2 / 23.7 / 48.6 / 4.5 & 22.3 / 23.1 / 50.0 / 4.5 & 0.31 (0.03) & 0.93 (0.01) \\
\midrule
Human labels (618, weighted) & n/a & 28.1 / 19.3 / 42.0 / 10.6 & 19.3 / 22.3 / 46.3 / 12.1 & 21.1 / 30.0 / 41.8 / 7.1 & 0.038 (0.13) & 0.37 (0.09) \\
\bottomrule
\end{tabular}%
}
\par\smallskip\parbox{\linewidth}{\footnotesize $^{a}$ Remaining share unclassified: 10.1\% (human-supervised), 4.3\% (Devin), 12.2\% (human). $^{b}$ 37 three-way splits left undecided.}
\end{table*}

\subsubsection*{Correcting for classification error}

Every classifier distorts the levels: compared with the human labels, the LLMs label more modifications adaptive and fewer perfective, and the keyword rules more corrective. Once each LLM's shares are corrected with its confusion matrix on the 618 labeled commits, the corrective shares of the three groups overlap under every LLM approach (Table~\ref{tab:rq4_corrected}). Under the majority vote, for example, the corrected corrective share is 24.0\% [19.8, 27.6] for human-supervised code, 24.4\% [19.5, 29.3] for Devin code, and 23.6\% [20.6, 26.7] for human code. After correcting for classifier mistakes, about one in four modifications fixes a defect, in agent and human code alike. No approach's accuracy differs by tier (all $p \ge 0.13$). Under the majority vote, for example, it is 83.0\% for human-supervised, 77.7\% for Devin, and 80.2\% for human commits. A separate confusion matrix per tier gives the same picture: under the majority vote, 24.8\% [20.0, 29.6] for human-supervised code, 26.1\% [16.5, 38.3] for Devin code, and 23.1\% [17.9, 27.4] for human code. The data therefore do not show that agent code is modified more often to fix defects, nor less often.

\begin{table}[t]
\centering
\caption{Corrective share (\%) per group after correcting each LLM approach with its confusion matrix on the 618 labeled commits, either one matrix for all commits or one matrix per tier. 95\% bootstrap CIs resample the labeled commits and the corpus.}
\label{tab:rq4_corrected}
\small
\begin{tabular}{llll}
\toprule
\textbf{Approach} & \textbf{Human-supervised} & \textbf{Fully-autonomous} & \textbf{Human baseline} \\
\midrule
\multicolumn{4}{l}{\textit{One confusion matrix}} \\
DeepSeek & 23.6 [19.4, 27.4] & 23.4 [18.4, 28.5] & 23.0 [19.6, 26.1] \\
gpt-oss & 25.6 [22.1, 29.0] & 26.8 [21.7, 31.8] & 25.2 [22.6, 28.4] \\
MiniMax & 23.7 [19.4, 27.5] & 23.6 [19.0, 28.7] & 23.5 [20.7, 26.4] \\
Majority vote & 24.0 [19.8, 27.6] & 24.4 [19.5, 29.3] & 23.6 [20.6, 26.7] \\
\midrule
\multicolumn{4}{l}{\textit{One confusion matrix per tier}} \\
DeepSeek & 23.0 [18.6, 27.7] & 27.0 [16.2, 41.7] & 23.2 [17.7, 28.4] \\
gpt-oss & 23.8 [18.4, 29.3] & 30.4 [22.1, 42.2] & 25.4 [21.7, 30.0] \\
MiniMax & 25.7 [21.1, 30.1] & 24.0 [13.3, 35.7] & 23.0 [18.2, 27.3] \\
Majority vote & 24.8 [20.0, 29.6] & 26.1 [16.5, 38.3] & 23.1 [17.9, 27.4] \\
\bottomrule
\end{tabular}
\end{table}

\subsection{Comparison with Prior Work}
These results partly agree with the methods literature on commit classification. Levin and Yehudai~\cite{levin2017boosting} found that classifiers using only the commit message are much weaker across projects than those that also see the code change. For LLM classifiers, we find otherwise: with the message alone they agree with human labels about as well as with the diff ($\kappa = 0.68$ to $0.74$), and far better than keyword rules. Our agreement results match those of Ahmed et al.~\cite{ahmed2025can}: the LLMs agree with each human about as well as the two humans agree with each other. For commit intent, they can therefore take the place of keyword rules and come close to a second human annotator. They also agree better than the LLM of Sazid et al.~\cite{sazid2024commit} ($\kappa = 0.62$), although the two studies differ in data and label sets. Herzig et al.~\cite{herzig2013not} and Berger et al.~\cite{berger2019impact} showed that label errors can create or remove effects. Here, they decide which tier appears to be modified for different reasons. Xia and Miller~\cite{xia2026these} validate their rule-first classifier at $\kappa = 0.94$ on 100 commits, far above the $\kappa = 0.38$ of our keyword rules, which suggests that well-engineered rules can do much better than ours. What our results add is that the verdict depends on the family of classifier. A high agreement on a small sample should therefore be complemented by checking the verdict under a second classifier. A verdict from human labels should be checked against the unlabeled commits, and the agreement between two human raters shows how far any classifier's agreement can be read as accuracy. Section~\ref{sec:synthesis} shows that this choice also moves their corrective result on our data.

\begin{tcolorbox}[colback=gray!10, colframe=gray!50, title=RQ4 Finding]
The intent classifier decides the verdict. LLM classifiers agree with the human labels about as well as two human raters agree with each other ($\kappa = 0.69$ to $0.72$), and only keyword rules ($\kappa = 0.38$) find that Devin code is modified for different reasons than human code.
\end{tcolorbox}

\section{RQ5: Do Measurement Choices Change the Verdict on Who Maintains Agent Code?}
\label{sec:rq5}

\noindent\textbf{Objective.}

RQ4 shows that the verdict on why agent code is modified depends on the intent classifier. A second claim that post-merge studies make about agent code concerns who does the modifying. This claim matters to practitioners because it decides whether adopting coding agents moves maintenance effort to the agents or leaves it with developers. It also matters for quality, because ownership research shows that who changes a component is related to its defects~\cite{bird2011don,greiler2015code}. However, the published answers disagree. Sawada et al.~\cite{sawada2026extent} report that human developers make about 83\% of the later commits to files created by agents, whereas Takerngsaksiri et al.~\cite{takerngsaksiri2026finishes} report that most follow-up fixes of agent PRs come from the same agent. The two studies count follow-up work differently. The first counts every commit to a file over six months, and the second counts follow-up fix PRs within 30 days. If the unit, the window, or the way a modifier is identified decides which picture a study reports, then claims about who maintains agent code cannot be compared across studies either. The goal of RQ5 is to measure who modifies the code of each tier after merge, and whether the measurement choices change the answer.

\subsection{Approach}

We classify the modifier of each of the 5,303 modification commits as human or agent, and identify the agent (Section~\ref{sec:modifier_method}). We then vary the unit of follow-up work (commit, line, or follow-up PR), the window (7 days, 30 days, or the full window), and the identification rule (author, or author with co-author trailers). For each tier and for human code, we report the share of modifications made by agents and the share made by the agent that wrote the code. For each agent self-modification, we measure the time since the merge of the original PR. To see what self-modifications are, the first author compared a sample of 20 Devin and 20 Copilot self-modification commits with the description of the PR that added the modified lines. Each was labeled as continuing the same task, as the next item of a series of similar PRs, or as later maintenance. Finally, we compare the intent of agent-made and human-made modifications under the seven classification approaches of RQ4 and under the human labels.

\subsection{Findings}

We first report who modifies the code of each tier, then how the unit, the window, and the identification rule change the answer, and finally why agents modify code.

\subsubsection*{Who modifies agent code}

Counted per modifying commit, maintaining agent code is mostly human work (Table~\ref{tab:modifier_type}). Human developers make 88.6\% of the modifications to human-supervised code and 85.7\% of those to Devin code, against 96.9\% of the modifications to human code. Over both tiers, they make 87.7\% of the modifications to agent code (1,816 of 2,070 modification commits). When an agent does modify agent code, it is almost always the agent that wrote it. For human-supervised code, GitHub Copilot makes 151 of the 166 agent modifications (91.0\%), and in 143 of these 151 commits the modified lines were written by Copilot itself. For Devin code, Devin makes 85 of the 88 agent modifications (96.6\%), and every one of them changes lines Devin wrote.

The two agents differ in when they modify their own code. Devin does so soon after merge (median 2.8 days), whereas Copilot does so weeks later (median 42.0 days after the merge of its own PR), and only 32 of Copilot's 143 self-modifications come within 7 days.

The samples of 20 show the same contrast. Of the Devin self-modifications, 14 belong to the work of the original PR rather than to later maintenance. Of these, 7 continue the same feature or fix, and 7 are the next item of a series of similar PRs, usually by the same requester. Of the Copilot ones, only 6 do: 4 continue the same feature or fix, and 2 are the next item of a series (Fisher's exact test~\cite{fisher1922interpretation}, $p = 0.026$ against Devin). The other 14 are later maintenance, such as a new feature, a deprecation, or a fix unrelated to the original PR. Both samples are small and clustered by repository: 12 of the 20 Copilot commits come from \texttt{dotnet/aspire}, including all 4 same-task ones, and 4 of the 20 Devin commits come from \texttt{near/nearcore}. Because Fisher's test treats the commits as independent, we read its $p$-value as indicative only, and the counts as descriptive of these samples rather than as precise shares of all self-modifications.



\begin{table}[t]
\centering
\caption{Who makes the modifications of each group's code (\% of modification commits), and which agent modifies agent code when an agent does.}
\label{tab:modifier_type}
\small
\resizebox{\linewidth}{!}{%
\begin{tabular}{lrrrl}
\toprule
\textbf{Code written by} & \textbf{Commits} & \textbf{Human} & \textbf{Agent} & \textbf{Most frequent agent modifier} \\
\midrule
Human-supervised agents & 1,453 & 88.6\% & 11.4\% & Copilot, 151 of 166 \\
Fully-autonomous (Devin) & 617 & 85.7\% & 14.3\% & Devin, 85 of 88 \\
Human developers & 3,230 & 96.9\% & 3.1\% & Copilot and Devin, 40 each of 100 \\
\bottomrule
\end{tabular}}
\end{table}

\begin{figure}[t]
    \centering
    \includegraphics[width=\textwidth]{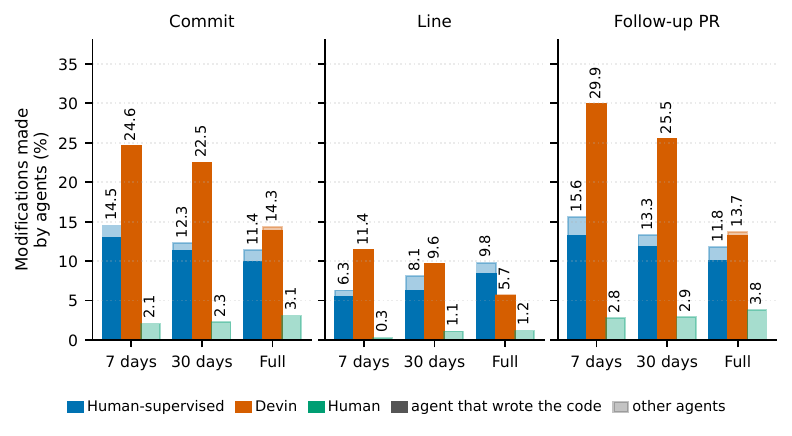}
    \caption{Share of modifications of each group's code made by agents (\%), by unit and window. Panels: the unit counted, either the modifying commit, the modified line, or the follow-up PR (the earliest merged PR, other than the original one, that contains the modifying commit). Windows: modifications within 7 or 30 days of merge, or within the full observation window. The darker part of each agent bar is the share made by the agent that wrote the code, and the lighter top the share made by other agents. The number above each bar is the total agent share. Human code has no authoring agent. Table~\ref{tab:modifier_variants} gives both shares.}
    \label{fig:rq5_share}
\end{figure}

\subsubsection*{The unit and the window decide whether agents look like their own maintainers}

Table~\ref{tab:modifier_variants} shows how the answer changes with the unit and the window. Per modifying commit over the full window, agents make 11.4\% of the modifications to human-supervised code and 14.3\% of those to Devin code. Within 30 days, these shares are 12.3\% and 22.5\%, and within 7 days 14.5\% and 24.6\%. Counted as follow-up PRs within 30 days, the unit of Takerngsaksiri et al.~\cite{takerngsaksiri2026finishes}, agents make 13.3\% of the follow-up PRs of human-supervised code and 25.5\% of those of Devin code. One in four follow-up PRs of Devin code within a month thus comes from an agent, almost always Devin itself, but three in four come from people. Shorter windows raise the agent share for Devin code, whose self-modifications come soon after merge, but hardly change it for human-supervised code. Counted per line over the full window, the order of the tiers reverses (9.8\% and 5.7\%). However, no unit or window makes agents the main maintainers: the largest agent share in Table~\ref{tab:modifier_variants} and Figure~\ref{fig:rq5_share} is 29.9\%, for follow-up PRs of Devin code within 7 days (37.7\% when each pair of original and follow-up PR counts once). Agents are the majority only among the follow-ups that agents make: within 30 days, the agent that wrote the code makes all 71 agent-made follow-up PRs of Devin code and 65 of the 73 of human-supervised code. Counting co-author trailers that name a coding agent changes the human share of modifications to agent code from 87.7\% to 83.7\%. Counting the 530 commits that modify lines of more than one group under every group they modify, or dropping them, changes no agent share by more than 2.7 percentage points.

\begin{table}[t]
\centering
\caption{Share of modifications of each group's code made by agents (\%), by unit and window, and, in parentheses, the share made by the agent that wrote the code. A follow-up PR is the earliest merged PR, other than the original one, that contains the modifying commit.}
\label{tab:modifier_variants}
\small
\resizebox{\linewidth}{!}{%
\begin{tabular}{llrrr}
\toprule
\textbf{Unit} & \textbf{Window} & \textbf{Human-supervised} & \textbf{Fully-autonomous (Devin)} & \textbf{Human baseline} \\
\midrule
Commit & 7 days & 14.5 (12.9) & 24.6 (24.6) & 2.1 \\
Commit & 30 days & 12.3 (11.3) & 22.5 (22.5) & 2.3 \\
Commit & Full & 11.4 (9.9) & 14.3 (13.8) & 3.1 \\
Line & 7 days & 6.3 (5.5) & 11.4 (11.4) & 0.3 \\
Line & 30 days & 8.1 (6.2) & 9.6 (9.6) & 1.1 \\
Line & Full & 9.8 (8.4) & 5.7 (5.6) & 1.2 \\
Follow-up PR & 7 days & 15.6 (13.2) & 29.9 (29.9) & 2.8 \\
Follow-up PR & 30 days & 13.3 (11.8) & 25.5 (25.5) & 2.9 \\
Follow-up PR & Full & 11.8 (10.1) & 13.7 (13.2) & 3.8 \\
\midrule
Commit, co-author rule & Full & 16.3 (12.9) & 16.2 (14.6) & 5.5 \\
\bottomrule
\end{tabular}}
\end{table}

\subsubsection*{Why agents modify code depends on the classifier}

Why agents modify code depends, again, on the classifier. Under keyword rules, agent-made modifications are more often corrective than human-made ones (46.9\% against 33.4\%, $p < 0.001$). Under every LLM approach, they are instead more often preventive (7.3\% to 9.3\% against 4.2\% to 4.7\%) and less often corrective, and their intent mix differs ($\chi^2$, $p = 0.043$ for gpt-oss, $p < 0.003$ for the other approaches). The human labels point the same way. The 618 labeled commits contain 60 agent-made modifications. These are more often preventive than human-made ones (16.7\% against 8.2\%) and less often perfective (13.3\% against 26.7\%), and the difference is significant ($p = 0.040$, $V = 0.12$). Split by tier, however, the LLM difference comes from Devin code. For Devin code, every LLM approach labels agent-made modifications less often corrective than human-made ones (13.6\% to 15.9\% against 23.3\% to 26.1\%), significantly so under DeepSeek, MiniMax, and the majority vote ($p = 0.012$, $0.005$, and $0.003$). The keyword rules find no difference for Devin code ($p = 0.14$). For human-supervised code, no LLM approach finds a difference ($p \ge 0.60$), whereas the keyword rules label agent-made modifications more often corrective (54.8\% against 33.8\%, $p < 0.001$). Per tier, the human labels contain too few agent-made modifications (28 and 18) to show a difference ($p \ge 0.64$).

\subsection{Comparison with Prior Work}
These results speak to two published findings. Sawada et al.~\cite{sawada2026extent} report that human developers do about 83\% of the maintenance of agent-generated files. We find 87.7\% of modification commits, and 88.2\% with their design on our data. Takerngsaksiri et al.~\cite{takerngsaksiri2026finishes}, in contrast, find that most follow-up fixes of agent PRs come from the same agent. We find this pattern only among the modifications that agents make: within 30 days, the agent that wrote the code makes all 71 agent-made follow-up PRs of Devin code and 65 of the 73 of human-supervised code. Among all follow-up PRs within 30 days, their unit and window, agents make only 25.5\% for Devin code and 13.3\% for human-supervised code. The unit and the window therefore change the agent share, by up to a factor of two for Devin code, but not who does most of the work. The difference from Takerngsaksiri et al.~\cite{takerngsaksiri2026finishes} more likely lies in their outcome, verified follow-up fixes, and in their denominator, fixes rather than all follow-up work, which our data cannot reproduce. Therefore, a study that reports who maintains agent code should state its unit, window, outcome, and denominator.

\begin{tcolorbox}[colback=gray!10, colframe=gray!50, title=RQ5 Finding]
Human developers make most modifications to agent code under every unit, window, and identification rule (87.7\% of the modifying commits). When an agent modifies agent code, it is almost always the agent that wrote it.
\end{tcolorbox}

\section{Synthesis: The Specification Curve and Published Designs Re-Run}
\label{sec:synthesis}

RQ1 to RQ3 varied the measurement choices of the survival verdict one or two at a time, starting from the units and tracker of our earlier study, and RQ4 and RQ5 examined its interpretation. In this section, we combine the retention choices. We first show the full specification curve, and then re-run the designs of five published post-merge studies on our data, to see where we agree with them and which single choice moves their headline result.

\subsection{The Specification Curve}

\begin{figure*}[t]
    \centering
    \includegraphics[width=\textwidth]{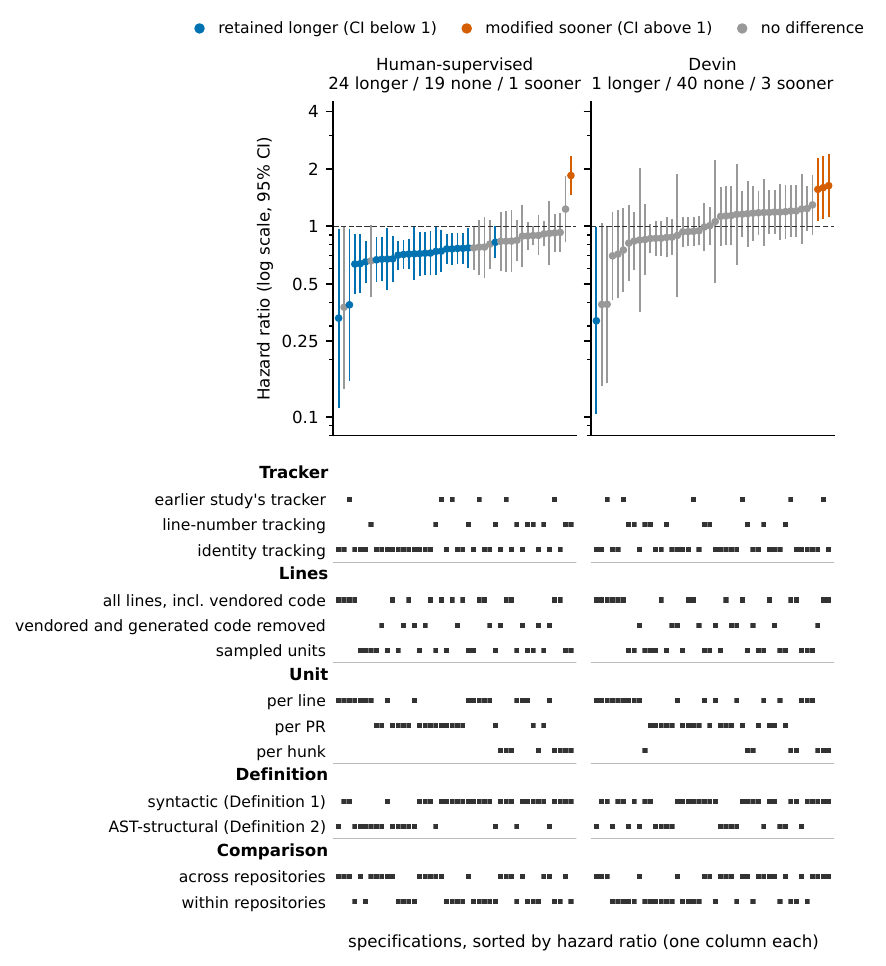}
    \caption{Specification curve over 88 specifications. Top: hazard ratio of each tier against human code (log scale, 95\% CI) for every specification, sorted by hazard ratio. The panel titles count the verdicts. Bottom: the choices behind each specification, grouped by dimension, one column per specification, with a square for each choice it uses. Sampled units: at most 20 files per PR and 100 lines per file. All intervals use analytic standard errors clustered by repository. The headline within-repository estimates in RQ3 use the 500-resample cluster bootstrap.}
    \label{fig:spec_curve}
\end{figure*}

Figure~\ref{fig:spec_curve} shows all 88 specifications, sorted by hazard ratio, with the choices behind each one marked below, and Table~\ref{tab:spec_summary} in RQ3 counts the verdicts under three variance estimators. With standard errors clustered by repository, our primary estimator, 1 specification reports Devin code retained longer than human code, 40 report no difference, and 3 report it modified sooner. Human-supervised code is reported retained longer in 24 specifications, no different in 19, and modified sooner in 1. On the 56 specifications that use only the three line sets without vendored code, the counts are 0, 27, and 1 for Devin, and 15, 12, and 1 for human-supervised code.

RQ3 showed that, for Devin, the variance estimator decides most of the verdicts on this curve and that they are not beyond chance, whereas the human-supervised verdicts are beyond chance in both scopes. Two further patterns concern the curve as a whole.

\noindent\textbf{The human-supervised verdict is stable in direction.} Apart from the two specifications that combine hunk units with line-number tracking, no specification reports human-supervised code modified sooner under any estimator, and 42 of its 44 point estimates lie below 1. The estimator decides only how many of these differences are significant. For Devin, in contrast, the direction itself changes: 6 of its 22 across-repository point estimates and 15 of its 22 within-repository ones lie below 1.

\noindent\textbf{The observation window moves Devin across repositories.} RQ1 shows that requiring eight months of follow-up removes 61.7\% of Devin's lines and moves its line-level estimate from 0.896 to 1.704, because it changes which PRs enter the cohort (Section~\ref{sec:rq1}). Over the whole curve, the same cohort makes all 22 across-repository specifications report Devin code modified sooner under model-based errors, and 8 under repository clustering (Table~\ref{tab:window_curve}). Within repositories, all 22 specifications report no difference in the four windows of the curve, with median hazard ratios from 0.91 to 1.06. Dropping the same 142 PRs and censoring the rest at 150 days gives 5 ``sooner'' verdicts across repositories, so on the curve, too, the cohort rather than the follow-up moves the verdict. Against permuted labels, the median across-repository hazard ratio is beyond chance in every variant ($p = 0.008$ to $0.040$) except the full window ($p = 0.052$). The count of ``sooner'' verdicts under repository clustering is beyond chance only with eight months of follow-up ($p = 0.046$, Figure~\ref{fig:perm_null}b). Within repositories, no variant differs from chance ($p \ge 0.21$ for the median).

The human-supervised tier cannot be tested with eight months, because only 3 of its PRs have that much follow-up, but 760 of its 1,055 PRs have six months. Censored at 180 days, this cohort gives 9 ``retained longer'' verdicts across repositories and 8 within them (median hazard ratios 0.71 and 0.88), and none ``modified sooner''. The direction of the human-supervised verdict therefore does not depend on the cohort.

For human-supervised code, a single pipeline can change how large the effect looks, but not its direction. For Devin, a single pipeline can report any of three verdicts, depending on its unit, its window, and its variance estimator. In the full window, only three hunk-level ``sooner'' verdicts and one ``longer'' verdict survive clustering by repository, as many as permuted labels produce.

\begin{table}[t]
\centering
\caption{Verdicts of the 22 specifications per comparison and observation window, as longer / none / sooner, with the median hazard ratio of the 22 specifications in parentheses. All rows but the last are Devin code. Clustered: standard errors clustered by repository. Model-based: units treated as independent.}
\label{tab:window_curve}
\small
\resizebox{\linewidth}{!}{%
\begin{tabular}{lcccc}
\toprule
\textbf{Window} & \textbf{PRs} & \textbf{Across, clustered} & \textbf{Across, model-based} & \textbf{Within, clustered} \\
\midrule
Full ($\ge$ 5 months) & 412 & 1 / 18 / 3 \enskip (1.16) & 5 / 2 / 15 & 0 / 22 / 0 \enskip (0.91) \\
$\ge$ 8 months & 270 & 0 / 14 / 8 \enskip (1.28) & 0 / 0 / 22 & 0 / 22 / 0 \enskip (1.06) \\
$\ge$ 8 months, censored at 240 days & 270 & 0 / 18 / 4 \enskip (1.24) & 0 / 1 / 21 & 0 / 22 / 0 \enskip (1.03) \\
$\ge$ 10 months, censored at 300 days & 125 & 0 / 17 / 5 \enskip (1.43) & 0 / 1 / 21 & 0 / 22 / 0 \enskip (0.98) \\
\midrule
Full, without pdfme \#1059 & 411 & 0 / 19 / 3 \enskip (1.18) & 3 / 1 / 18 & 0 / 22 / 0 \enskip (0.91) \\
Full, without the 142 PRs, censored at 150 days & 270 & 0 / 17 / 5 \enskip (1.29) & 3 / 1 / 18 & 0 / 21 / 1 \enskip (1.09) \\
\midrule
Human-supervised, $\ge$ 6 months, censored at 180 days & 760 & 9 / 13 / 0 \enskip (0.71) & 20 / 1 / 1 & 8 / 14 / 0 \enskip (0.88) \\
\bottomrule
\end{tabular}}
\end{table}

\subsection{Published Designs Re-Run on Our Data}
\label{sec:emulation}

We re-run the design of each post-merge study in Table~\ref{tab:prior_designs} on our data, as closely as our data allow. These are approximations, not reproductions: the studies use other datasets, windows, and in two cases other outcomes. Table~\ref{tab:emulation} summarizes the results.

\begin{table*}[t]
\centering
\caption{Published post-merge designs re-run on our data. ``Reported'' gives each study's own headline result, ``On our data'' the result of its design here, and the last column the single change that moves it.}
\label{tab:emulation}
\footnotesize
\begin{tabular}{@{}>{\raggedright\arraybackslash}p{1.7cm}>{\raggedright\arraybackslash}p{2.6cm}>{\raggedright\arraybackslash}p{2.4cm}>{\raggedright\arraybackslash}p{2.8cm}>{\raggedright\arraybackslash}p{2.8cm}@{}}
\toprule
\textbf{Study} & \textbf{Design as re-run} & \textbf{Reported} & \textbf{On our data} & \textbf{Single change that moves it} \\
\midrule
Xia and Miller~\cite{xia2026these} & Agents pooled, line units, within-repository Cox & HR 1.11, not significant; corrective HR 1.49 & HR 0.92, $p = 0.68$: we agree & Corrective outcome as a cause-specific hazard, as in their study: HR 0.68 with keyword-first labels, 1.37 with LLM labels ($p = 0.22$ and $0.14$) \\
Sawada et al.~\cite{sawada2026extent} & File level, months 1 to 6, human share of later edits & Humans do about 83\% of maintenance & Humans make 88.2\%; Devin files 0.156 modifying commits per month, human 0.140, human-supervised 0.102 & Direction agrees; the share depends on the unit and window \\
Popescu et al.~\cite{popescu2026investigating} & Per agent, survival at 3 days, 1 week, 3 weeks, pooled across repositories; at the line level, since our data do not identify each PR's first commit & Human commits keep their added lines more often than every agent's (small effects) & At 3 weeks: Cursor 77.2\%, Claude Code 81.8\%, Devin 88.6\%, human 91.7\%, Codex 93.2\%, Copilot 96.1\%: of the four agents we share with them (Jules is not in AIDev), we agree for Claude Code and Devin; Cursor, which they do not study, is also below & Within repositories, only Copilot differs, and it is retained longer (HR 0.65, $p = 0.015$) \\
Kraishan~\cite{kraishan2026agents} & Per agent, PR churn on the most-changed file within 90 days, against same-repository humans & Claude Code lowest churn; Copilot lower; Codex and Devin at human level & Cliff's $\delta$~\cite{cliff1993dominance}: Codex $-0.11$, Copilot $-0.10$, Devin $+0.08$, Claude Code $+0.05$, Cursor $+0.04$: we agree on Copilot & With a repository model, no agent differs ($p \ge 0.38$) \\
Takerng\-saksiri et al.~\cite{takerngsaksiri2026finishes} & PR units, 30 days, Mantel-Haenszel OR~\cite{mantel1959statistical} over repositories with both groups; our outcome is any modification, not a verified fix & OR 1.62 for a follow-up fix; most fixes by the same agent & Mantel-Haenszel OR, agents pooled 0.86 ($p = 0.095$); Devin 0.77, human-supervised 0.92 & Not comparable (different outcome). The raw Devin gap (51.7\% vs.\ 46.0\% of PRs modified) reverses within the 31 shared repositories (crude, unstratified OR 0.86 for Devin) \\
\bottomrule
\end{tabular}
\par\smallskip\parbox{\linewidth}{\footnotesize Cause-specific hazard: a Cox model in which only corrective modifications count as events, and other modifications censor the line~\cite{therneau2000modeling}. Cliff's $\delta$: the probability that a value from one group exceeds a value from the other, minus the reverse, from $-1$ to $+1$~\cite{cliff1993dominance}. Mantel-Haenszel OR: an odds ratio pooled over repositories~\cite{mantel1959statistical}.}
\end{table*}

Where our data allow a comparison, we reproduce the direction of several published results. Like Xia and Miller~\cite{xia2026these}, we find no significant difference for pooled agent code within repositories (HR\,=\,0.92 here, 1.11 there). We agree with Sawada et al.~\cite{sawada2026extent} that humans do most of the maintenance. We agree with Popescu et al.~\cite{popescu2026investigating} that lines from two of the four agents we share with them (Claude Code and Devin) are modified sooner than human lines in a pooled comparison. We also agree with Kraishan~\cite{kraishan2026agents} that Copilot code churns less than human code. Each of these results, however, moves when a single choice changes. Xia and Miller's corrective result depends on the intent classifier. With keyword-first labels, pooled agent code receives less corrective maintenance than human code (HR\,=\,0.68), and with LLM labels more (HR\,=\,1.37, against their 1.49), although neither estimate is significant on our data. The per-agent differences reported by Popescu et al. and Kraishan disappear when agents are compared with humans in the same repositories, except that Copilot code is retained longer than human code. For Takerngsaksiri et al.~\cite{takerngsaksiri2026finishes}, the outcome differs, so the numbers are not comparable. The raw difference between Devin and human PRs, however, reverses within the repositories that contain both.

\begin{tcolorbox}[colback=gray!10, colframe=gray!50, title=Synthesis Finding]
Across all 88 specifications, the direction of the human-supervised verdict holds, whereas Devin code can be reported retained longer, no different, or modified sooner, depending above all on the variance estimator (15 or 3 of 22 ``sooner'' verdicts across repositories). Each published design changes its headline result with a single choice.
\end{tcolorbox}

\section{Discussion}
\label{sec:discussion}

RQ1 to RQ5 and the specification curve show that the verdict of a post-merge study of agent code depends on how the study measures it. In this section, we explain why the verdicts move, state what holds across all specifications, turn the results into a checklist for future studies, and discuss what they imply for researchers and practitioners.

\subsection{Why the Verdicts Move}

Five mechanisms explain most of the movement we observe.

\paragraph{What is counted as a modification.}
Line-number tracking records a line as modified when its position changes, for example when lines are inserted above it or its file is renamed. On the 276,314 units that all three trackers cover, it records about twice as many modifications as identity tracking, more so for human code than for Devin code, which moves Devin's estimate from 1.29 to 0.99. For human-supervised code, the tracker moves the estimate only from 0.67 to 0.77, and both are significant. Vendored and generated files inflate the count in the same way before any tracker is applied (46.5\% of human lines modified with them, 25.0\% without), and 18.8\% of content modifications touch only whitespace (RQ2).

\paragraph{Whose code dominates the sum.}
When each line counts once, a few large PRs decide the estimate. One Devin PR, \texttt{pdfme/pdfme\#1059}, holds 41.7\% of all Devin lines, and only 4.8\% of them are modified within the window. Devin's estimate is therefore 0.90 per line, 1.19 per PR, 1.56 per hunk, and 1.29 when lines are sampled within PRs (RQ1). Human-supervised code is less concentrated, since its five largest PRs hold 23.9\% of its lines, and its estimate stays below 1 under every weighting (0.89 per line, 0.72 per PR, and 0.83 per hunk). None of these weightings is wrong. They answer different questions (about lines, about contributions, or about blocks of code), and a study should state which one it asks.

\paragraph{Which code is followed for how long.}
The observation window decides which PRs enter the cohort. Requiring eight months of follow-up removes 61.7\% of Devin's lines, including the PR above, and turns Devin's line-level estimate across repositories from 0.90 into 1.70. The change comes from the cohort, not from the longer follow-up: dropping the same PRs and censoring the rest at 150 days gives 1.72, and dropping pdfme PR \#1059 alone gives 1.42. The window therefore acts on the same few large PRs as the unit. For human-supervised code, a six-month cohort keeps the direction of the verdict.

\paragraph{Where the code lives, and how uncertainty is estimated.}
The two tiers come from largely different repositories: 74.9\% of the repositories contain only one tier. Repositories with human-supervised code do not differ from the rest on any characteristic we measured, whereas repositories with Devin code have more contributors. A comparison across repositories is then partly a comparison of repository populations: on the 31 repositories that contain both Devin and human code, the pooled Devin estimate is 0.62 rather than 0.90. The same clustering governs the uncertainty. Lines of one repository are not independent, and a model that treats 644,177 lines as independent reports intervals about 50 times too narrow. With such intervals, 15 of Devin's 22 across-repository specifications report it modified sooner. With intervals clustered by repository, 3 do, about as many as permuted labels produce. For human-supervised code, 13 of 22 ``retained longer'' verdicts remain after clustering, more than chance would give.

\paragraph{How intent is inferred.}
Keyword rules classify a commit by the words in its message, and mostly add corrective labels: 584 commits that they label corrective are labeled adaptive or perfective by all three LLMs, against 57 in the other direction. For human-supervised code, the only difference comes from the human labels, and no classifier reproduces it on the unlabeled commits. For Devin code, keyword labels show a difference from human code, whereas no LLM approach does. The same mechanism moves a published result: Xia and Miller's~\cite{xia2026these} corrective hazard ratio becomes 0.68 with keyword-first labels and 1.37 with LLM labels on our data.

\subsection{What Holds Across Specifications}
\label{sec:disc_holds}

Some conclusions survive every choice we vary, and they are the ones we recommend that readers take from this study.

\textit{Human-supervised code is retained longer than human code, mostly because of Copilot.} With errors clustered by repository, it is retained longer in 13 of 22 across-repository and 11 of 22 within-repository specifications, more than chance would give in both scopes ($p = 0.008$), and 42 of its 44 point estimates lie below 1. Within repositories, the difference is significant under the structural definition (HR\,=\,0.717, 95\% CI [0.477, 0.921]), also after a Bonferroni correction, when lines are followed past leaf-only edits, and with file covariates (0.742, [0.487, 0.967]). Under the syntactic definition with each line counted once, it is not significant. Per agent, only Copilot's interval excludes 1 (0.653, [0.459, 0.954]).

\textit{Devin code shows no difference that survives the variance estimator.} Across repositories, its estimate on all lines with each line counted once is not significant (Figure~\ref{fig:rq2_def}), and its ``modified sooner'' verdicts under repository clustering are no more frequent than under permuted labels. Within repositories, all 22 specifications, in every observation window and with file covariates, report no difference, and 15 of their 22 point estimates lie below 1. With 31 shared repositories, however, the design detects only hazard ratios below 0.32 or above 3.13, so this is the absence of a detectable difference, not evidence that Devin code is retained as long as human code. The early-churn signal for Devin code is the one result we found that depends on a window chosen after seeing the data: it replicates in none of 5 held-out splits, and we treat it as a hypothesis.

\textit{Per modifying commit, humans maintain agent code.} Human developers make 87.7\% of the modifying commits to agent code (88.6\% for human-supervised and 85.7\% for Devin code), in line with Sawada et al.~\cite{sawada2026extent} and with their design re-run on our data (88.2\%). When an agent modifies agent code, it is almost always the agent that wrote it, as Takerngsaksiri et al.~\cite{takerngsaksiri2026finishes} find, and for Devin it usually comes within days of the merge. The two results are compatible: when an agent follows up, it is usually the same agent, but most follow-up work is done by people under every unit and window we test (agents make at most 29.9\% in Table~\ref{tab:modifier_variants}).

\textit{The data do not show that agent code needs more defect fixes.} No classifier's accuracy differs by tier. Once the LLM shares are corrected for classification error, with one confusion matrix or one per tier, the corrective shares of the three groups overlap (for example 24.8\%, 26.1\%, and 23.1\% with one matrix per tier).

\subsection{An Operationalization Checklist}
\label{sec:checklist}

Table~\ref{tab:checklist} turns our results into ten recommendations for post-merge studies of agent code. Each recommendation addresses a choice that moved a verdict or an effect size on our data. The last column gives the size of that movement, so that a reader can judge how much a study that departs from the recommendation may be affected.

\begin{table*}[p]
\centering
\caption{Operationalization checklist for post-merge studies of agent code, with the effect of each choice on our data.}
\label{tab:checklist}
\footnotesize
\setlength{\tabcolsep}{4pt}
\begin{tabular}{@{}>{\raggedright\arraybackslash}p{2.2cm}>{\raggedright\arraybackslash}p{5.4cm}>{\raggedright\arraybackslash}p{5.6cm}@{}}
\toprule
\textbf{Choice} & \textbf{Recommendation} & \textbf{Effect of the choice on our data} \\
\midrule
1. Line tracking & Follow line identity rather than line location (for example, reverse blame with rename and copy detection along the first-parent chain), verify each modification in its diff, and report the estimate under a second tracker. & Recorded modifications about double under line-number tracking, and only 35.9\% of its ``modified'' claims correct in a manual check, against 98.7\% under identity tracking; Devin HR 1.29 with identity tracking, 0.99 with line-number tracking. \\
2. Files, sampling, and weighting & Remove vendored and generated files. Do not cap or sample lines within a PR. Report line-weighted and PR-weighted estimates side by side, and the share of lines from the largest PRs. & HR 0.38 to 0.39 with vendored code, about 0.9 without; Devin HR 0.90 per line, 1.19 per PR, 1.56 per hunk, 1.29 with caps; one PR holds 41.7\% of Devin lines. \\
3. Modification definition & Report modification events by type (content change, line deleted, file deleted, whitespace only), with an AST-structural definition as a sensitivity check. & Devin HR 0.90 syntactic, 0.85 structural (across); human-supervised within 0.81 (not significant) and 0.72 (significant); 18.8\% of content modifications are whitespace only. \\
4. Intent classifier & Prefer LLM classifiers over keyword rules when the commits are too many to label by hand. Validate the classifier against a stratified human sample, check that its accuracy does not differ by tier, report $\kappa$ with a CI, label by hand the commits on which the models disagree, correct shares for classification error, and check any human-label difference on the unlabeled commits. & $\kappa$ 0.38 for keyword rules, 0.69 to 0.72 for LLMs with the diff, 0.68 to 0.74 with the message alone; unanimous LLM labels 86\% correct, two-to-one majorities 54\%; keyword rules alone find that Devin code differs. \\
5. Comparison & Compare within repositories: stratify by repository, use cluster-bootstrap intervals, and report the number of contributing repositories and the minimum detectable effect. & Human-supervised HR 0.89 across, 0.81 within (0.72 structural); Devin HR 0.90 across, 1.06 within, and 0.62 pooled on the 31 shared repositories; within repositories, only hazard ratios beyond 0.32 and 3.13 are detectable for Devin. \\
6. Standard errors & Cluster standard errors by repository in every model, also across repositories, and test a specification curve against permuted labels. & Human-supervised across: 21 of 22 ``longer'' with model-based errors, 13 clustered by repository. Devin across: 15 of 22 ``sooner'' with model-based errors, 9 clustered by PR, 3 clustered by repository; line-level CI [0.882, 0.910] becomes [0.425, 1.889]. \\
7. Observation window & Report the follow-up window and the cohort it implies, and the estimate under a longer window. & Eight months of follow-up drops 61.7\% of Devin lines; Devin line HR 0.90 becomes 1.70; ``sooner'' specifications across repositories 3 become 8 (clustered). Dropping the same PRs and censoring the rest at 150 days gives 1.72, so the change comes from the cohort. \\
8. Maintainer & State the unit and window of follow-up work and the modifier rule (author or co-author trailer), and report the self-modification share separately. & Humans make 87.7\% of modification commits to agent code (83.7\% with co-author trailers); agents make 13.3\% (human-supervised) and 25.5\% (Devin) of follow-up PRs within 30 days. \\
9. Time & Check proportional hazards; if it fails, report piecewise hazard ratios with clustered errors, and fix or test the time windows on held-out repositories. & Human-supervised HR 1.20 in the first 30 days and 0.45 (CI [0.23, 0.90]) from 30 to 90 days; Devin early churn replicates in none of 5 splits. \\
10. Reporting & Publish a specification curve over these choices rather than a single pipeline, with the estimator stated for every interval. & Devin across repositories: 1 / 18 / 3 (clustered) or 5 / 2 / 15 (model-based) longer / none / sooner. \\
\bottomrule
\end{tabular}
\end{table*}

On our data, items 2, 5, 6, and 7 matter most for the survival verdict, because each of them alone changes the Devin estimate or its verdict. Items 3 and 4 change the size of the retention effect, whether the within-repository human-supervised effect is detected, and the verdict on intent. Item 1 changes the point estimate for Devin, though not a significant verdict once errors are clustered. Item 8 decides whether agents appear to maintain their own code (RQ5). Items 9 and 10 do not change any single estimate, but they show the reader how much the estimate depends on choices that a single pipeline hides.

\subsection{Implications for Researchers}

Firstly, disagreements between post-merge studies should be read through their measurement choices. The published studies differ in all of these choices at once (Table~\ref{tab:prior_designs}). Our re-runs show that, on one dataset, their designs reproduce several of their own headline results, and that each result moves with a single choice (Table~\ref{tab:emulation}). Before a disagreement is read as a difference between coding agents or datasets, the choices in Table~\ref{tab:checklist} should be compared.

Secondly, the variance estimator should be treated as a measurement choice. Line-level data contain hundreds of thousands of units but only a few hundred repositories, and with model-based errors almost any difference becomes significant. On our data, clustering by repository makes every across-repository estimate on all lines, with each line counted once, non-significant, and it leaves Devin with no more ``sooner'' verdicts than permuted labels produce. Studies should state the estimator for each interval and cluster by repository.

Thirdly, collaboration modes should be separated, and the fully-autonomous tier scoped to its tools. Human-supervised and fully-autonomous agents differ in how their code is weighted, where it lives, and who follows it up, and within the human-supervised tier the effect is carried by one agent, Copilot. Because AIDev's fully-autonomous tier contains only Devin, findings about this tier are findings about Devin until other autonomous agents are observed.

Fourthly, LLM classifiers can replace keyword rules for commit intent, if they are validated. On our data, three LLMs agree with human labels at $\kappa = 0.69$ to $0.72$, about as well as two human raters agree with each other, against 0.38 for keyword rules. Their accuracy does not differ by tier, and they reach the same agreement from the commit message alone. Keyword rules, in contrast, create a Devin difference that no LLM and no human label confirms. Studies that cannot label every commit by hand should therefore prefer validated LLM classifiers. They should label a stratified sample by hand, report $\kappa$ against it, use more than one model, and correct the shares for classification error (Table~\ref{tab:checklist}, item~4). They should also label by hand the commits on which the models disagree. On our data, a label on which all three LLMs agree matches the human label for 86\% of commits, and a two-to-one majority for only 54\%.

Finally, power should be reported together with null results. A within-repository comparison is limited by the number of repositories that contain both groups: 128 for human-supervised code and 31 for Devin code in our cohort. Studies should report this number, the interval, and the minimum detectable effect, so that ``no difference'' is not read as ``equal''. Our Devin interval, [0.409, 2.076], is consistent with a 59\% lower and a twofold higher hazard.

\subsection{Implications for Practitioners}

Firstly, practitioners should ask how a retention or churn figure for AI code was measured. Such figures, whether in studies or in tools, depend on the tracker, the weighting, the window, and the comparison. On our data, the same Devin code can be reported as retained longer or as modified sooner than human code. Such figures should be compared only with human code from the same project, measured in the same way. Within one project the comparison can go either way: in \texttt{wandb/weave}, 17.2\% of 332 human-supervised lines are modified against 93.2\% of 296 human lines, while in \texttt{lyqht/mini-qr} 73.3\% of 30 human-supervised lines are modified against 1.3\% of 1,068 human lines.

Secondly, teams should plan for human maintenance of agent code. Adopting coding agents does not move maintenance to the agents: people make almost nine in ten of the later modifying commits to agent code, so review and maintenance effort for agent code should be budgeted as for human code.

Thirdly, the follow-up work of autonomous agents deserves the same review as their first contribution. Devin's modifications of its own code come quickly (median 2.8 days after merge, with 30 of 85 commits within a day). In a sample of 20, 14 belong to the work of the original PR, either the same feature or fix or the next item of a series of similar PRs. Copilot's self-modifications come weeks later (median 42.0 days), and in a sample of 20, 14 are later maintenance rather than part of the original task. We conjecture that a follow-up from the same agent may be reviewed less carefully than a new contribution.

\section{Threats to Validity}
\label{sec:threats}

We organize threats following the construct, internal, external, and conclusion validity framework~\cite{wohlin2012experimentation}, and group related threats within each category. Table~\ref{tab:scope} summarizes the scope of our claims.

\begin{table}[tbp]
\centering
\caption{Scope of our claims.}
\label{tab:scope}
\footnotesize
\begin{tabular}{@{}>{\raggedright\arraybackslash}p{2.2cm}>{\raggedright\arraybackslash}p{5.0cm}>{\raggedright\arraybackslash}p{4.4cm}@{}}
\toprule
\textbf{Claim} & \textbf{Applies to} & \textbf{Not claimed} \\
\midrule
Sensitivity of verdicts & Line-level post-merge designs in general; one dataset suffices to show that a choice can change a verdict. & That every dataset is equally sensitive to every choice. \\
Agent effects & Merged AIDev PRs in 215 repositories, merged before August 2025 and followed until December 2025, without vendored and generated files. & Effects of agents in general, or of fully-autonomous agents other than Devin. \\
No difference & The specification for which it is reported, with its interval. & That agent code and human code are retained equally long. \\
Intent & Relative differences between tiers. & Exact intent shares of any group. \\
\bottomrule
\end{tabular}
\end{table}

\subsection{Construct Validity}

Construct validity concerns whether our measures capture the concepts they are meant to capture: who wrote and who modified a line, whether and how a line was modified, and why.

\paragraph{Attribution of authors and modifiers.}
We acknowledge that AIDev~\cite{li2025rise} identifies agent-attributed PRs from metadata (e.g., branch prefixes, bot accounts, and co-author lines), not from ground-truth authorship, so the human baseline may contain AI-assisted code. To mitigate this, following prior work that uses AIDev attribution~\cite{xia2026these,sawada2026extent}, we compare explicitly agent-attributed contributions with the standard developer workflow. We also quantify the residual. In 200 human-baseline PRs, of which two raters labeled 100, an estimated 5.5\% (95\% CI [2.8, 8.9]) show that an AI tool wrote the code or the description although AIDev did not flag them. A sensitivity analysis covers nine scenarios: three contamination levels crossed with three assumptions about how often contaminated code is modified. It keeps the approximate hazard ratio between 0.78 and 0.87 for human-supervised code and between 0.84 and 0.93 for Devin code. No scenario therefore changes the direction of the across-repository point estimate. We also acknowledge that vendor names inside logins can label human staff accounts as agents. To reduce this risk, we classify modifiers with an explicit rule (i.e., GitHub App accounts, \texttt{-bot} accounts, and a list of known coding-agent and automation accounts) and audited every distinct modifier login by hand.

\paragraph{Tracking.}
We acknowledge that reverse blame can confuse a line with an identical line elsewhere in the file, and that line-level blame attributes some lines to the wrong commit~\cite{german2019cregit}. Therefore, we verify every modification in the diff of the modifying commit, exclude the 680 lines (0.11\%) whose next commit does not visibly change them, and use merge commits that lie within one day of AIDev's merge time for every PR. We also compare our tracker with two others on the same units and check all three by hand (Section~\ref{sec:operationalization}). None of the three trackers is a ground truth, which is why we report the verdict under each of them. Identity tracking slightly understates how often lines are modified. The manual check also excludes the long and large file histories in which trackers are most likely to fail, so the accuracy of every tracker may be lower on those files.

\paragraph{Defining modification.}
We acknowledge that neither the syntactic nor the AST-structural definition is a perfect proxy for a meaningful change. To reduce its impact, following work that separates non-essential from structural changes~\cite{kawrykow2011non,fluri2007change}, we report both definitions and report modification events by type. We also follow lines past leaf-only edits, and show that dropping the 16.1\% to 19.8\% of modified lines without an AST status leaves the result unchanged (RQ2). A manual check of 383 modifications agrees with the AST rule in 93.7\% of cases ($\kappa = 0.60$). The rule is less reliable for anonymous functions and module-level statements, where reformatting can be counted as structural (Section~\ref{sec:operationalization}).

\paragraph{Classifying intent.}
We acknowledge that keyword rules operationalize Swanson's taxonomy~\cite{swanson1976dimensions} through message vocabulary, which need not reflect a commit's primary purpose, and that LLM classifiers carry their own biases. To mitigate this, following the validation practice of prior commit-classification work~\cite{levin2017boosting}, we evaluate every classifier against 618 human-labeled commits of the corpus (Cohen's $\kappa$~\cite{cohen1960coefficient}: 0.38 for keyword rules, 0.69 to 0.72 for the LLM approaches). We report results under all approaches rather than designating one. We test whether their accuracy differs by tier, and correct each LLM's shares with its confusion matrix, pooled and per tier. We also acknowledge that the human labels come from the first author and the LLMs agree with the first author at $\kappa = 0.63$ to $0.65$. The human labels, not the classifiers, therefore limit how precisely intent can be measured. Small differences in the human labels, such as the human-supervised one in RQ4 ($p = 0.038$), are fragile. To reduce its impact, the first author followed a written guide with the same category definitions as the LLM prompt and had no access to classifier output. The labels cover both strata of the corpus with a worst-case margin of error of $\pm$4.87\% at 99\% confidence, and the guide, labels, and notes are released for re-annotation. We therefore interpret intent results as relative differences between tiers, not as exact shares.


\paragraph{Validation samples.}
We acknowledge that two of our validation samples were drawn before the final cohort was mined. The review of 200 human-baseline PRs and the check of 383 AST classifications come from an earlier mining of 142 of the cohort's 215 repositories. Their error rates may therefore differ in the other 73 repositories. The intent labels do not have this limitation, because we topped them up from the new repositories and checked that every labeled commit belongs to the corpus. Finally, the first author coded the samples of agent self-modifications in RQ5 by comparing each commit with the description of the original PR. 

\subsection{Internal Validity}

Internal validity concerns whether the differences we attribute to tiers and measurement choices could be explained by other factors.

\paragraph{Confounding, selection, and tier composition.}
We acknowledge that a comparison of tiers is observational. Following Xia and Miller~\cite{xia2026these}, we stratify by repository, which controls for every repository-level characteristic, measured or not. Within repositories, tiers may still handle different kinds of files and tasks: human-supervised code sits more often in new and test files, which are modified later. Therefore, we add file age, prior file activity, new-file, and test-file covariates to the within-repository models, and the human-supervised structural estimate stays significant (RQ3), but the type of task remains unmeasured. Furthermore, our cohort contains merged PRs only, and agent PRs are rejected more often than human PRs~\cite{ehsani2026aicodingagentsfail}, so the agent code we observe is a positively selected subset. Finally, the fully-autonomous tier consists of Devin only, in 31 repositories, so tool-level and tier-level effects cannot be separated. AIDev also does not record how closely a developer supervised each of the four human-supervised agents (GitHub Copilot, Cursor, Claude Code, and OpenAI Codex). To reduce the impact of these threats, we limit our claims to merged contributions and to associations, report each tier and each agent separately, and scope fully-autonomous findings to Devin (Table~\ref{tab:scope}).

\paragraph{The specification curve.}
We acknowledge that a specification curve can be made to favor a conclusion by the choice of its dimensions~\cite{simonsohn2020specification}. To mitigate this, the dimensions we vary are exactly the choices made by the published studies in Table~\ref{tab:prior_designs} and by our earlier study, plus the sampling caps of the position-based miner (Section~\ref{sec:operationalization}) and the observation window. We report every specification we fitted, under three variance estimators, together with a permutation test of the whole curve. The curve covers measurement choices, not modeling choices such as covariates.

\subsection{External Validity}

External validity concerns how far our findings generalize beyond the studied repositories, coding agents, and period.

\paragraph{Dataset, repositories, and languages.}
Firstly, we cannot guarantee the generalizability of our agent effects. All our data come from one dataset, AIDev, and from 215 public GitHub repositories in which TypeScript (25.6\%) and Python (20.0\%) dominate. Closed-source projects, other languages, and later versions of rapidly evolving coding agents may behave differently. In particular, our fully-autonomous tier is a single agent from the AIDev period. Since then, other agents to which developers delegate whole tasks, such as Jules~\cite{julesAgent} and Kiro~\cite{kiroAgent}, have started to contribute to GitHub projects in large numbers. We counted the merged PRs opened by each agent's GitHub account from January to June 2026 with the GitHub search API on 8 November 2026: about 76,000 for Jules and 11,000 for Kiro. Our cohort cannot simply be extended in time to cover them: from January to June 2026, Devin opened PRs in only 6 of our 215 repositories, and 724 of its 780 PRs went to one of them. PR metadata also cannot replace the tier. In the same count, the share of an agent's PRs opened by the agent's own account ranges from 0\% for OpenAI Codex to 98\% for GitHub Copilot, so it does not tell how autonomously a PR was produced. The replication package describes how we counted.
To reduce this threat, we adapt established criteria for selecting engineered software projects~\cite{munaiah2017curating,xiao2025self} and compare tiers within repositories, which holds language and project practices fixed. Our main claim, that measurement choices change verdicts, needs only one dataset to be demonstrated. Whether the agent effects themselves generalize needs replication on other datasets, such as those of Xia and Miller~\cite{xia2026these}.

\paragraph{Observation window.}
Secondly, we acknowledge that every line has at least five months of follow-up, which is short for code that may live for years. Longer windows can be tested only for Devin code, because they restrict the cohort to PRs merged earlier. They change the across-repository Devin verdict (Section~\ref{sec:rq1}, Step~5, and Section~\ref{sec:synthesis}): with eight months of follow-up, its line-level hazard ratio is 1.70 instead of 0.90, because the restriction removes 61.7\% of its lines. Dropping the same PRs and censoring the rest at 150 days gives 1.72, so the change comes from the cohort rather than from the follow-up length. Within repositories, no window changes the Devin verdict. Only 3 human-supervised PRs have eight months of follow-up, but a six-month cohort of 760 of its PRs keeps the direction of the human-supervised verdict. We added the window dimension after seeing the other results, so we test its variants against permuted labels.

\subsection{Conclusion Validity}
Conclusion validity concerns whether our statistical conclusions are justified by the data and could be reproduced.

\paragraph{Statistical assumptions and power.}
We acknowledge that Schoenfeld residual tests~\cite{schoenfeld1982partial} reject proportional hazards in every model, and that the hazard ratio changes with the time since merge. To mitigate this, we interpret Cox hazard ratios as averages over the window, report piecewise models with clustered errors, and test window-specific effects on held-out repositories (Section~\ref{sec:rq3}). We also acknowledge that units are not independent and that clusters are few. Only 128 repositories contribute to the within-repository human-supervised estimate and 31 to the Devin estimate, and clustered standard errors become unreliable with few clusters~\cite{cameron2015practitioner}. Therefore, we cluster every interval by repository, and report 500-resample cluster-bootstrap intervals for the headline within-repository estimates. We apply a Bonferroni correction for the two tiers ($\alpha = 0.025$), test the specification curve against permuted labels, and report the minimum detectable hazard ratio. With 31 repositories, the Devin comparison can detect only hazard ratios below 0.32 or above 3.13, so we report ``no detectable difference'' and never ``equal''.

\paragraph{Reproducibility of LLM labels.}
We acknowledge that serverless models can change or be withdrawn by the provider, and that temperature 0 does not guarantee identical outputs across runs. To mitigate this, we use three models from different families, set the temperature to 0, and record the model strings and the dates on which we called them. We also release every raw model output, finish reason, and token count, so that all reported intent results can be recomputed without calling the models again.

\section{Conclusion}
\label{sec:conclusion}

In this paper, we revisited the survival of agent-attributed code and measured how much measurement choices decide the verdict of post-merge studies. We started from the 854,139 lines that 2,802 merged PRs added to 215 open-source repositories of the AIDev dataset in our earlier study, and removed vendored and generated code, which left 644,177 lines. We then varied nine measurement choices: line tracking, the files kept, the unit and its weighting, the definition of a modification, the intent classifier, the comparison scope, the variance estimator, the observation window, and how the modifier is identified. Finally, we re-ran the designs of five published studies on our data.

Our results show that the measurement can decide the verdict, but not for every tier. Code from human-supervised agents, mostly from GitHub Copilot, is retained longer than human code in more specifications than chance would give, both across and within repositories. For this code, the choices change only how large the effect looks.

For Devin code, in contrast, defensible pipelines report it retained longer, no different, or modified sooner than human code. With model-based standard errors, 15 of its 22 across-repository specifications report it modified sooner, but with errors clustered by repository only 3 do, about as many as permuted labels produce. Within repositories, no specification detects a difference for Devin, although the 31 shared repositories can detect only large ones. Re-run on our data, each published design changes its headline result with a single choice.

The intent classifier decides the intent conclusion. Keyword rules find that Devin code is modified for different reasons than human code, whereas LLM classifiers, which agree with the human labels about as well as two human raters agree with each other, do not. LLM classifiers can therefore replace keyword rules where commits are too many to label by hand. Human developers make most modifications to agent code under every choice we test (87.7\% of the modifying commits), and when an agent modifies agent code, it is almost always the agent that wrote it.

We therefore recommend that post-merge studies follow the checklist in Table~\ref{tab:checklist}, cluster their standard errors by repository, and report a specification curve rather than a single pipeline. In the future, we plan to replicate the agent effects on other datasets and with other agents to which developers delegate whole tasks, such as Jules and Kiro. This first requires a way to measure how autonomously each PR was produced. We also plan to follow the code over longer windows, and to use controlled studies to separate the roles of generation quality, review, and task selection in the retention of human-supervised code.

\section*{Data Availability}
\label{sec:replication}

The replication package, with the scripts, data, and annotation guides needed to reproduce every result in this paper, is available at~\cite{replicationpackage}.

\section*{Acknowledgement}

We used AI-enabled tools (Claude and Grammarly) to assist with manuscript revisions. All AI-assisted content was reviewed and verified for accuracy by the first author.

\bibliographystyle{abbrv}
\bibliography{references}

\end{document}